\documentclass[opre,nonblindrev]{informs3_modified1}

\DoubleSpacedXI 

\usepackage{endnotes}
\usepackage{bm}
\usepackage{algorithm}
\usepackage{algorithmic}
\usepackage{diagbox}
\usepackage{multirow}
\usepackage{subcaption}

\let\footnote=\endnote
\let\enotesize=\normalsize
\def\notesname{Endnotes}%
\def\makeenmark{\hbox to1.275em{\theenmark.\enskip\hss}}
\def\enoteformat{\rightskip0pt\leftskip0pt\parindent=1.275em
  \leavevmode\llap{\makeenmark}}

\newcommand{\R}{\mathbb{R}}

\newcommand{\ceil}[1]{\lceil #1\rceil}
\newcommand{\tabincell}[2]{\begin{tabular}{@{}#1@{}}#2\end{tabular}}

\usepackage{natbib}
 \bibpunct[, ]{(}{)}{,}{a}{}{,}%
 \def\bibfont{\small}%
\TheoremsNumberedThrough     
\ECRepeatTheorems

\EquationsNumberedThrough    

\begin{document}


\RUNAUTHOR{Lam and Qian}

\RUNTITLE{Subsampling for Input Uncertainty Quantification}

\TITLE{Subsampling to Enhance Efficiency in Input Uncertainty Quantification}

\ARTICLEAUTHORS{%
\AUTHOR{Henry Lam}
\AFF{Department of Industrial Engineering and Operations Research, Columbia University, New York, NY 10027, \EMAIL{khl2114@columbia.edu}} 
\AUTHOR{Huajie Qian}
\AFF{Department of Industrial Engineering and Operations Research, Columbia University, New York, NY 10027, \EMAIL{h.qian@columbia.edu}}
} 

\ABSTRACT{%
In stochastic simulation, input uncertainty refers to the output variability arising from the statistical noise in specifying the input models. This uncertainty can be measured by a variance contribution in the output, which, in the nonparametric setting, is commonly estimated via the bootstrap. However, due to the convolution of the simulation noise and the input noise, the bootstrap consists of a two-layer sampling and typically requires substantial simulation effort. This paper investigates a subsampling framework to reduce the required effort, by leveraging the form of the variance and its estimation error in terms of the data size and the sampling requirement in each layer. We show how the total required effort can be reduced from an order bigger than the data size in the conventional approach to an order independent of the data size in subsampling. We explicitly identify the procedural specifications in our framework that guarantee relative consistency in the estimation, and the corresponding optimal simulation budget allocations. We substantiate our theoretical results with numerical examples.
}%



\maketitle

%


\section{Introduction}
\label{sec:intro}

Stochastic simulation is one of the most widely used analytic tools in operations research. It provides a flexible means to approximate complex models and to inform decisions. See, for instance, \cite{law1991simulation} for applications in manufacturing, revenue management, service and operations systems etc. In practice, the simulation platform relies on input models that are typically observed or calibrated from data. These statistical noises can propagate to the output analysis, leading to significant errors and suboptimal decision-making. In the literature, this problem is commonly known as input uncertainty or extrinsic uncertainty.

In conventional simulation output analysis where the input model is completely pre-specified, the statistical errors come solely from the Monte Carlo noises, and it suffices to account only for such noises in analyzing the output variability. When input uncertainty is present, such an analysis will undermine the actual variability. One common approach to quantify the additional uncertainty is to estimate the variance in the output that is contributed from the input noises (e.g., \cite{song2014advanced}); for convenience, we call this the \emph{input variance}. This quantity acts as an uncertainty measure which, when added together with the Monte Carlo variance, gives rise to the overall variance in the outputs. A refined decomposition of input variance across multiple input sources can be used to identify models that are overly ambiguous and flag the need of more data collection (e.g., \cite{song2014advanced}). Input variance also provides a building block to construct valid output confidence intervals (CIs) that account for combined input and simulation errors (e.g., \cite{cheng2004calculation}). Motivated by its central role in quantifying input uncertainty, this paper aims to study the efficient estimation of input variance.


In the literature, bootstrap resampling is a common approach for the above purpose. This applies most prominently in the nonparametric regime, namely when no assumptions are placed on the input parametric family. It could also be used in the parametric case (where more alternatives are available). For example, \cite{cheng1997sensitivity} proposes the variance bootstrap, and \cite{song2015quickly}  studies the consistency of this strategy on a random-effect model that describes the uncertainty propagation. A bottleneck with using bootstrap resampling in estimating input variances, however, is the need to ``outwash" the simulation noise, which often places substantial burden on the required simulation effort. More precisely, to handle both the input and the simulation noises, the bootstrap procedure typically comprises a two-layer sampling that first resamples the input data (i.e., \emph{outer} sampling), followed by running simulation replications using each resample (i.e., \emph{inner} replications). Due to the reciprocal relation between the magnitude of the input variance and the input data, the input variance becomes increasingly small as the input data size increases. This deems the control of the relative estimation error increasingly expensive, and requires either a large outer bootstrap size or inner replication size to extinguish the effect of simulation noises.




The main goal of this paper is to investigate \emph{subsampling} as a simulation saver for input variance estimation. This means that, instead of creating distributions by resampling a data set of the full size, we only resample (with or without replacement) a set of smaller size. We show that a judicious use of subsampling can reduce the total simulation effort from an order bigger than the data size in the conventional two-layer bootstrap to an order independent of the data size, while retaining the estimation accuracy. This approach leverages the interplay between the form of the input variance and its estimation error, in terms of the data size and the sampling effort in each layer of the bootstrap. On a high level, the subsample is used to estimate an input variance as if less data are available, followed by a correction of this discrepancy in the data size by properly rescaling the input variance. We call this approach \emph{proportionate subsampled variance bootstrap}. We explicitly identify the procedural specifications in our approach that guarantee estimation consistency, including the minimally required simulation effort in each layer. We also study the theoretical behavior of our estimation error, in relation to the simulation effort allocation in these layers as well as the input data and subsample sizes, which in turn reveals the optimal configurations and provides implementation guidance.



In the statistics literature, subsampling has been used as a remedy for situations where the full-size bootstrap does not apply, due to a lack (or undeterminability) of uniform convergence required for its statistical consistency, which relates to the functional smoothness or regularity of the estimators (e.g.,  \cite{politis1994large}). Subsampling has been used in time series and dependent data~(e.g., \cite{politis1999,hall1995blocking,datta1995bootstrap}), extremal estimation (e.g., \cite{bickel2008choice}), shape-constrained estimation (e.g., \cite{sen2010inconsistency}) and other econometric contexts (e.g., \cite{abadie2008failure,andrews2009validity,andrews2010asymptotic}). In contrary to these works, our subsampling approach is introduced to reduce the simulation effort faced by the two-layer sampling necessitated from the presence of both the input and simulation noises. In other words, we are not concerned about the issue of uniform convergence, but instead, we aim to distort the relation between the required simulation effort and data size in a way that allows more efficient deconvolution of the effects of the two noises. We also note that, as we will use resampling with replacement (instead of without replacement), our approach is closer to the so-called $m$ out of $n$ bootstrap (\cite{bickel1997resampling,bickel2008choice}). For coherence, throughout the paper we use the term subsampling broadly to indicate a bootstrap with a smaller resample size than the original data size.

We close this introduction with a brief review of other related work in input uncertainty. In the nonparametric regime (the focus of this paper), besides \cite{cheng1997sensitivity} and \cite{song2015quickly} that study bootstrap-based estimation of the input variance, \cite{barton1993uniform} and \cite{barton2001resampling} investigate the percentile bootstrap to construct CIs (i.e., the CI limits are determined from the quantiles of the bootstrap distributions). Like variance bootstrap, percentile bootstrap also encounters two-layer sampling that requires substantial simulation efforts. \cite{yi2017efficient} investigates adaptive budget allocation policies based on ranking and selection to reduce simulation cost in the percentile bootstrap, and empirically shows the computational advantage of their approach. On the other hand, contrary to this paper, they do not investigate the required simulation efforts in relation to the input data size. \cite{lam2016empirical,lam2017optimization} study the use of empirical likelihood as an optimization-based alternative to the percentile bootstrap, which requires simulation efforts to estimate the gradient information that remain substantial. Beyond the frequentist regime considered in this paper, \cite{xlz18} studies nonparametric Bayesian methods based on Dirichlet process mixtures to estimate the variance contributed from input uncertainty and construct CIs. \cite{glasserman2014robust,hu2012robust,lam2016robust} and \cite{ghosh2019robust} study input uncertainty from a robust optimization viewpoint, where they compute worst-case bounds subject to constraints or so-called uncertainty sets that represent partial beliefs on unknown distributions. In the parametric regime, \cite{barton2013quantifying} and \cite{xie2016multivariate} investigate the basic bootstrap with a metamodel built in advance, a technique known as the metamodel-assisted bootstrap. \cite{xie2016multivariate} and \cite{biller2011accounting} study multivariate input uncertainty assuming a parametric dependency structure in the form of product-moment correlations. \cite{cheng1997sensitivity} studies the delta method, and \cite{cheng1998two,cheng2004calculation} reduce its computation burden via the so-called two-point method. \cite{lin2015single} and \cite{song2019input} study regression approaches to estimate sensitivity coefficients which are used to apply the delta method, generalizing the gradient estimation method in \cite{wieland2006stochastic}. \cite{zhu2020risk} studies risk criteria and computation to quantify parametric uncertainty. Finally, \cite{chick2001input}, \cite{zouaoui2003accounting}, \cite{zouaoui2004accounting} and \cite{xie2014bayesian} study variance estimation and interval construction from a Bayesian perspective. We comment that although the exposition in this paper focuses on the nonparametric setting, the same idea of subsampling can be adapted naturally to the parametric setting, with similar advantages in computational efficiency. For general surveys on input uncertainty, readers are referred to \cite{barton2002panel}, \cite{henderson2003input}, \cite{chick2006bayesian}, \cite{barton2012tutorial}, \cite{song2014advanced}, \cite{lam2016advanced}, and \cite{nelson2013foundations} Chapter 7.

The remainder of the paper is as follows. Section \ref{sec:motivation} introduces the input uncertainty problem and explains the simulation complexity bottleneck in the existing bootstrap schemes. Section \ref{sec:overview} presents our subsampling idea, procedures and the main statistical results. Section \ref{sec:var boot} discusses the key steps in our theoretical developments. Section \ref{sec:numerics} reports our numerical experiments. Section \ref{sec:conclusion} concludes the paper. All proofs are relegated to the Appendix.

\section{Problem Motivation}\label{sec:motivation}
This section describes the problem and our motivation. Section \ref{sec:input} first describes the input uncertainty problem, Section \ref{sec:bootstrap} presents the existing bootstrap approach, and Section \ref{sec:barrier} discusses its computational barrier, thus motivating our subsampling investigation. We aim to provide intuitive explanations in this section, and defer mathematical details to later sections.

\subsection{The Input Uncertainty Problem}\label{sec:input}
Suppose there are $m$ independent input processes driven by input distributions $F_1,F_2,\ldots,F_m$. We consider a  generic performance measure $\psi(F_1,\dots,F_m)$ that is simulable, i.e., given the input distributions, independent unbiased replications of $\psi$ can be generated in a computer. As a primary example, think of $F_1$ and $F_2$ as the interarrival and service time distributions in a queue, and $\psi$ is some output measure such as the mean queue length averaged over a time horizon. Our study also applies when the $F_i$'s are multivariate distributions.

The input uncertainty problem arises in situations where the input distributions $F_1,\ldots,F_m$ are unknown but real-world data are available. One then has to use their estimates $\widehat{F}_1,\ldots,\widehat{F}_m$ to drive the simulation. Denote a point estimate of $\psi(F_1,\ldots,F_m)$ as $\bar\psi(\widehat F_1,\ldots,\widehat F_m)$, where typically we take
$$\bar\psi(\widehat F_1,\ldots,\widehat F_m)=\frac{1}{q}\sum_{r=1}^{q}\hat\psi_r(\widehat F_1,\ldots,\widehat F_m)$$
with $\hat\psi_r(\widehat F_1,\ldots,\widehat F_m)$ being a conditionally unbiased simulation replication driven by $\widehat F_1,\ldots,\widehat F_m$. This point estimate is affected by both the input statistical noises and the simulation noises. By conditioning on the estimated input distributions (or viewing the point estimate as a random effect model with uncorrelated input and simulation noises), the variance of $\bar\psi(\widehat F_1,\ldots,\widehat F_m)$ can be expressed as
$$\mathrm{Var}[\bar\psi(\widehat F_1,\ldots,\widehat F_m)]=\sigma^2_I+\sigma^2_S$$
where
\begin{equation}\label{input_var}
\sigma_I^2=\mathrm{Var}[\psi(\widehat F_1,\ldots,\widehat F_m)]
\end{equation}
is the input variance, and
$$\sigma^2_S=\frac{\mathbb E[\mathrm{Var}[\hat\psi_r(\widehat F_1,\ldots,\widehat F_m)|\widehat F_1,\ldots,\widehat F_m]]}{q}$$
is the variance contributed from the simulation noises. Assuming that the estimates $\widehat F_i$'s are consistent in estimating $F_i$'s, then, as input data sizes grow, $\sigma_S^2$ is approximately $\mathrm{Var}[\hat\psi_r(F_1,\ldots,F_m)]/q$ and can be estimated by taking the sample variance of all simulation replications (see, e.g., \cite{cheng1997sensitivity}). On the other hand, $\sigma^2_I$ signifies the output variance contributed solely from the input data noises, assuming a fully accurate evaluation of the performance measure $\psi$. Estimating $\sigma_I^2$ is the key and the challenge in quantifying input uncertainty, which is the focus of this paper.
%

Before going into details, we discuss two conceptual properties on $\sigma_I^2$ that would be relevant in motivating and pinpointing our study. First, suppose further that for each input model $i$, we have $n_i$ i.i.d. data $\{X_{i,1},\dots,X_{i,n_i}\}$ generated from the distribution $F_i$. When $n_i$'s are large, typically the overall input variance $\sigma_I^2$ is decomposable into
\begin{equation}
\sigma_I^2\approx\sum_{i=1}^m\frac{\sigma_i^2}{n_i}\label{var decompose}
\end{equation}
where $\sigma_i^2/n_i$ is the variance contributed from the data noise for model $i$, with $\sigma_i^2$ being a constant. In the parametric case where $\widehat F_i$ comes from a parametric family containing the estimated parameters, this decomposition is well known from the delta method (\cite{asmussen2007stochastic}, Chapter 3). Here, $\sigma_i^2/n_i$ is typically $\nabla_i\psi'\Sigma_i\nabla_i\psi$, where $\nabla_i\psi$ is the collection of sensitivity coefficients, i.e., the gradient, with respect to the parameters in model $i$, and $\Sigma_i$ is the asymptotic estimation variance of the point estimates of these parameters (scaled reciprocally with $n_i$). In the nonparametric case where the empirical distribution $\widehat{F}_i(x):=\sum_{j=1}^{n_i}\delta_{X_{i,j}}(x)/n_i$ is used (where $\delta_{X_{i,j}}$ denotes the delta measure at $X_{i,j}$), \eqref{var decompose} still holds under mild conditions (e.g., Propositions \ref{input_var_order} and \ref{input_var:tight} in the sequel). In this setting the quantity $\sigma_i^2$ is equal to $\mathrm{Var}_{F_i}[g_i(X_i)]$, where $g_i(\cdot)$ is the influence function (\cite{hampel1974influence}) of $\psi$ with respect to the distribution $F_i$, whose domain is the value space of the input variate $X_i$, and $\mathrm{Var}_{F_i}[\cdot]$ denotes the variance under $F_i$. The influence function can be viewed as a functional derivative taken with respect to the probability distributions $F_i$'s (see \cite{serfling2009approximation}, Chapter 6), and dictates the first-order asymptotic behavior of the plug-in estimate of $\psi$. Although the mathematical form of $\sigma_i^2$'s is known, it relies on gradient information that needs to be estimated via simulation itself. Moreover, in the nonparametric case, the gradient dimension in a sense grows with the data size. Thus directly using the delta method in this case could be challenging. In our subsequent developments, we focus on the nonparametric case, both because this is more challenging, and also that this can be viewed as a generalization of the parametric case by viewing the ``parameter" simply as a function of $F_i$'s.


Second, under further regularity conditions, a Gaussian approximation holds for $\bar\psi(\widehat F_1,\ldots,\widehat F_m)$ so that
\begin{equation}\label{CLT1}
\bar\psi(\widehat F_1,\ldots,\widehat F_m)\pm z_{1-\alpha/2}\sqrt{\sigma_I^2+\sigma_S^2}
\end{equation}
is an asymptotically tight $(1-\alpha)$-level CI for $\psi(F_1,\ldots,F_m)$, where $z_{1-\alpha/2}$ is the standard normal $1-\alpha/2$ quantile. This CI, which provides a bound-based alternative to quantify input uncertainty, again requires a statistically valid estimate of $\sigma_I^2$ or $\sum_{i=1}^m\sigma_i^2/n_i$ (and $\sigma_S^2$). In this paper we primarily focus on the estimation of $\sigma_I^2$ and how our proposed approach substantially improves upon previous methods in this regard. Naturally, the improved estimate of $\sigma_I^2$ also translates into a better CI when using \eqref{CLT1}. We caution, however, that an optimal procedural configuration to estimate $\sigma_I^2$ does not necessarily correspond to an optimal configuration in constructing the CI, as the performance of the latter is measured by different criteria such as coverage or half-width (such a difference in optimally estimating variance versus CI has also been observed in other contexts such as time series (\cite{sun2008optimal})). Nonetheless, we will show that a direct plug-in of our new estimator of $\sigma_I^2$ into \eqref{CLT1} is already enough to significantly outperform conventional bootstrap-based CIs suggested in the literature, both theoretically and also supported by consistent empirical evidence.


Next we will discuss bootstrap resampling, the commonest estimation technique that forms the basis of our comparison.

\subsection{Bootstrap Resampling}\label{sec:bootstrap}
Let $\widehat{F}_i^*$ represent the empirical distribution constructed using a bootstrap resample from the original data $\{X_{i,1},\ldots,X_{i,n_i}\}$ for input $F_i$, i.e., $n_i$ points drawn by uniformly sampling with replacement from $\{X_{i,1},\ldots,X_{i,n_i}\}$. The bootstrap variance estimator is $\mathrm{Var}_*[\psi(\widehat{F}^*_1,\ldots,\widehat{F}^*_m)]$, where $\mathrm{Var}_*[\cdot]$ denotes the variance over the bootstrap resamples from the data, conditional on $\widehat{F}_1,\ldots,\widehat{F}_m$.

The principle of bootstrap entails that $\mathrm{Var}_*[\psi(\widehat{F}^*_1,\ldots,\widehat{F}^*_m)]\approx \mathrm{Var}[\psi(\widehat{F}_1,\ldots,\widehat{F}_m)]=\sigma_I^2$. Here $\mathrm{Var}_*[\psi(\widehat{F}^*_1,\ldots,\widehat{F}^*_m)]$ is obtained from a (hypothetical) infinite number of bootstrap resamples and simulation runs per resample. In practice, however, one would need to use a finite bootstrap size and a finite simulation size. This comprises $B$ conditionally independent bootstrap resamples of $\{\widehat F_1^*,\ldots,\widehat F_m^*\}$, and $R$ simulation replications driven by each realization of the resampled input distributions. This generally incurs two layers of Monte Carlo errors.

Denote $\hat\psi_r(\widehat F_1^b,\ldots,\widehat F_m^b)$ as the $r$-th simulation run driven by the $b$-th bootstrap resample $\{\widehat F_1^b,\ldots,\widehat F_m^b\}$. Denote $\bar\psi^b$ as the average of the $R$ simulation runs driven by the $b$-th resample, and $\bar{\bar\psi}$ as the grand sample average from all the $BR$ runs. An unbiased estimator for $\mathrm{Var}_*[\psi(\widehat{F}^*_1,\ldots,\widehat{F}^*_m)]$ is given by
\begin{equation}
\frac{1}{B-1}\sum_{b=1}^B(\bar{\psi}^b-\bar{\bar{\psi}})^2-\frac{V}{R}\label{ANOVA estimate}
\end{equation}
where
$$V=\frac{1}{B(R-1)}\sum_{b=1}^B\sum_{r=1}^R(\hat{\psi}_r(\widehat{F}_1^b,\ldots,\widehat{F}_m^b)-\bar{\psi}^b)^2.$$
To explain, the first term in \eqref{ANOVA estimate} is an unbiased estimate of the variance of $\bar\psi^b$, which is $\mathrm{Var}_*[\psi(\widehat{F}^*_1,\ldots,\widehat{F}^*_m)]+(1/R)\mathbb E_*[\mathrm{Var}[\hat\psi_r(\widehat{F}^*_1,\ldots,\widehat{F}^*_m)|\widehat{F}^*_1,\ldots,\widehat{F}^*_m]]$ (where $\mathbb E_*[\cdot]$ denotes the expectation on $\widehat F_i^*$'s conditional on $\widehat F_i$'s), since $\bar\psi^b$ incurs both the bootstrap noise and the simulation noise. In other words, the variance of $\bar\psi^b$ is upward biased for $\mathrm{Var}_*[\psi(\widehat{F}^*_1,\ldots,\widehat{F}^*_m)]$. The second term in \eqref{ANOVA estimate}, namely $V/R$, removes this bias. This bias adjustment can be derived by viewing $\mathrm{Var}_*[\psi(\widehat{F}^*_1,\ldots,\widehat{F}^*_m)]$ as the variance of a conditional expectation. Alternately, $\hat\psi_r(\widehat F_1^*,\ldots,\widehat F_m^*)$ can be viewed as a random effect model where each ``group" corresponds to each realization of $\widehat F_1^*,\ldots,\widehat F_m^*$, and \eqref{ANOVA estimate} estimates the ``between-group" variance in an analysis-of-variance (ANOVA). Formula \eqref{ANOVA estimate} has appeared in the input uncertainty literature, e.g., \cite{cheng1997sensitivity}, \cite{song2015quickly}, \cite{lin2015single}, and also in \cite{zouaoui2004accounting} in the Bayesian context. Algorithm \ref{anova} summarizes the procedure.
\begin{algorithm}
\caption{ANOVA-based Variance Bootstrap}
\label{anova}
\begin{algorithmic}
\STATE {Given: $B\geq 2,R\geq 2$; data $=\{X_{i,j}:i=1,\ldots,m, j=1,\ldots,n_i\}$}
\FOR{$b=1$ \TO $B$ }
\STATE {For each $i$, draw a sample $\{X_{i,1}^b,\ldots,X_{i,n_i}^b\}$ uniformly with replacement from the data to obtain a resampled empirical distribution $\widehat{F}_i^b$}
\FOR{$r=1$ \TO $R$}
\STATE {Simulate $\hat{\psi}_r(\widehat{F}_1^b,\ldots,\widehat{F}_m^b)$}
\ENDFOR
\STATE {Compute $\bar{\psi}^b_{BV}=\frac{1}{R}\sum_{r=1}^R\hat{\psi}_r(\widehat{F}_1^b,\ldots,\widehat{F}_m^b)$}
\ENDFOR
\STATE {Compute $V=\frac{1}{B(R-1)}\sum_{b=1}^B\sum_{r=1}^R(\hat{\psi}_r(\widehat{F}_1^b,\ldots,\widehat{F}_m^b)-\bar{\psi}^b_{BV})^2$ and $\bar{\bar{\psi}}_{BV}=\frac{1}{B}\sum_{b=1}^B\bar{\psi}^b_{BV}$}
\STATE {Output $\hat{\sigma}_{BV}^2=\frac{1}{B-1}\sum_{b=1}^B(\bar{\psi}^b_{BV}-\bar{\bar{\psi}}_{BV})^2-\frac{V}{R}$}
\end{algorithmic}
\end{algorithm}

More generally, to estimate the variance contribution from the data noise of model $i$ only, namely $\sigma_i^2/n_i$, one can bootstrap only from $\{X_{i,1},\ldots,X_{i,n_i}\}$ and keep other input distributions $\widehat F_j,j\neq i$ fixed. Then $\widehat F_i^*$ and $\widehat F_j,j\neq i$ are used to drive the simulation runs. With this modification, the same formula \eqref{ANOVA estimate} or Algorithm \ref{anova} is an unbiased estimate for $\mathrm{Var}_*[\psi(\widehat F_1,\ldots,\widehat F_{i-1},\widehat F_i^*,\widehat F_{i+1},\ldots,\widehat F_m)]$, which is approximately $\mathrm{Var}[\psi(F_1,\ldots,F_{i-1},\widehat F_i,F_{i+1},\ldots,F_m)]$ by the bootstrap principle, in turn asymptotically equal to $\sigma_i^2/n_i$ introduced in \eqref{var decompose}. This observation appeared in, e.g., \cite{song2014advanced}; in Section \ref{sec:var boot} we give further justifications.

In subsequent discussions, we use the following notations. For any sequences $a$ and $b$, both depending on some parameter, say, $n$, we say that $a=O(b)$ if $|a/b|\leq C$ for some constant $C>0$ for all sufficiently large $n$, and $a=o(b)$ if $a/b\to0$ as $n\to\infty$. Alternately, we say $a=\Omega(b)$ if $|a/b|\geq C$ for some constant $C>0$ for all sufficiently large $n$, and $a=\omega(b)$ if $|a/b|\to\infty$ as $n\to\infty$. We say that $a=\Theta(b)$ if $\underline C\leq|a/b|\leq \overline C$ as $n\to\infty$ for some constants $\underline C,\overline C>0$. We use $A=O_p(b)$ to represent a random variable $A$ that has stochastic order at least $b$, i.e., for any $\epsilon>0$, there exists $M,N>0$ such that $P(|A/b|\leq M)>1-\epsilon$ for $n>N$. We use $A=o_p(b)$ to represent a random variable $A$ that has stochastic order less than $b$, i.e., $A/b\stackrel{p}{\to}0$. We use $A=\Theta_p(b)$ to represent a random variable $A$ that has stochastic order exactly at $b$, i.e., $A$ satisfies $A=O_p(b)$ but not $A=o_p(b)$.

\subsection{A Complexity Barrier}\label{sec:barrier}
We explain intuitively the total number of simulation runs needed to ensure that the variance bootstrap depicted above can meaningfully estimate the input variance. For convenience, we call this number the \emph{simulation complexity}. This quantity turns out to be of order bigger than the data size. On a high level, it is because the input variance scales reciprocally with the data size (recall \eqref{var decompose}). Thus, when the data size increases, the input variance becomes smaller and increasingly difficult to estimate with controlled relative error. This in turn necessitates the use of more simulation runs.



To explain more concretely, denote $n$ as a scaling of the data size, i.e., we assume $n_i$ all grow linearly with $n$, which in particular implies that $\sigma_I^2$ is of order $1/n$. We analyze the error of $\hat\sigma_{BV}^2$ from Algorithm \ref{anova} in estimating $\sigma_I^2$. Since $\hat\sigma_{BV}^2$ is unbiased for $\mathrm{Var}_*[\psi(\widehat F_1^*,\ldots,\widehat F_m^*)]$ which is in turn close to $\sigma_I^2$, roughly speaking it suffices to focus on the variance of $\hat\sigma_{BV}^2$. To analyze this later quantity, we denote a generic simulation run in our procedure, $\hat\psi_r(\widehat{F}_1^*,\ldots,\widehat{F}_m^*)$, as
$$\hat\psi_r(\widehat{F}_1^*,\ldots,\widehat{F}_m^*)=\psi(\widehat{F}_1,\ldots,\widehat{F}_m)+\delta+\xi$$
where
\begin{align*}
\delta:=\psi(\widehat{F}_1^*,\ldots,\widehat{F}_m^*)-\psi(\widehat{F}_1,\ldots,\widehat{F}_m),\ \xi:=\hat\psi_r(\widehat{F}_1^*,\ldots,\widehat{F}_m^*)-\psi(\widehat{F}_1^*,\ldots,\widehat{F}_m^*).
\end{align*}
are the errors arising from the bootstrap of the input distributions and the simulation respectively. If $\psi$ is sufficiently smooth, $\delta$ elicits a central limit theorem and is of order $\Theta_p(1/\sqrt n)$. On the other hand, the simulation noise $\xi$ is of order $\Theta_p(1)$.

Via an ANOVA-type analysis as in \cite{sun2011efficient}, we have
\begin{align}
\nonumber\mathrm{Var}_*[\hat{\sigma}_{BV}^2]=&\frac{1}{B}(\mathbb E_*[\delta^4]-(\mathbb E_*[\delta^2])^2)+\frac{2}{B(B-1)}(\mathbb E_*[\delta^2])^2+\frac{2}{B^2R^2(B-1)}(\mathbb E_*[\xi^2])^2+\frac{2}{B^2R^3}\mathbb E_*[\xi^4]\\
\nonumber&+\frac{2(B+1)}{B^2R(B-1)}\mathbb E_*[\delta^2]\mathbb E_*[\xi^2]+\frac{2(BR^2+R^2-4R+3)}{B^2R^3(R-1)}\mathbb E_*[(\mathbb E[\xi^2\vert \widehat{F}_1^*,\ldots,\widehat{F}_m^*])^2]\\
&+\frac{4B+2}{B^2R}\mathbb E_*[\delta^2\xi^2]+\frac{4}{B^2R^2}\mathbb E_*[\delta\xi^3].\label{var formula}
\end{align}
Now, putting $\delta=\Theta_p(1/\sqrt n)$ and $\xi=\Theta_p(1)$ formally into \eqref{var formula}, and ignoring constant factors, results in
\begin{equation*}
\mathrm{Var}_*[\hat{\sigma}_{BV}^2]=O_p\left(\frac{1}{Bn^2}+\frac{1}{B^2n^2}+\frac{1}{B^3R^2}+\frac{1}{B^2Rn}+\frac{1}{B^2R^3}+\frac{1}{BR^2}+\frac{1}{BRn}+\frac{1}{B^2R^2\sqrt n}\right)
\end{equation*}
or simply
\begin{equation}
O_p\left(\frac{1}{Bn^2}+\frac{1}{BR^2}\right)\label{mse simplified}
\end{equation}
The two terms in \eqref{mse simplified} correspond to the variances coming from the bootstrap resampling and the simulation runs respectively.

Since $\sigma_I^2$  is of order $1/n$, meaningful estimation of $\sigma_I^2$ needs measured by the relative error.  In other words, we want to achieve $\hat\sigma_{BV}^2/\sigma_I^2\stackrel{p}{\to}1$ as the simulation budget grows. This property, which we call relative consistency, requires $\hat\sigma_{BV}^2$ to have a variance of order $o(1/n^2)$ in order to compensate for the decreasing order of $\sigma_I^2$.

We argue that this implies unfortunately that the total number of simulation runs, $BR$, must be $\omega(n)$, i.e., of order higher than the data size. To explain, note that the first term in \eqref{mse simplified} forces one to use $B=\omega(1)$, i.e., the bootstrap size needs to grow with $n$, an implication that is quite natural. The second term in \eqref{mse simplified}, on the other hand, dictates also that $BR^2=\omega(n^2)$. Suppose, for the sake of contradiction, that $B$ and $R$ are chosen such that $BR=O(n)$. Then, because we need $BR\times R=BR^2=\omega(n^2)$, $R$ must be $\omega(n)$ which, combining with $B=\omega(1)$, implies that $BR=\omega(n)$ and leads to a contradiction.

We summarize the above with the following result. Let $N$ be the total simulation effort, and recall $n$ as the scaling of the data size. We have:
\begin{theorem}[Simulation complexity of the variance bootstrap]\label{rough lower}
Under Assumptions \ref{size_data}-\ref{anova_mu4_sim} to be stated in Section \ref{sec:assumptions}, the required simulation budget to achieve relative consistency in estimating $\sigma_I^2$ by Algorithm \ref{anova}, i.e., $\hat{\sigma}_{BV}^2/\sigma_I^2\stackrel{p}{\to}1$, is $N=\omega(n)$.
\end{theorem}


Though out of the scope of this paper, there are indications that such a computational barrier occurs in other types of bootstrap. For instance, the percentile bootstrap studied in \cite{barton1993uniform,barton2001resampling} appears to also require an inner replication size large enough compared to the data size in order to obtain valid quantile estimates (the authors actually used one inner replication, but \cite{barton2012tutorial} commented that more is needed). \cite{yi2017efficient} provides an interesting approach based on ranking and selection to reduce the simulation effort, though they do not investigate the order of the needed effort relative to the data size. The empirical likelihood framework studied in \cite{lam2017optimization} requires a similarly higher order of simulation runs to estimate the influence function. Nonetheless, in this paper we focus only on how to reduce computation load in variance estimation.


\section{Procedures and Guarantees in the Subsampling Framework}\label{sec:overview}
This section presents our methodologies and results on subsampling. Section \ref{sec:psvb} first explains the rationale and the subsampling procedure. Section \ref{sec:main guarantees} then presents our main theoretical guarantees, deferring some elaborate developments to Section \ref{sec:var boot}.


\subsection{Proportionate Subsampled Variance Bootstrap}\label{sec:psvb}
As explained before, the reason why the $\hat{\sigma}_{BV}^2$ in Algorithm \ref{anova} requires a huge simulation effort, as implied by its variance \eqref{mse simplified}, lies in the small scale of the input variance. In general, in order to estimate a quantity that is of order $1/n$, one must use a sample size more than $n$ so that the estimation error relatively vanishes. This requirement manifests in the inner replication size in constructing $\hat{\sigma}_{BV}^2$.

To reduce the inner replication size, we leverage the relation between the form of the input variance and the estimation variance depicted in \eqref{mse simplified} as follows. The approximate input variance contributed from model $i$, with data size $n_i$, has the form $\sigma_i^2/n_i$. If we use the variance bootstrap directly as in Algorithm \ref{anova}, then we need an order more than $n$ total simulation runs due to \eqref{mse simplified}. Now, pretend that we have fewer data, say $s_i$, then the input variance will be  $\sigma_i^2/s_i$, and the required simulation runs is now only of order higher than $s_i$. An estimate of $\sigma_i^2/s_i$, however, already gives us enough information in estimating $\sigma_i^2/n_i$, because we can rescale our estimate of $\sigma_i^2/s_i$ by $s_i/n_i$ to get an estimate of $\sigma_i^2/n_i$. Estimating $\sigma_i^2/s_i$ can be done by subsampling the input distribution with size $s_i$. With this, we can both use fewer simulation runs and also retain correct estimation via multiplying by a $s_i/n_i$ factor.

To make the above argument more transparent, the bootstrap principle and the asymptotic approximation of the input variance imply that
$$\mathrm{Var}_*[\psi(\widehat{F}^*_1,\ldots,\widehat{F}^*_m)]=\sum_{i=1}^m\frac{\sigma_i^2}{n_i}(1+o_p(1)).$$
The subsampling approach builds on the observation that a similar relation holds for
$$\mathrm{Var}_*[\psi(\widehat F_{s_1,1}^*,\ldots,\widehat F_{s_m,m}^*)]=\sum_{i=1}^m\frac{\sigma_i^2}{s_i}(1+o_p(1))$$
where $\widehat F_{s_i,i}^*$ denotes a bootstrapped input distribution of size $s_i$ (i.e., an empirical distribution of size $s_i$ that is uniformly sampled with replacement from $\{X_{i,1},\ldots,X_{i,n_i}\}$). If we let $s_i=\lfloor\theta n_i\rfloor$ for some $\theta>0$ so that $s_i\to\infty$ (where $\lfloor\cdot\rfloor$ is the floor function, i.e.~the largest integer less than or equal to $\cdot$), then we have
\begin{equation*}
\mathrm{Var}_*[\psi(\widehat{F}_{\lfloor\theta n_1\rfloor,1}^*,\ldots,\widehat{F}_{\lfloor\theta n_m\rfloor,m}^*)]=\sum_{i=1}^m\frac{\sigma_i^2}{\theta n_i}(1+o_p(1)).
\end{equation*}
Multiplying both sides with $\theta$, we get
\begin{equation*}
\theta\mathrm{Var}_*[\psi(\widehat{F}_{\lfloor\theta n_1\rfloor,1}^*,\ldots,\widehat{F}_{\lfloor\theta n_m\rfloor,m}^*)]=\sum_{i=1}^m\frac{\sigma_i^2}{n_i}(1+o_p(1)).
\end{equation*}
Note that the right hand side above is the original input variance of interest. This leads to our \emph{proportionate subsampled variance bootstrap}: We repeatedly subsample collections of input distributions from the data, with size $\lfloor\theta n_i\rfloor$ for model $i$, and use them to drive simulation replications. We then apply the ANOVA-based estimator in \eqref{ANOVA estimate} on these replications, and multiply it by a factor of $\theta$ to obtain our final estimate. We summarize this procedure in Algorithm \ref{anova sub}. The term ``proportionate" refers to the fact that we scale the subsample size for all models with a single factor $\theta$. For convenience, we call $\theta$ the \emph{subsample ratio}.

\begin{algorithm}
\caption{Proportionate Subsampled Variance Bootstrap}
\label{anova sub}
\begin{algorithmic}
\STATE {Parameters: $B\geq 2,R\geq 2,0<\theta\leq 1$; data $=\{X_{i,j}:i=1,\ldots,m, j=1,\ldots,n_i\}$}
\STATE Compute $s_i=\lfloor\theta n_i\rfloor$ for all $i$
\FOR{$b=1$ \TO $B$ }
\STATE {For each $i$, draw a subsample $\{X_{i,1}^b,\ldots,X_{i,s_i}^b\}$ uniformly with replacement from the data, which forms the empirical distribution $\widehat F_{s_i,i}^b$}
\FOR{$r=1$ \TO $R$}
\STATE {Simulate $\hat{\psi}_r(\widehat F_{s_1,1}^b,\ldots,\widehat F_{s_m,m}^b)$}
\ENDFOR
\STATE {Compute $\bar{\psi}^b=\frac{1}{R}\sum_{r=1}^R\hat{\psi}_r(\widehat F_{s_1,1}^b,\ldots,\widehat F_{s_m,m}^b)$}
\ENDFOR
\STATE {Compute $V=\frac{1}{B(R-1)}\sum_{b=1}^B\sum_{r=1}^R(\hat{\psi}_r(\widehat F_{s_1,1}^b,\ldots,\widehat F_{s_m,m}^b)-\bar{\psi}^b)^2$ and $\bar{\bar{\psi}}=\frac{1}{B}\sum_{b=1}^B\bar{\psi}^b$}
\STATE {Output $\hat{\sigma}_{SVB}^2=\theta(\frac{1}{B-1}\sum_{b=1}^B(\bar{\psi}^b-\bar{\bar{\psi}})^2-\frac{V}{R})$}
\end{algorithmic}
\end{algorithm}

Similar ideas apply to estimating the individual variance contribution from each input model, namely $\sigma_i^2/n_i$. Instead of subsampling all input distributions, we only subsample the distribution, say $\widehat F_{s_i,i}^*$ whose uncertainty is of interest, while fixing all the other distributions as the original empirical distributions, i.e., $\widehat F_j,j\neq i$. All the remaining steps in Algorithm \ref{anova sub} remain the same (thus the ``proportionate" part can be dropped). This procedure is depicted in Algorithm \ref{anova sub individual}.

\begin{algorithm}
\caption{Subsampled Variance Bootstrap for Variance Contribution from the $i$-th Input Model}
\label{anova sub individual}
\begin{algorithmic}
\STATE {Parameters: $B\geq 2,R\geq 2,0<\theta\leq 1$; data $=\{X_{i,j}:i=1,\ldots,m, j=1,\ldots,n_i\}$}
\STATE Compute $s_i=\lfloor\theta n_i\rfloor$
\FOR{$b=1$ \TO $B$ }
\STATE {Draw a subsample $\{X_{i,1}^b,\ldots,X_{i,s_i}^b\}$ uniformly with replacement from the $i$-th input data set, which forms the empirical distribution $\widehat F_{s_i,i}^b$}
\FOR{$r=1$ \TO $R$}
\STATE {Simulate $\hat{\psi}_r(\widehat{F}_{1},\ldots,\widehat F_{i-1},\widehat F_{s_i,i}^b,\widehat F_{i+1},\ldots,\widehat{F}_{m})$}
\ENDFOR
\STATE {Compute $\bar{\psi}^b=\frac{1}{R}\sum_{r=1}^R\hat{\psi}_r(\widehat{F}_{1},\ldots,\widehat F_{i-1},\widehat F_{s_i,i}^b,\widehat F_{i+1},\ldots,\widehat{F}_{m})$}
\ENDFOR
\STATE {Compute $V=\frac{1}{B(R-1)}\sum_{b=1}^B\sum_{r=1}^R(\hat{\psi}_r(\widehat{F}_{1},\ldots,\widehat F_{i-1},\widehat F_{s_i,i}^b,\widehat F_{i+1},\ldots,\widehat{F}_{m})-\bar{\psi}^b)^2$ and $\bar{\bar{\psi}}=\frac{1}{B}\sum_{b=1}^B\bar{\psi}^b$}
\STATE {Output $\hat{\sigma}_{SVB,i}^2=\theta(\frac{1}{B-1}\sum_{b=1}^B(\bar{\psi}^b-\bar{\bar{\psi}})^2-\frac{V}{R})$}
\end{algorithmic}
\end{algorithm}

\subsection{Statistical Guarantees}\label{sec:main guarantees}
Algorithm \ref{anova sub} provides the following guarantees. Recall that $N=BR$ is the total simulation effort, and $n$ is the scaling of the data size.
We have the following result:
\begin{theorem}[Procedural configurations to achieve relative consistency]\label{consis:anova}
Under Assumptions \ref{size_data}-\ref{anova_mu4_sim} to be stated in Section \ref{sec:assumptions}, if the parameters $B,R,\theta$ of Algorithm \ref{anova sub} are chosen such that
\begin{equation}\label{parameter:psvb}
B=\omega(1),\;BR^2=\omega\big( (\theta n)^2\big),\;\theta=\omega\big(\frac{1}{ n}\big)
\end{equation}
then the variance estimate $\hat{\sigma}_{SVB}^2$ is relatively consistent, i.e.~$\hat{\sigma}_{SVB}^2/\sigma_I^2\stackrel{p}{\to}1$.
\end{theorem}
Theorem \ref{consis:anova} tells us what orders of the bootstrap size $B$, inner replication size $R$ and subsample ratio $\theta$ would guarantee a meaningful estimation of $\sigma_I^2$. Note that $\theta\approx s_i/n_i$ for each $i$, so that $\theta=\omega(1/n)$ is equivalent to setting the subsample size $s_i=\omega(1)$. In other words, we need the natural requirement that the subsample size grows with the data size, albeit can have an arbitrary rate.

Given a subsample ratio $\theta$ specified according to \eqref{parameter:psvb}, the configurations of $B$ and $R$ under \eqref{parameter:psvb} that achieve the minimum overall simulation budget is $B=\omega(1)$ and $R=\Omega(\theta n)$. This is because to minimize $N=BR$ while satisfying the second requirement in \eqref{parameter:psvb}, it is more economical to allocate as much budget to $R$ instead of $B$. This is stated precisely as:

\begin{corollary}[Minimum configurations to achieve relative consistency]\label{consis:anova optimal}
Under the same conditions of Theorem \ref{consis:anova}, given $\theta=\omega(n^{-1})$, the values of $B$ and $R$ to achieve \eqref{parameter:psvb} and hence relative consistency that requires the least order of effort are $B=\omega(1)$ and $R=\Omega(\theta n)$, leading to a total simulation budget $N=\omega(\theta n)$.
\end{corollary}

Note that $\theta n$ is the order of the subsample size. Thus Corollary \ref{consis:anova optimal} implies that the required simulation budget must be of higher order than the subsample size. However, since the subsample size can be chosen to grow at an arbitrarily small rate, this implies that the total budget can also grow arbitrarily slowly. Therefore, we have:



\begin{corollary}[Simulation complexity of proportionate subsampled variance bootstrap]
Under the same conditions of Theorem \ref{consis:anova}, the minimum required simulation budget to achieve relative consistency in estimating $\sigma_I^2$ by Algorithm \ref{anova sub}, i.e., $\hat{\sigma}_{SVB}^2/\sigma_I^2\stackrel{p}{\to}1$, is $N=\omega(1)$ by using $\theta=\omega(n^{-1})$.\label{rough sub var}
\end{corollary}

Compared to Theorem \ref{rough lower}, Corollary \ref{rough sub var} stipulates that our subsampling approach reduces the required simulation effort from a higher order than $n$ to an arbitrary order, i.e., independent of the data size. This is achieved by using a subsample size that grows with $n$ at an arbitrary order, or equivalently a subsample ratio $\theta$ that grows faster than $1/n$.

The following result describes the configurations of our scheme when a certain total simulation effort is given. In particular, it shows, for a given total simulation effort, the range of subsample ratio for which Algorithm \ref{anova sub} can possibly generate valid variance estimates by appropriately choosing $B$ and $R$:
\begin{theorem}[Valid subsample ratio given total budget]\label{sub ratio}
Assume the same conditions of Theorem \ref{consis:anova}. Given a total simulation budget $N=\omega(1)$, if the subsample ratio satisfies $\omega(1/n)\leq \theta \leq o(N/n)\wedge 1$, then the bootstrap size $B$ and the inner  replication size $R$ can be appropriately chosen according to criterion \eqref{parameter:psvb} to achieve relative consistency, i.e., $\hat{\sigma}_{SVB}^2/\sigma_I^2\stackrel{p}{\to}1$.
\end{theorem}


The next result is on the optimal configurations of our scheme in minimizing the Monte Carlo error. To proceed, define
\begin{equation}\label{psvb:estimator}
\sigma_{SVB}^2=\theta\mathrm{Var}_*[\psi(\widehat{F}_{\lfloor\theta n_1\rfloor,1}^*,\ldots,\widehat{F}_{\lfloor\theta n_m\rfloor,m}^*)]
\end{equation}
as the perfect form of our proportionate subsampled variance bootstrap introduced in Section \ref{sec:psvb}, namely without any Monte Carlo noises, and $0<\theta\leq 1$ is the subsample ratio. We have:
\begin{theorem}\label{opt_allocation:anova}
Assume the same conditions of Theorem \ref{consis:anova}. Given a simulation budget $N$ and a subsample ratio $\theta$ such that $N=\omega(\theta n)$ and $\theta=\omega(n^{-1})$, the optimal outer and inner sizes that minimize the order of the conditional mean squared error $\mathbb E_*[(\hat{\sigma}_{SVB}^2-\sigma_{SVB}^2)^2]$ are
\begin{equation*}
B^*= \frac{N}{R^*},\;R^*=\Theta(\theta n)
\end{equation*}
giving a conditional mean squared error $\mathbb E_*[(\hat{\sigma}_{SVB}^2-\sigma_{SVB}^2)^2]=\Theta(\theta/(Nn))(1+o_p(1))$.
\end{theorem}
Note that the mean squared error, i.e.~$\mathbb E_*[(\hat{\sigma}_{SVB}^2-\sigma_{SVB}^2)^2]$, of the Monte Carlo estimate $\hat{\sigma}_{SVB}^2$ is random because the underlying resampling is conditioned on the input data, therefore the bound at the end of Theorem \ref{opt_allocation:anova} contains a stochastically vanishing term $o_p(1)$.

We next present the optimal tuning of the subsample ratio. This requires a balance of the trade-off between the input statistical error and the Monte Carlo simulation error. To explain, the overall error of $\hat{\sigma}_{SVB}^2$ by Algorithm \ref{anova sub} can be decomposed as
\begin{equation}\label{error_decompose}
\hat{\sigma}_{SVB}^2-\sigma_I^2=(\hat{\sigma}_{SVB}^2-\sigma_{SVB}^2)+(\sigma_{SVB}^2-\sigma_I^2).
\end{equation}
The first term is the Monte Carlo error for which the optimal outer size $B$, inner size $R$ and the resulting mean squared error are governed by Theorem \ref{opt_allocation:anova}. In particular, the mean squared error there shows that under a fixed simulation budget $N$ and the optimal allocation $R=\Theta(\theta n)$, the Monte Carlo error gets larger as $\theta$ increases. The second term is the statistical errors due to the finiteness of input data and $\theta$. Since $\theta$ measures the amount of data contained in the resamples, we expect this second error to become smaller as $\theta$ increases. The optimal tuning of $\theta$ relies on balancing such a trade-off between the two sources of errors.

We have the following optimal configurations of $B$, $R$ and $\theta$ altogether given a budget $N$:
\begin{theorem}[Optimal subsample size and budget allocation]\label{optimal_allocation}
Suppose Assumptions \ref{size_data}, \ref{smoothness_truth}-\ref{anova_mu4_sim} in Section \ref{sec:assumptions} and Assumptions \ref{third-differentiability}-\ref{3smoothness_empirical} in Section \ref{sec:opt_subsample} hold. For a given simulation budget $N=\omega(1)$, if the subsample ratio $\theta$ and outer and inner sizes $B,R$ for Algorithm \ref{anova sub} are set to
\begin{align}
&\begin{cases}
\theta^*=\Theta\big(N^{1/3} n^{-1}\big)&\text{ if }1\ll N\leq  n^{3/2}\\
\Theta(n^{-1/2})\leq \theta^*\leq \Theta\big(N n^{-2}\wedge 1\big)&\text{ if }N>n^{3/2}
\end{cases}\label{opt_sub:PSBV}\\
&R^*=\Theta(\theta^* n),\;B^*=\frac{N}{R^*}\label{opt_inner:PSBV}
\end{align}
then the gross error $\hat{\sigma}_{SVB}^2-\sigma_I^2=\mathcal E+o_p(N^{-1/3}n^{-1}+n^{-3/2})$, where the leading term has a mean squared error
\begin{equation}\label{min var}
\mathbb E[\mathcal E^2]=O\big(\frac{1}{N^{2/3}n^2}+\frac{1}{n^3}\big).
\end{equation}
Moreover, if $\mathcal R=\Theta ((ns)^{-1})$ and at least one of the $\Sigma_i$'s are positive definite, where $\mathcal R$ and $\Sigma_i$ are as defined in Lemma \ref{PSBV:error}, then \eqref{min var} holds with an exact order (i.e., $O(\cdot)$ becomes $\Theta(\cdot)$) and the configuration \eqref{opt_sub:PSBV}, \eqref{opt_inner:PSBV} is optimal in the sense that no configuration gives rise to a gross error $\hat{\sigma}_{SVB}^2-\sigma_I^2=o_p\big(N^{-1/3}n^{-1}+n^{-3/2}\big)$.
\end{theorem}

Note from \eqref{min var} that, if the budget $N=\omega(1)$, our optimal configurations guarantee the estimation mean squared error decays faster than $1/n^2$. Recall that the input variance is of order $1/n$, and thus an estimation error of order higher than $1/n^2$ ensures that the estimator is relatively consistent in the sense $\hat{\sigma}_{SVB}^2/\sigma_I^2\stackrel{p}{\to}1$. This recovers the result in Corollary \ref{rough sub var}. We also comment that the algorithmic configuration given in Theorem \ref{optimal_allocation} is chosen to optimize the mean squared error of the input variance estimate, but does not necessarily generates the most accurate CI. There exists evidence (e.g., \cite{sun2008optimal}) that the optimal choice to minimize the mean squared error of the variance estimate can be different from the one that is optimal for statistical inference, although in our experiments they seem to match closely with each other.



We comment that all the results in this section hold if one estimates the individual variance contribution from each input model $i$, namely by using Algorithm \ref{anova sub individual}. In this case we are interested in estimating the variance $\sigma_i^2/n_i$, and relative consistency means $\hat{\sigma}_{SVB,i}^2/(\sigma_i^2/n_i)\stackrel{p}{\to}1$. The data size scaling parameter $n$ can be replaced by $n_i$ in all our results.

Finally, we also comment that the complexity barrier described in Section \ref{sec:barrier} and our framework presented in this section applies in principle to the parametric regime, i.e., when the input distributions are known to lie in parametric families with unknown parameters. The assumptions and mathematical details would need to be catered to that situation, which could be done naturally by viewing the ``parameter" as a function of $F_i$'s.


\section{Developments of Theoretical Results}\label{sec:var boot}
We present our main developments leading to the algorithms and results in Section \ref{sec:overview}. Section \ref{sec:assumptions} first states in detail our assumptions on the performance measure. Section \ref{sec:psvb theory} presents the theories leading to estimation accuracy, simulation complexity and optimal budget allocation in the proportionate subsampled variance bootstrap. Section \ref{sec:opt_subsample} investigates optimal subsample sizes that lead to overall best configurations.

%
%
%
%
%
%
%
%
%

\subsection{Regularity Assumptions}\label{sec:assumptions}
We first assume that the data sets for all input models are of comparable size.
\begin{assumption}[Balanced data]\label{size_data}
$\limsup_{\text{all }n_i\to\infty}\frac{\max_in_i}{\min_in_i}<\infty$ as all $n_i\to \infty$.
\end{assumption}
Recall in Sections \ref{sec:motivation} and \ref{sec:overview} that we have denoted $n$ as a scaling of the data size. More concretely, we take $n=(1/m)\sum_{i=1}^mn_i$ as the average input data size under Assumption \ref{size_data}.

We next state a series of general assumptions on the performance measure $\psi$. These assumptions hold for common finite-horizon measures, as we will present. For each $i$ let $\Xi_i$ be the support of the $i$-th true input model $F_i$, and the collection of distributions $\mathcal P_i$ be the convex hull spanned by $F_i$ and all Dirac measures on $\Xi_i$, i.e.
$$\mathcal P_i=\big\{\nu_1F_i+\sum_{k=2}^l\nu_k\mathbf{1}_{x_k}:\sum_{k=1}^l\nu_k=1,\nu_k\geq 0,l<\infty,x_k\in \Xi_i\text{ for all }k\big\}.$$
We assume the following differentiability of the performance measure.
\begin{assumption}[First order differentiability]\label{first-differentiability}
For any distributions $P_i,Q_i\in\mathcal P_i$, denote $P_i^{\nu_i}=(1-\nu_i)P_i+\nu_i Q_i$ for $\nu_i\in [0,1]$. Assume there exist functions $g_i(P_1,\ldots,P_m;\cdot):\Xi_i\to \R$ such that $\mathbb E_{P_i}[g_i(P_1,\ldots,P_m;X_i)]=0$ for $i=1,\ldots,m$ and as all $\nu_i$'s approach zero
\begin{equation}\label{expansion}
\psi(P_1^{\nu_1},\ldots,P_m^{\nu_m})-\psi(P_1,\ldots,P_m)=\sum_{i=1}^m\nu_i\int g_i(P_1,\ldots,P_m;x)d(Q_i-P_i)(x)+o\Big(\sqrt{\sum_{i=1}^m\nu_i^2}\Big).
\end{equation}

\end{assumption}
The differentiability described above is defined with respect to a particular direction, namely $Q_i-P_i$, in the space of probability measures, and is known as Gateaux differentiability or directional differentiability (e.g.,~\cite{serfling2009approximation,van2000asymptotic}). Assumption \ref{first-differentiability} therefore requires the performance measure $\psi$ to be Gateaux differentiable when restricted to the convex set $\mathcal P_1\times\cdots \times\mathcal P_m$. The functions $g_i$'s are also called the influence functions (e.g., \cite{hampel1974influence}) that play analogous roles as standard gradients in the Euclidean space. The condition of $g_i$'s having vanishing means is without loss of generality since such a condition can always be achieved by centering, i.e., subtracting the mean. Note that doing this does not make any difference to the first term of expansion \eqref{expansion} because both $Q_i$ and $P_i$ are probability measures. Taking each $\nu_i=1$ in \eqref{expansion}, one informally obtains the Taylor expansion of $\psi$ around $P_i$'s
\begin{equation*}
\psi(Q_1,\ldots,Q_m)-\psi(P_1,\ldots,P_m)\approx\sum_{i=1}^m\int g_i(P_1,\ldots,P_m;x)d(Q_i-P_i)(x).
\end{equation*}
When each $P_i$ is set to be the true input model $F_i$ and $Q_i$ to be the empirical input model $\widehat F_i$, the above linear expansion is expected to be a reasonably good approximation as the data size grows. The next assumption imposes a moment bound on the error of this approximation:
\begin{assumption}[Smoothness at true input models]\label{smoothness_truth}
Denote by $g_i(\cdot):=g_i(F_1,\ldots,F_m;\cdot)$ the influence functions at the true input distributions $F_i,i=1,\ldots,m$. Assume that the remainder in the Taylor expansion of the performance measure
\begin{equation}\label{taylor_expansion}
\psi(\widehat F_1,\ldots,\widehat F_m)=\psi(F_1,\ldots,F_m)+\sum_{i=1}^m\int g_i(x)d(\widehat F_i-F_i)(x)+\epsilon
\end{equation}
satisfies $\mathbb E[\epsilon^2]=o(n^{-1})$, and the influence functions $g_i$'s are non-degenerate, i.e.~$\sigma_i^2:=\mathrm{Var}_{F_i}[g_{i}(X_i)]>0$, and have finite fourth moments, i.e.~$\mathbb E_{F_i}[g_{i}^4(X_i)]<\infty$.
\end{assumption}
Assumption \ref{smoothness_truth} entails that the error of the linear approximation formed by influence functions is negligible in the asymptotic sense. Indeed, the linear term in \eqref{taylor_expansion} is asymptotically of order $\Theta_p(n^{-1/2})$ by the central limit theorem, whereas the error $\epsilon$ is implied by Assumption \ref{smoothness_truth} to be $o_p(n^{-1/2})$. Hence the variance of the linear term contributes dominantly to the overall input variance as $n_i$'s are large. Then, thanks to the independence among the input models, the input variance can be expressed in the additive form described in \eqref{var decompose} together with a negligible error.
\begin{proposition}\label{input_var_order}
Under Assumptions \ref{size_data}-\ref{smoothness_truth}, the input variance $\sigma_I^2$ defined in \eqref{input_var} takes the form
\begin{equation*}
\sigma_I^2=\sum_{i=1}^m\frac{\sigma_i^2}{n_i}+o\big(\frac{1}{n}\big)
\end{equation*}
where each $\sigma_i^2=\mathrm{Var}_{F_i}[g_{i}(X_i)]$ is the variance of the $i$-th influence function.
\end{proposition}
As the higher order $o\big(1/n\big)$ error suggests, the additive decomposition $\sum_{i=1}^m\frac{\sigma_i^2}{n_i}$ is guaranteed to be accurate only in the large-sample regime. Note that this decomposition is used solely as a theoretical vehicle for asymptotic analysis rather than the actual input variance estimator in our procedure, the latter using bootstrapping schemes that could exhibit better finite-sample performances.

As mentioned before, consistent estimation of input variance $\sigma_I^2$ relies on the bootstrap principle, for which we make the following additional assumptions. The assumption states that the error of the linear approximation \eqref{taylor_expansion} remains small when the underlying distributions $F_i$ are replaced by the empirical input distributions $\widehat F_i$, hence can be viewed as a bootstrapped version of Assumption \ref{smoothness_truth}.
\begin{assumption}[Smoothness at empirical input models]\label{smoothness_empirical}
Denote by $\hat g_i(\cdot):=g_i(\widehat F_1,\ldots,\widehat F_m;\cdot)$ the influence functions at the empirical input distributions $\widehat F_i,i=1,\ldots,m$. Assume the empirical influence function converges to the truth in the sense that $\mathbb E[(\hat g_i-g_i)^4(X_{i,1})]\to 0$. For each $i$ let $\overline F_i$ be either the $i$-th empirical input model $\widehat F_i$ or the resampled model $\widehat F_{s_i,i}^*$. For every $(\overline F_1,\ldots,\overline F_m)\in\prod_{i=1}^m\{\widehat F_{i},\widehat F_{s_i,i}^*\}$, assume the remainder in the Taylor expansion
\begin{equation}\label{taylor_expansion_empirical}
\psi(\overline F_1,\ldots,\overline F_m)=\psi(\widehat F_1,\ldots,\widehat F_m)+\sum_{i=1}^m\int\hat g_i(x)d(\overline F_i-\widehat F_i)(x)+{\epsilon}^*
\end{equation}
satisfies $\mathbb E_*[({\epsilon}^*)^4]=o_p\big(s^{-2}\big)$.
\end{assumption}
As the data sizes $n_i$'s grow, the empirical input distributions $\widehat F_i$ converge to the true ones $F_i$. Hence the empirical influence functions $\hat g_i$'s are expected to approach the influence functions $g_i$'s associated with the true input distributions, which explains the convergence condition in Assumption \ref{smoothness_empirical}. The fourth moment condition on the remainder $\epsilon^*$ is needed for controlling the variance of our variance estimator. Since the fourth moment is with respect to the resampling measure and thus depends on the underlying input data, the condition is described in terms of stochastic order. Note that we require \eqref{taylor_expansion_empirical} to hold not just when $\overline F_i=\widehat F_{s_i,i}^*$ for all $i$ but also when some $\overline F_i=\widehat F_i$. This allows us to estimate the variance contributed from an arbitrary group of input models and in particular an individual input model.

Assumptions \ref{first-differentiability}-\ref{smoothness_empirical} are on the performance measure $\psi$ itself. Next we impose assumptions on the simulation noise, i.e.~the stochastic error $\hat\psi_r-\psi$ where $\hat\psi_r$ is an unbiased simulation replication for $\psi$. We denote by $\tau^2(P_1,\ldots,P_m)$ the variance of $\hat\psi_r$ when simulation is driven by arbitrary input models $P_1,\ldots,P_m$, i.e.
\begin{equation*}
\tau^2(P_1,\ldots,P_m)=\mathbb E_{P_1,\ldots,P_m}[(\hat{\psi}_r-\psi(P_1,\ldots,P_m))^2].
\end{equation*}
Similarly we denote by $\mu_4(P_1,\ldots,P_m)$ the fourth central moment of $\hat{\psi}_r$ under the input models $P_1,\ldots,P_m$
\begin{equation*}
\mu_4(P_1,\ldots,P_m)=\mathbb E_{P_1,\ldots,P_m}[(\hat{\psi}_r-\psi(P_1,\ldots,P_m))^4].
\end{equation*}
In particular, for convenience we write $\tau^2=\tau^2(F_1,\ldots,F_m)$ for the variance of $\hat\psi$ under the true input models, and $\hat\tau^2=\tau^2(\widehat F_1,\ldots,\widehat F_m)$ for that under the empirical input models.

The assumptions on the simulation noise are:
\begin{assumption}[Convergence of empirical variance]\label{var_emp}
$\hat{\tau}^2\stackrel{p}{\to}\tau^2$.
\end{assumption}
\begin{assumption}[Convergence of bootstrapped variance]\label{var_sim}
For every $(\overline F_1,\ldots,\overline F_m)\in\prod_{i=1}^m\{\widehat F_{i},\widehat F_{s_i,i}^*\}$, it holds that $\mathbb E_*[(\tau^2(\overline F_1,\ldots,\overline F_m)-\hat\tau^2)^2]=o_p(1)$.
\end{assumption}
\begin{assumption}[Boundedness of the fourth moment]\label{anova_mu4_sim}
For every $(\overline F_1,\ldots,\overline F_m)\in\prod_{i=1}^m\{\widehat F_{i},\widehat F_{s_i,i}^*\}$, it holds that $\mathbb E_*[\mu_4(\overline F_1,\ldots,\overline F_m)]=O_p(1)$.
\end{assumption}
Assumptions \ref{var_emp} and \ref{var_sim} stipulate that the variance of the simulation replication $\hat\psi_r$ as a functional of the underlying input models is smooth enough in the inputs. Conceptually Assumption \ref{var_emp} is in line with Assumption \ref{smoothness_truth} in the sense that both concern smoothness of a functional around the true input models, whereas Assumption \ref{var_sim} is similar to Assumption \ref{smoothness_empirical} since both are about smoothness property around the empirical input models. Assumption \ref{anova_mu4_sim} is a fourth moment condition like in Assumption \ref{smoothness_empirical} used to control the variance of the variance estimator. Similar to Assumption \ref{smoothness_empirical}, we impose Assumptions \ref{var_sim} and \ref{anova_mu4_sim} for each $\overline F_i=\widehat F_i\text{ or }\widehat F_{s_i,i}^*$ so that the same guarantees remain valid when estimating input variances from individual input models, i.e., Algorithm \ref{anova sub individual}.

Although the above assumptions may look complicated, they can be verified, under minimal conditions, for generic finite-horizon performance measures in the form
\begin{equation}\label{finite_horizon:pm}
\psi(F_1,\ldots,F_m)=\mathbb E_{F_1,\ldots,F_m}[h(\mathbf X_1,\ldots,\mathbf X_m)]
\end{equation}
where $\mathbf X_i=(X_i(1),\ldots,X_i(T_i))$ represents the $i$-th input process consisting of $T_i$ i.i.d.~variables distributed under $F_i$, each $T_i$ being a deterministic time, and $h$ is a performance function. An unbiased simulation replication $\hat{\psi}_r$ of the performance measure is $h(\mathbf X_1,\ldots,\mathbf X_m)$.


Suppose we have the following conditions for the performance function $h$:
\begin{assumption}\label{var_pos:finite}
For each $i$, $0<\mathrm{Var}_{F_i}[\sum_{t=1}^{T_i}\mathbb E_{F_1,\ldots,F_m}[h(\mathbf X_1,\ldots,\mathbf X_m)\vert X_i(t)=X_i]]<\infty$.
\end{assumption}
\begin{assumption}[Parameter $k$]\label{moment8:finite}
For each $i$ let $I_i=(I_i(1),\ldots,I_i(T_i))$ be a sequence of indices such that $1\leq I_i(t)\leq t$, and $\mathbf X_{i,I_i} =(X_i(I_i(1)),\ldots,X_i(I_i(T_i)))$. Assume
$$\max_{I_1,\ldots,I_m}\mathbb E_{F_1,\ldots,F_m}[\lvert h(\mathbf X_{1,I_1},\ldots,\mathbf X_{m,I_m})\rvert^k]<\infty.$$
\end{assumption}
The conditional expectation in Assumption \ref{var_pos:finite} is in fact the influence function of the performance measure \eqref{finite_horizon:pm} under the true input models. So Assumption \ref{var_pos:finite} is precisely the non-degenerate variance condition in Assumption \ref{smoothness_truth}. All other parts of Assumptions \ref{first-differentiability}-\ref{anova_mu4_sim} are consequences of the moment condition in Assumption \ref{moment8:finite}:
\begin{theorem}\label{finitetime_justification}
Under Assumptions \ref{size_data}, \ref{var_pos:finite} and Assumption \ref{moment8:finite} with $k=4$, we have Assumptions \ref{first-differentiability}-\ref{anova_mu4_sim} hold for the finite-horizon performance measure $\psi$ given by \eqref{finite_horizon:pm}.
\end{theorem}

\subsection{Simulation Complexity and Allocation}\label{sec:psvb theory}
This section presents theoretical developments on our proportionate subsampled variance bootstrap. We first establish relative consistency assuming infinite computation resources. Recall \eqref{psvb:estimator} as the proportionate subsampled variance bootstrap estimator without any Monte Carlo errors. The following theorem gives a formal statement on the performance of this estimator discussed in Section \ref{sec:psvb}.



\begin{theorem}\label{consis:AV}
Under Assumptions \ref{size_data}-\ref{smoothness_empirical}, if the subsample ratio $\theta=\omega( n^{-1})$, then the proportionate subsampled variance bootstrap without Monte Carlo error, namely \eqref{psvb:estimator}, is relatively consistent as $n_i\to \infty$, i.e.
\begin{equation*}
\sigma_{SVB}^2/\sigma_I^2\stackrel{p}{\to}1.
\end{equation*}
\end{theorem}
The requirement $\theta=\omega(n^{-1})$ implies that $s_i\to \infty$, which is natural as one needs minimally an increasing subsample size to ensure the consistency of our estimator. It turns out that this minimal requirement is enough to ensure consistency even relative to the magnitude of $\sigma_I^2$.

Now we turn to the discussion of the Monte Carlo estimate of the bootstrap variance generated from Algorithm \ref{anova sub}. The following lemma characterizes the amount of Monte Carlo noise in terms of mean squared error.
\begin{lemma}\label{MC_MSE:anova}
The output $\hat{\sigma}_{SVB}^2$ of Algorithm \ref{anova sub} is unbiased for the proportionate subsampled variance bootstrap without Monte Carlo errors, namely $\sigma_{SVB}^2$. Furthermore, under Assumptions \ref{size_data}-\ref{anova_mu4_sim}, if
\begin{equation}\label{minimal_B}
B=\omega(1),\;\theta=\omega\big(\frac{1}{ n}\big)
\end{equation}
and $R$ is arbitrary, then the conditional mean squared error
\begin{equation}\label{mse:anova sub}
\mathbb E_*[(\hat{\sigma}_{SVB}^2-\sigma_{SVB}^2)^2]=\frac{2}{B}\Big(\sum_{i=1}^m\frac{\sigma_i^2}{n_i}+\frac{\tau^2\theta}{R}\Big)^2(1+o_p(1)).
\end{equation}
\end{lemma}
In addition to the condition $\theta=\omega( n^{-1})$ which has appeared in Theorem \ref{consis:AV}, we also require $B=\omega(1)$ in Lemma \ref{MC_MSE:anova}. As the proof reveals, with such a choice of $B$, we can extract the leading term of the conditional mean squared error shown in \eqref{mse:anova sub}, which takes a neat form and is easy to analyze.

Note that $\sigma_I^2$ here is of order $n^{-1}$ by Proposition \ref{input_var_order}. Hence the Monte Carlo noise of the variance estimate output by our algorithm has to vanish faster than $n^{-1}$ in order to achieve relative consistency. Combining Theorem \ref{consis:AV} and Lemma \ref{MC_MSE:anova}, we obtain the simulation complexity of $\hat{\sigma}_{SVB}^2 $ in Theorem \ref{consis:anova}.
To establish the theoretical optimal allocation on the outer and inner sizes $B$, $R$, for given data sizes $n_i$, subsample ratio $\theta$, and total simulation budget $N$, we minimize the conditional mean square error \eqref{mse:anova sub} subject to the budget constraint $BR=N$. This gives rise to the following result that gives a more precise (theoretical) statement than Theorem \ref{opt_allocation:anova}.
\begin{theorem}\label{opt_allocation_detail:anova}
Suppose Assumptions \ref{size_data}-\ref{anova_mu4_sim} hold. Given a simulation budget $N$ and a subsample ratio $\theta$ such that $N=\omega(\theta n)$ and $\theta=\omega(n^{-1})$, the optimal outer and inner sizes that minimize the conditional mean squared error $\mathbb E_*[(\hat{\sigma}_{SVB}^2-\sigma_{SVB}^2)^2]$ are
\begin{equation*}
B^*= \frac{N}{R^*},\;R^*=\frac{\theta\tau^2}{\sum_{i=1}^m\sigma_i^2/n_i}
\end{equation*}
which gives a conditional mean squared error
\begin{equation}\label{opt_mse:anova}
\mathbb E_*[(\hat{\sigma}_{SVB}^2-\sigma_{SVB}^2)^2]=\frac{8\theta \tau^2}{N}\sum_{i=1}^m\frac{\sigma_i^2}{n_i}(1+o_p(1)).
\end{equation}
\end{theorem}

Theorem \ref{opt_allocation_detail:anova} gives the exact choices of $B$ and $R$ that minimize the Monte Carlo error. However, this is more of theoretical interest because the optimal $R^*$ involves the desired input variance $\sum_{i=1}^m\sigma_i^2/n_i$. Having said that, we can conclude from the theorem that the optimal inner size $R$ is of order $\Theta(\theta  n)$, the same as the subsample size, because the input variance is of order $\Theta(1/n)$ by Proposition \ref{input_var_order} and $\tau^2$ is a constant. This results in Theorem \ref{opt_allocation:anova} in Section \ref{sec:main guarantees}.





\subsection{Optimal Subsample Ratio}\label{sec:opt_subsample}
In this section we further establish the optimal subsample ratio $\theta$ or equivalently subsample sizes $s_i$ that balance the two sources of errors in \eqref{error_decompose}. For this, we need more regularity conditions on the performance measure. The first assumption we need is third order Gateaux differentiability in the convex set $\mathcal P_1\times\cdots\times \mathcal P_m$:
\begin{assumption}[Third order differentiability]\label{third-differentiability}
Using the same notations $P_i,Q_i,P_i^{\nu_i}$ as in Assumption \ref{first-differentiability}, assume that there exist second order influence functions $g_{i_1i_2}(P_1,\ldots,P_m;\cdot):\Xi_{i_1}\times \Xi_{i_2}\to \R$ and third order influence functions $g_{i_1i_2i_3}(P_1,\ldots,P_m;\cdot):\Xi_{i_1}\times \Xi_{i_2}\times \Xi_{i_3}\to \R$ for $i_1,i_2,i_3=1,\ldots,m$ which are symmetric under permutations, namely
\begin{align*}
&g_{i_1i_2}(P_1,\ldots,P_m;x_1,x_2)=g_{i_2i_1}(P_1,\ldots,P_m;x_2,x_1)\\
&g_{i_1i_2i_3}(P_1,\ldots,P_m;x_1,x_2,x_3)=g_{i_2i_1i_3}(P_1,\ldots,P_m;x_2,x_1,x_3)=g_{i_1i_3i_2}(P_1,\ldots,P_m;x_1,x_3,x_2).
\end{align*}
and for all $x,y$ satisfy
\begin{align*}
\mathbb E_{P_{i_2}}[g_{i_1i_2}(P_1,\ldots,P_m;x,X_{i_2})]=0,\;\mathbb E_{P_{i_3}}[g_{i_1i_2i_3}(P_1,\ldots,P_m;x,y,X_{i_3})]=0.
\end{align*}
Moreover, as all $\nu_i$'s approach zero the following Taylor expansion holds
\begin{align*}
&\psi(P_1^{\nu_1},\ldots,P_m^{\nu_m})-\psi(P_1,\ldots,P_m)\\
=&\sum_{i=1}^m\nu_i\int g_i(P_1,\ldots,P_m;x)d(Q_i-P_i)(x)+\frac{1}{2}\sum_{i_1,i_2=1}^m\nu_{i_1}\nu_{i_2}\int g_{i_1i_2}(P_1,\ldots,P_m;x_1,x_2)\prod_{k=1}^2d(Q_{i_k}-P_{i_k})(x_k)\\
&+\frac{1}{6}\sum_{i_1,i_2,i_3=1}^m\nu_{i_1}\nu_{i_2}\nu_{i_3}\int g_{i_1i_2i_3}(P_1,\ldots,P_m;x_1,x_2,x_3)\prod_{k=1}^3d(Q_{i_k}-P_{i_k})(x_k)+o\Big(\big(\sum_{i=1}^m\nu_i^2\big)^{\frac{3}{2}}\Big).
\end{align*}
\end{assumption}
Assumption \ref{third-differentiability} complements and strengthens Assumption \ref{first-differentiability} in that it imposes stronger differentiability property. Similarly, the following two assumptions strengthen Assumptions \ref{smoothness_truth} and \ref{smoothness_empirical} respectively by considering cubic expansions.
\begin{assumption}[Third order smoothness at true input models]\label{3smoothness_truth}
Denote by $g_{i_1i_2}(\cdot):=g_{i_1i_2}(F_1,\ldots,F_m;\cdot)$ and $g_{i_1i_2i_3}(\cdot):=g_{i_1i_2i_3}(F_1,\ldots,F_m;\cdot)$ the second and third order influence functions under the true input models. Assume the remainder in the Taylor expansion of the plug-in estimator $\psi(\widehat F_1,\ldots,\widehat F_m)$
\begin{align*}
\psi(\widehat F_1,\ldots,\widehat F_m)=&\psi(F_1,\ldots,F_m)+\sum_{i=1}^m\int g_i(x)d(\widehat F_i-F_i)(x)+\frac{1}{2}\sum_{i_1,i_2=1}^m\int g_{i_1i_2}(x_1,x_2)\prod_{k=1}^2d(\widehat F_{i_k}-F_{i_k})(x_k)\\
&+\frac{1}{6}\sum_{i_1,i_2,i_3=1}^m\int g_{i_1i_2i_3}(x_1,x_2,x_3)\prod_{k=1}^3d(\widehat F_{i_k}-F_{i_k})(x_k)+\epsilon_3
\end{align*}
satisfies $\mathbb E[\epsilon_3^2]=o(n^{-3})$, and the high order influence functions satisfy the moment conditions
\begin{align*}
\mathbb E[g^4_{i_1i_2}(X_{i_1,1},X_{i_2,j_2})]<\infty,\;\mathbb E[g^2_{i_1i_2i_3}(X_{i_1,1},X_{i_2,j_2},X_{i_3,j_3})]<\infty
\end{align*}
for all $i_1, i_2, i_3$ and $j_2\leq 2,j_3\leq 3$, where $X_{i,j}$ is the $j$-th data point from the $i$-th input model.
\end{assumption}

Similar to the remainder $\epsilon$ in Assumption \ref{smoothness_truth}, the moment condition on $\epsilon_3$ here is used to control the error of the cubic approximation of $\psi$ formed by up to third order influence functions. With these additional assumptions, the error term in Proposition \ref{input_var_order} can be refined as follows:
\begin{proposition}\label{input_var:tight}
Under Assumptions \ref{size_data}, \ref{smoothness_truth} and \ref{third-differentiability}-\ref{3smoothness_truth}, the overall input variance, as defined in \eqref{input_var}, can be expressed as
\begin{equation*}
\sigma_I^2=\sum_{i=1}^{m}\frac{\sigma_i^2}{n_i}+O\big(\frac{1}{ n^2}\big).
\end{equation*}
\end{proposition}

We also need third order differentiability around the empirical input models:
\begin{assumption}[Third order smoothness at empirical input models]\label{3smoothness_empirical}
Denote by $\hat g_{i_1i_2}(\cdot):=g_{i_1i_2}(\widehat F_1,\ldots,\widehat F_m;\cdot)$ and $\hat g_{i_1i_2i_3}(\cdot):=g_{i_1i_2i_3}(\widehat F_1,\ldots,\widehat F_m;\cdot)$ the second and third order influence functions under the empirical input models. Assume that the remainder in the Taylor expansion of the bootstrapped performance measure $\psi(\widehat F_{s_1,1}^*,\ldots,\widehat F_{s_m,m}^*)$
\begin{align*}
\psi(\widehat F_{s_1,1}^*,\ldots,\widehat F_{s_m,m}^*)=&\psi(\widehat F_1,\ldots,\widehat F_m)+\int \hat g_i(x)d(\widehat F_{s_i,i}^*-\widehat F_i)(x)+\frac{1}{2}\sum_{i_1,i_2=1}^m\int \hat g_{i_1i_2}(x_1,x_2)\prod_{k=1}^2d(\widehat F_{s_{i_k},i_k}^*-\widehat F_{i_k})(x_k)\\
&+\frac{1}{6}\sum_{i_1,i_2,i_3=1}^m\int \hat g_{i_1i_2i_3}(x_1,x_2,x_3)\prod_{k=1}^3d(\widehat F_{s_{i_k},i_k}^*-\widehat F_{i_k})(x_k)+\epsilon_3^*
\end{align*}
satisfies $\mathbb E_*[(\epsilon_3^*)^2]=o_p(s^{-3})$. In addition, assume the high order empirical influence functions $\hat g_{i_1i_2}$ and $\hat g_{i_1i_2i_3}$ converge in mean square error, i.e.
\begin{align*}
\mathbb E[(\hat g_{i_1i_2}-g_{i_1i_2})^2(X_{i_1,1},X_{i_2,j_2})]\to 0,\;\mathbb E[(\hat g_{i_1i_2i_3}-g_{i_1i_2i_3})^2(X_{i_1,1},X_{i_2,j_2},X_{i_3,j_3})]\to 0
\end{align*}
for all $i_1,i_2,i_3$ and $j_2\leq 2,j_3\leq 3$, where $X_{i,j}$ is the $j$-th data point from the $i$-th input model. For the first order influence function $\hat g_i$, assume the remainder in the Taylor expansion
\begin{equation*}
\hat g_i(X_{i,1})=g_i(X_{i,1})+\sum_{i'=1}^m\int g_{ii'}(X_{i,1},x)d(\widehat F_{i'}-F_{i'})(x)-\int g_{i}(x)d(\widehat F_{i}-F_{i})(x)+\epsilon_g
\end{equation*}
satisfies $\mathbb E[\epsilon_g^2]=o(n^{-1})$.
\end{assumption}

As for Assumptions \ref{smoothness_truth} and \ref{smoothness_empirical}, finite-horizon performance measures under mild conditions satisfy the above two assumptions:
\begin{theorem}\label{finite_horizon_3smoothness}
Under Assumptions \ref{size_data}, \ref{var_pos:finite} and Assumption \ref{moment8:finite} with $k=4$, we have Assumptions \ref{third-differentiability}-\ref{3smoothness_empirical} hold for the finite-horizon performance measure $\psi$ given by \eqref{finite_horizon:pm}.
\end{theorem}
With Assumptions \ref{3smoothness_truth} and \ref{3smoothness_empirical}, we can identify the statistical error of our variance estimator assuming infinite computation resources, which we summarize in the following lemma.
\begin{lemma}\label{PSBV:error}
Under Assumptions \ref{size_data}, \ref{smoothness_truth}-\ref{smoothness_empirical} and \ref{third-differentiability}-\ref{3smoothness_empirical}, the statistical error of the proportionate subsampled bootstrap variance is characterized by
\begin{equation}\label{stat_error:PSBV}
\sigma_{SVB}^2-\sigma_I^2=\mathcal Z+\mathcal R+o_p(\frac{1}{ n^{3/2}}+\frac{1}{ ns})
\end{equation}
where $\mathcal Z$ is a random variable such that
\begin{equation*}
\mathbb E[\mathcal Z]=0,\;\mathrm{Var}[\mathcal Z]=\sum_{i=1}^m\frac{\lambda_i^T\Sigma_i\lambda_i}{n_i}
\end{equation*}
with $\lambda_i=(1/n_i,2/n_1,\ldots,2/n_m)^T$ and
\begin{equation*}
\Sigma_i=\text{covariance matrix of }(g^2_i(X_i),\mathbb E_{X'_1}[g_1(X'_1)g_{1i}(X'_1,X_i)],\ldots,\mathbb E_{X'_m}[g_m(X'_m)g_{mi}(X'_m,X_i)]).
\end{equation*}
$\mathcal R$ is defined as
\begin{eqnarray*}
\nonumber\mathcal R&=&\sum_{i=1}^m\frac{1}{n_is_i}\mathrm{Cov}(g_i(X_{i}),g_{ii}(X_{i},X_{i}))+\sum_{i,i'=1}^m\frac{1}{n_is_{i'}}\mathrm{Cov}(g_i(X_{i}),\mathbb E_{X'_{i'}}[g_{ii'i'}(X_{i},X'_{i'},X'_{i'})])\\
&&+\sum_{i=1}^m\frac{\mathrm{frac}(\theta n_i)\sigma_i^2}{n_is_i}+\sum_{i, i'=1}^m\frac{\mathrm{Var}[g_{ii'}(X_{i},X_{i'}')]}{4n_{i}s_{i'}}
\end{eqnarray*}
where $\mathrm{frac}(x):=x-\lfloor x \rfloor$ denotes the fraction part of $x\in \R$, and for each $i$, $X_i,X'_i$ are independent copies of the random variable distributed under $F_i$.
\end{lemma}


Combining the statistical error \eqref{stat_error:PSBV}, and the minimal Monte Carlo error \eqref{opt_mse:anova} under the optimal budget allocation into the trade-off \eqref{error_decompose}, we obtain the overall error of the output $\hat\sigma_{SVB}^2$ of Algorithm \ref{anova sub}:
\begin{theorem}[Overall error of the variance estimate]\label{overall_error}
Suppose Assumptions \ref{size_data}, \ref{smoothness_truth}-\ref{anova_mu4_sim} and \ref{third-differentiability}-\ref{3smoothness_empirical} hold. Given a simulation budget $N$ and a subsample ratio $\theta$ such that $N=\omega(\theta n)$ and $\theta=\omega(n^{-1})$, if outer and inner sizes $B,R$ for Algorithm \ref{anova sub} are chosen to be $R=\Theta(\theta n),B=N/R$, then the gross error of our Monte Carlo estimate $\hat{\sigma}_{SVB}^2-\sigma_I^2=\mathcal E+o_p(\theta^{1/2}(Nn)^{-1/2}+\theta^{-1}n^{-2}+n^{-3/2})$, where the leading term has a mean squared error
\begin{equation}\label{gross_error}
\mathbb E[\mathcal E^2]= \Theta\big(\frac{\theta}{Nn}+\mathcal R^2+\sum_{i=1}^m\frac{\lambda_i^T\Sigma_i\lambda_i}{n_i}\big)
\end{equation}
where $\mathcal R$, $\lambda_i$'s and $\Sigma_i$'s are defined in Lemma \ref{PSBV:error}.
\end{theorem}
It is clear from their definitions in Lemma \ref{PSBV:error} that $\mathcal R=O(\theta^{-1}n^{-2})$ and each $(\lambda_i^T\Sigma_i\lambda_i)/n_i=O(n^{-3})$, hence the mean squared error \eqref{gross_error} is in general of order $O(\theta(Nn)^{-1}+\theta^{-2}n^{-4}+n^{-3})$. When $\mathcal R$ and at least one of the $\lambda_i^T\Sigma_i\lambda_i$'s satisfy the non-degeneracy condition in Theorem \ref{optimal_allocation}, this bound becomes tight in order, and the optimal subsample ratio can be established by minimizing the order of the leading overall error $\mathcal E$.

\section{Numerical Experiments}\label{sec:numerics}
This section reports our experimental findings. We consider two examples with different scales and complexities:

\textbf{M/M/1 queue:} The first example we consider is an M/M/1 queue that has true arrival rate $0.5$ and service rate $1$. Suppose the system is empty at time zero. The performance measure of interest is the probability that the waiting time of the $20$-th arrival exceeds $2$ units of time, whose true value is approximately $0.182$. Specifically, the system has two input distributions, i.e., the inter-arrival time distribution $F_1=\mathrm{Exp}(0.5)$ and the service time distribution $F_2=\mathrm{Exp}(1)$, for which we have $n_1$ and $n_2$ i.i.d.~data available respectively. If $A_t$ is the inter-arrival time between the $t$-th and $(t+1)$-th arrivals, and $S_t$ is the service time for the $t$-th arrival, then the system output
\begin{equation*}
\psi(F_1,F_2)=\mathbb E_{F_1,F_2}[\mathbf{1}\{W_{20}>2\}]
\end{equation*}
where the waiting time $W_{20}$ is calculated by the Lindley recursion $W_{t+1}=\max \{W_t+S_t-A_t,0\}$ for $t=1,\ldots,19$ and $W_1=0$. To test the proposed approach under different levels of utilization, we also consider true arrival rate $0.9$ and service rate $1$, for which case the target performance measure is taken to be the probability that the waiting time of the $20$-th arrival exceeds $6$ units of time (true value $0.190$). The data sizes $n_1,n_2$ are chosen so that $n_1=2n_2$ in the experiments, so only the minimum $\min_in_i$ is reported for convenience.

\begin{figure}[h] 
   \begin{center}
   \includegraphics[width=3.5in]{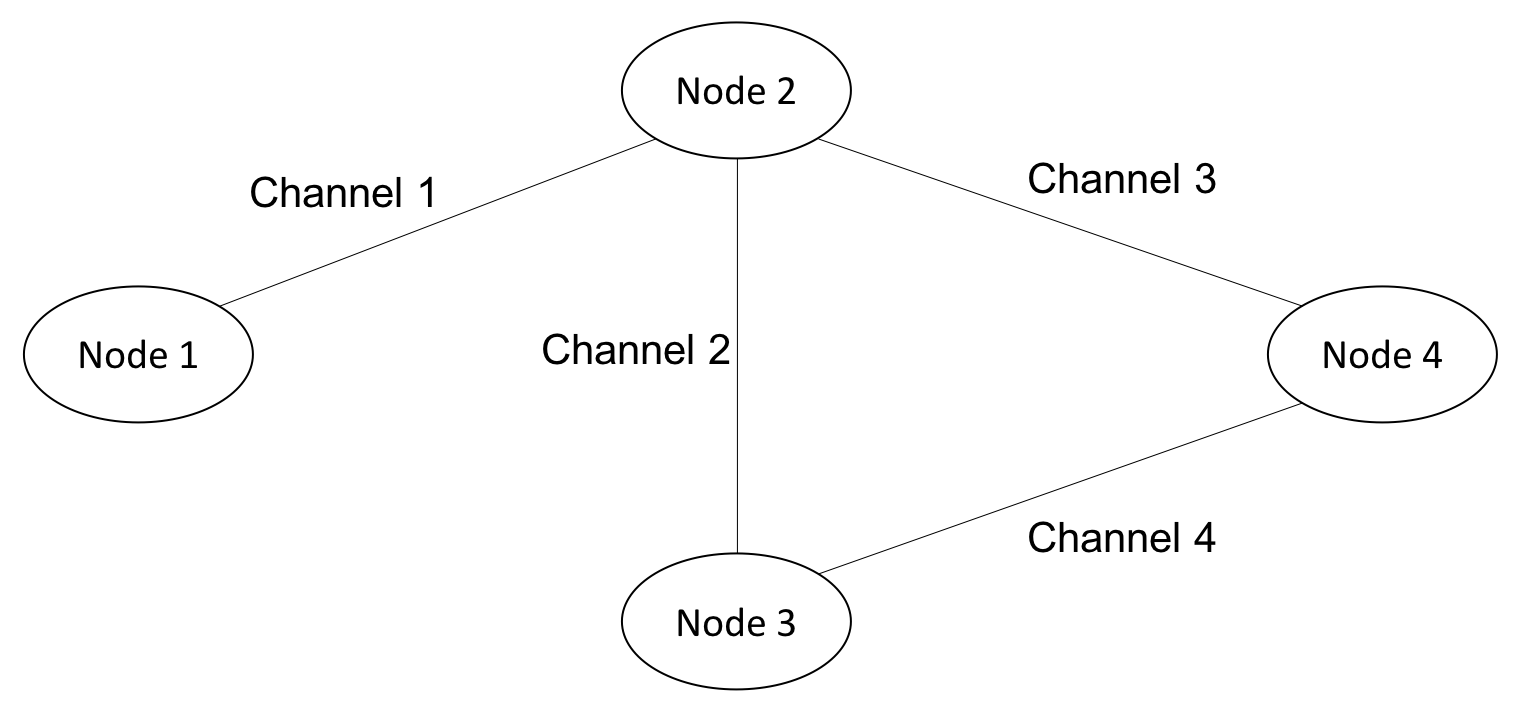}
   \end{center}
   \caption{A computer network with four nodes and four channels.}
   \label{network}
\end{figure}

\textbf{Computer network: }We also consider a computer communication network borrowed from \cite{cheng1997sensitivity} and \cite{lin2015single}. The structure of the system is characterized by the undirected graph in Figure \ref{network}: Four message-processing units, which correspond to the nodes, are connected by four transport channels that are represented by the edges.
For every pair $i,j$ of processing units with $i\neq j$, there are external messages that enter into unit $i$ and are to be transmitted to unit $j$ through a fixed path, and their arrival follows a Poisson process with rate $\lambda_{i,j}$. The specific values for $\lambda_{i,j}$'s are summarized in Table \ref{network: true rates}.
\begin{table}[h]
\begin{center}
\renewcommand{\arraystretch}{0.75}
\begin{tabular}{|c|cccc|}
\hline
\diagbox[width=20mm, height=13mm]{node $i$}{node $j$}&1&2&3&4\\\hline
1&n.a.&40&30&35\\
2&50&n.a.&45&15\\
3&60&15&n.a.&20\\
4&25&30&40&n.a.\\\hline
\end{tabular}
\end{center}
\caption{True arrival rates $\lambda_{i,j}$ of messages to be transmitted from node $i$ to node $j$.}
\label{network: true rates}
\end{table}
Each unit takes a constant time of $0.001$ seconds to process a message, and has unlimited storage capacity. The messages have lengths that are independent and follow an exponential distribution with mean $300$ bits, and each channel has a capacity of $275000$ bits, therefore there are queuing and transmission delays. The messages travel through the channels with a velocity of $150000$ miles per second, and the $i$-th channel has a length of $100 \cdot i$ miles for $i=1,2,3,4$, leading to a propagation delay of $\frac{100\cdot i}{150000}$ seconds along the $i$-th channel. The total time that a message of length $l$ bits occupies the $i$-th channel is therefore $\frac{l}{275000} + \frac{100\cdot i}{150000}$ seconds. Suppose the system is empty at time zero. The performance measure of interest is the average delay of the first $30$ messages that arrive to the system, or mathematically, $\mathbb E[\frac{1}{30}\sum_{k=1}^{30}D_k]$, where $D_k$ is the time for the $k$-th message to be transmitted from its entering node to destination node. The true value of the performance measure is approximately $6.91\times 10^{-3}$ seconds. In the experiment, we assume that the arrival rates of the different types of messages, as well as the distribution of the message length, are unknown, therefore there are $13$ input models in total. Like in the example of M/M/1 queue, the data sizes across different input models are kept proportional to each other and only the minimum size is reported.

In the experiments we investigate the simulation efforts needed for our subsampling procedure to generate accurate estimates of the input variance, the impacts of the procedural parameters $\theta,B,R$ on the estimation accuracy, and practical guidelines on optimal choices of these parameters. Regarding performance metrics of the method, we primarily focus on the mean squared error of the obtained input variance estimate. In addition, note that our estimated input variance can also be used to construct CIs by plugging into formula \eqref{CLT1}. We also examine the quality of these CIs, measured by coverage accuracy and width, as impacted by the estimation accuracy of the input variance.

We compare our subsampling approach with the variance bootstrap depicted in Algorithm \ref{anova} and the percentile bootstrap suggested by \cite{barton1993uniform,barton2001resampling}. The percentile bootstrap adopts the same nested simulation structure as in variance bootstrap, but does not estimate the input variance and instead directly outputs order statistics of the resampled performance measures to construct CIs. Specifically, after obtaining $B$ bootstrapped performance measure estimates $\bar{\psi}^b:=\frac{1}{R}\sum_{r=1}^R\hat{\psi}_r(\widehat{F}_1^b,\ldots,\widehat{F}_m^b)$, each averaged over $R$ i.i.d. replications, the percentile bootstrap outputs the $\frac{\alpha}{2}(B+1)$-th and $(1-\frac{\alpha}{2})(B+1)$-th order statistics of $\{\bar{\psi}^b:b=1,\ldots,B\}$ as a $(1-\alpha)$-level CI.


In converting our subsampled input variance estimate to CI, we also investigate the use of a ``splitting" versus a ``non-splitting" approach. In most part of this section, we use the splitting approach that divides the budget into two portions with one used to estimate the input variance and the other to compute the point estimator. To describe it in detail, suppose we have a total budget of $N$ simulation runs. We allocate $R_v$ simulation runs to estimate $\sigma_I^2$ using either Algorithm \ref{anova} or \ref{anova sub}, and the remaining $R_e=N-R_v$ simulation runs driven by the empirical input distributions to compute the point estimator $\bar\psi(\widehat F_1,\ldots,\widehat F_m)$. When constructing the CI in \eqref{CLT1}, the simulation variance $\sigma_S^2$ is calculated as $\frac{\tilde\tau^2}{R_e}$, where $\tilde\tau^2$ is the sample variance computed from the $R_e$ simulation replications. The second, ``non-splitting", approach invests all the $N$ simulation runs in estimating $\sigma_I^2$, and constructs the point estimator by averaging all the replications, i.e., $\bar\psi=\frac{1}{B}\sum_{b=1}^B\bar{\psi}^b$, where $\bar{\psi}^b$ is the performance measure estimate for the $b$-th resample from Algorithm \ref{anova sub}. The simulation variance $\sigma_S^2$ in this case is taken to be the sample variance of all the $\bar\psi^b$'s divided by the bootstrap size $B$. The rationale for this approach is that, when the subsample size $\theta n$ is large, $\mathbb E_*[\bar{\psi}]$ should accurately approximate the plug-in estimator $\psi(\widehat F_1,\ldots,\widehat F_m)$ with an error that is negligible relative to the input variability. Using the former as a surrogate for the latter avoids splitting the budget; however, we will see later that this may introduce too much bias to maintain the desired coverage level when the subsample size is relatively small.

The rest of this section is organized as follows. Section \ref{sec: configuration} investigates practical guidelines for choosing the algorithmic parameters in our procedure. Using these guidelines, in Section \ref{sec: comparison} we compare the proposed procedure with the variance bootstrap and the percentile bootstrap. Section \ref{sec: split vs nonsplit} studies further the conversion of input variance estimate into CI, and compares the associated splitting and non-splitting approaches.

\subsection{Guidelines for Algorithmic Configuration}\label{sec: configuration}
\begin{figure}[h] 
   \begin{subfigure}{0.5\textwidth}
      \centering
      \includegraphics[width=2.5in]{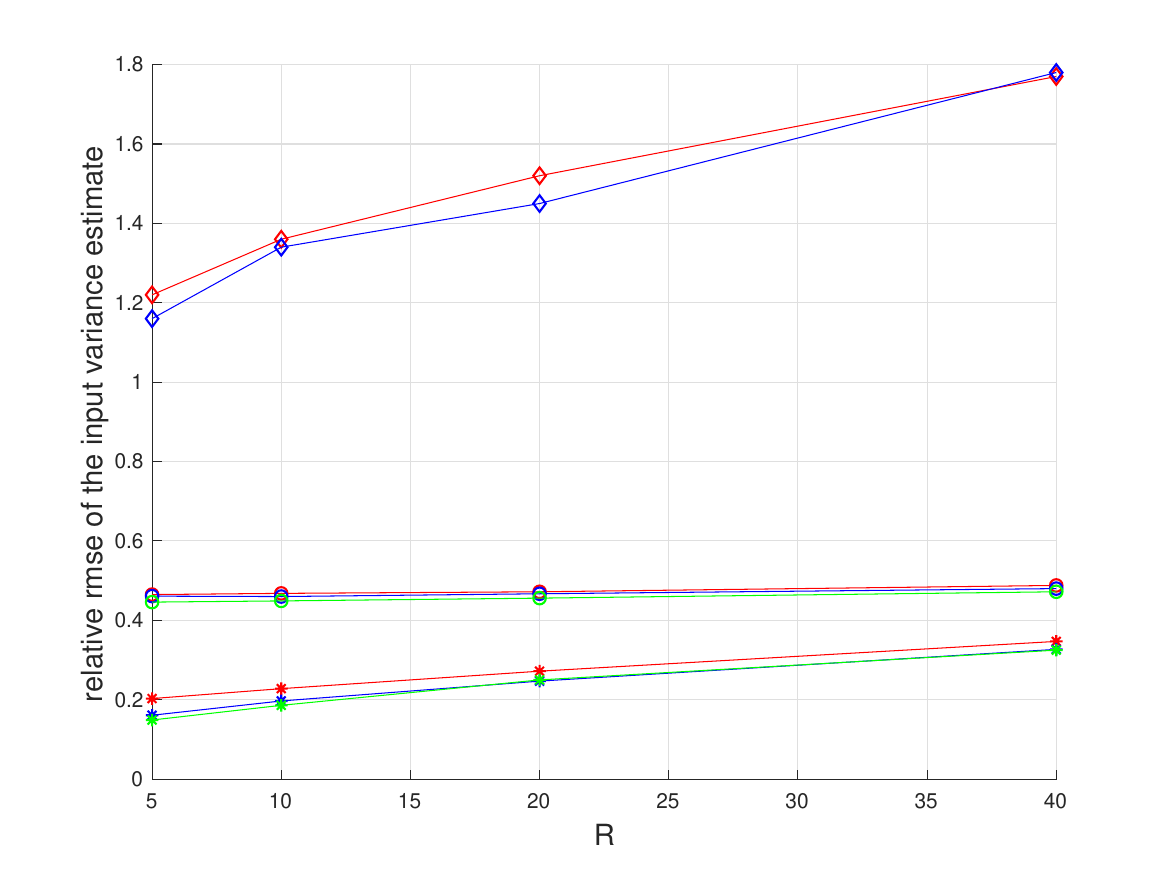}
      \caption{$\theta \min_in_i=5$.}
      \label{BR_choice:5}
   \end{subfigure}
   \begin{subfigure}{0.5\textwidth}
      \centering
      \includegraphics[width=2.5in]{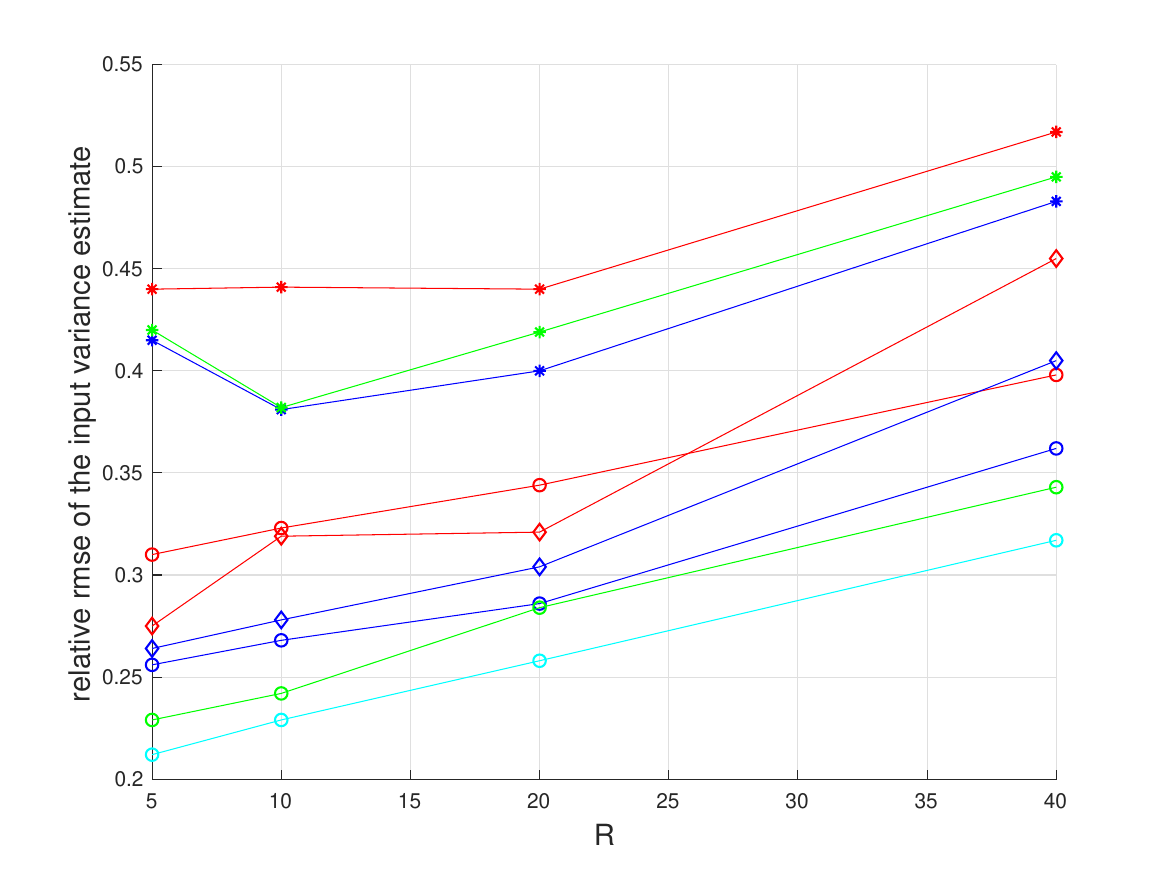}
      \caption{$\theta \min_in_i=30$.}
      \label{BR_choice:30}
   \end{subfigure}\\
   \begin{subfigure}{\textwidth}
      \centering
      \includegraphics[width=4in]{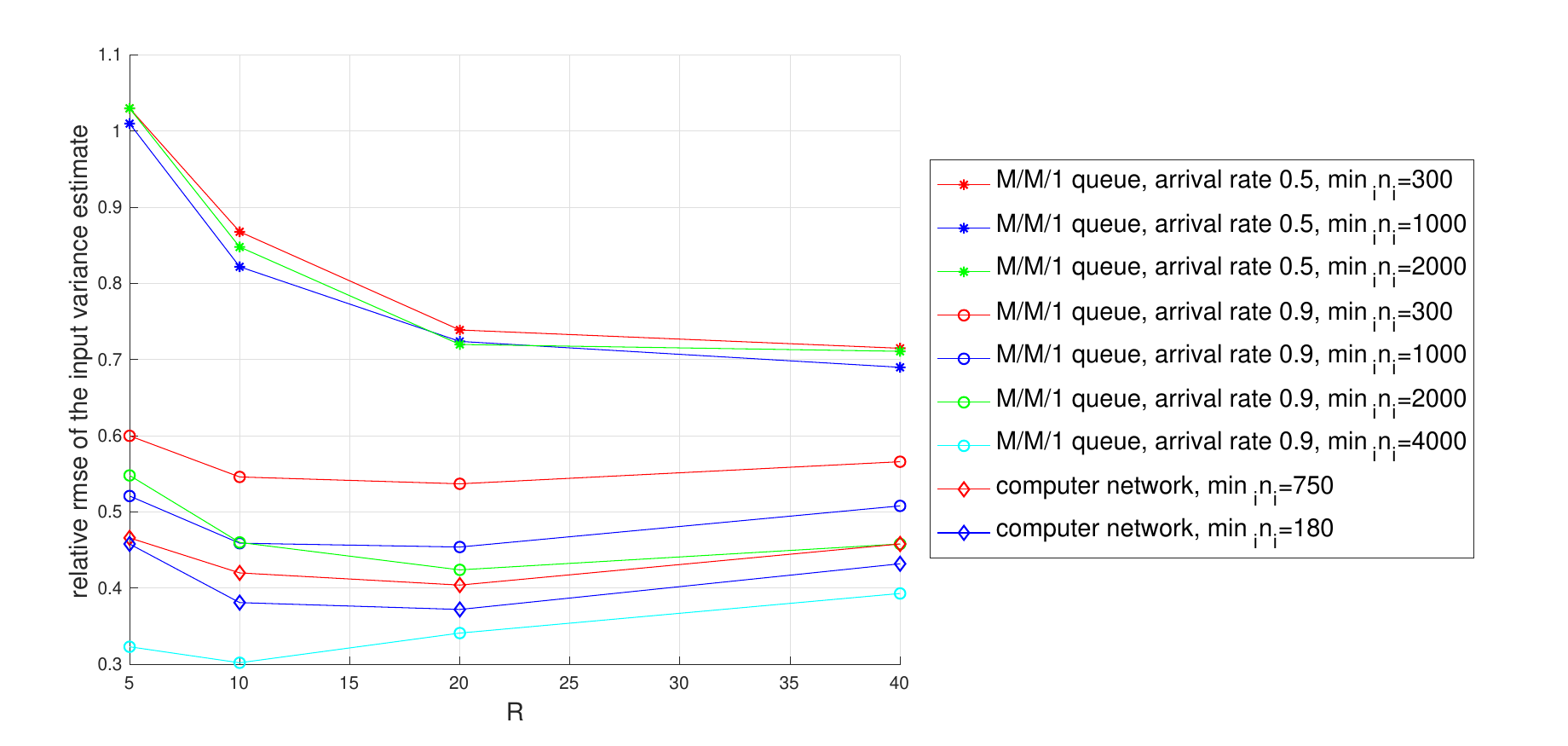}
      \caption{$\theta \min_in_i=120$.}
      \label{BR_choice:120}
   \end{subfigure}
   \caption{Input variance estimation accuracy under different configurations of $B,R$ such that $BR=1000$.}
   \label{result:BR_choice}
\end{figure}


We examine the performances using a wide range of parameter choices for  $\theta,B,R$. For each of the two considered examples, and input data sizes from $30$ to $2000$, we test our subsampling approach at various combinations of $\theta,B,R$ where the subsample size $\theta \min_in_i\in \{5,15,30,60,120\}$ and the budget allocation parameters $(B,R)\in\{(25,40),(50,20),(100,10),(200,5)\}$ (a total of $1000$ simulation runs). To calculate the mean square error of the input variance estimate, we perform $1000$ independent runs of the procedure, each on an independently generated input data set, and then take the average of the squared errors. The reported error metric is the relative root mean squared error (rmse) which can be expressed as $\frac{\sqrt{\mathbb E[(\hat\sigma_I^2-\sigma_I^2)^2]}}{\sigma_I^2}$ where $\hat\sigma_I^2$ and $\sigma_I^2$ are the estimated and true input variances respectively.

We first study and establish guidelines for the outer size $B$ and inner size $R$ for a given subsample size. Figure \ref{result:BR_choice} shows how the estimation error changes as the inner replication size $R$ grows from $5$ to $40$ (correspondingly the outer size $B$ drops from $200$ to $25$) and the subsample size $\theta \min_in_i$ is fixed at a certain value. Each curve represents the results for one of the considered examples under a particular input data size. Although the precise optimal choice for $B,R$ varies from one example to another even when the subsample size is chosen the same, the estimation error appears robust to the parameter choices, with a range of values that only slightly underperform the optimal. In particular, compared to the unknown optimal choice, an $R$ between $\frac{1}{6}\theta\min_in_i$ and $\frac{1}{3}\theta\min_in_i$ seems to achieve a comparable accuracy level in the variance estimation, hence is recommended as a general choice.
\begin{figure}[h] 
   \centering
   \includegraphics[width=4in]{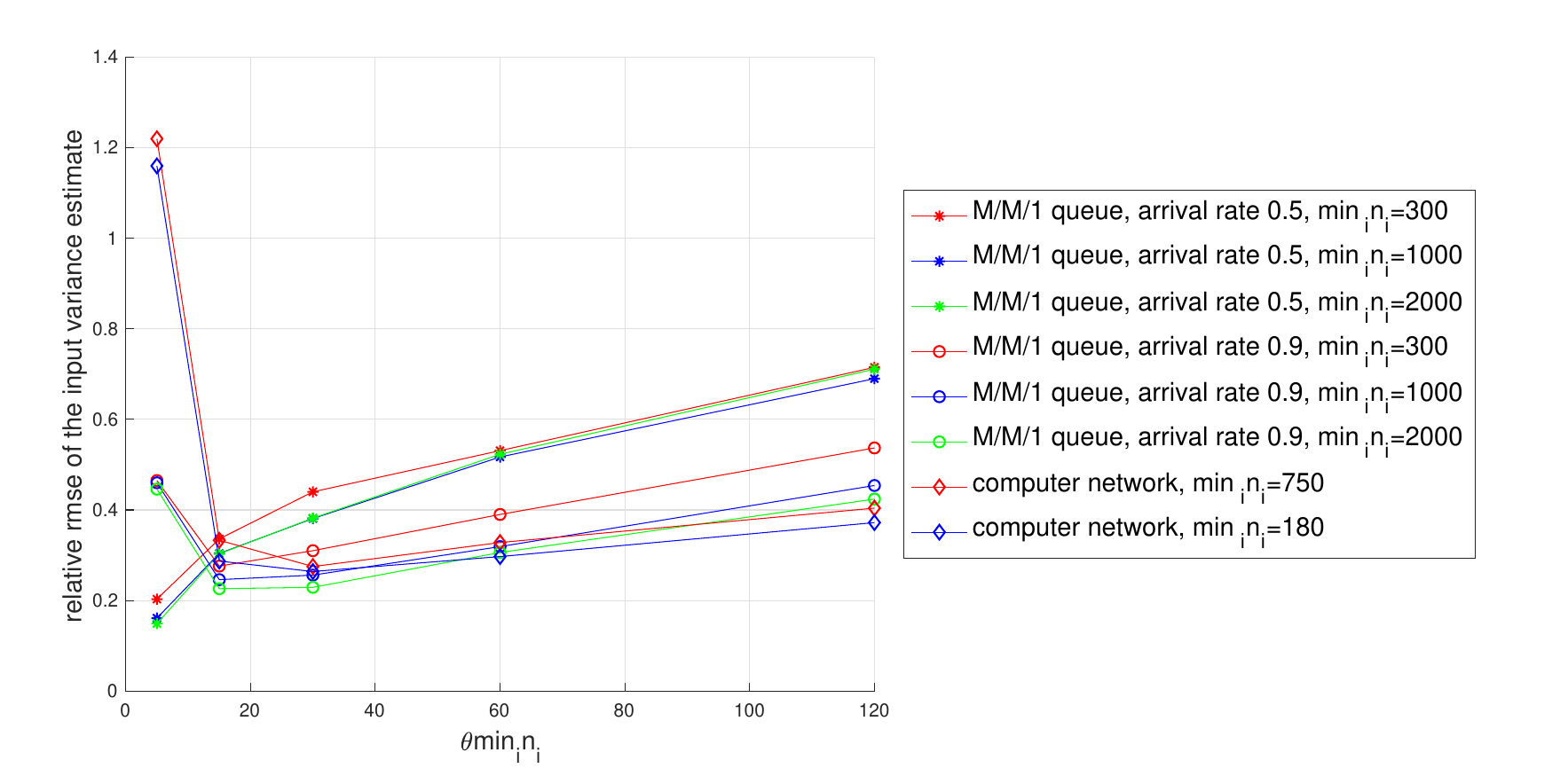}
   \caption{Input variance estimation accuracy under different subsample sizes with $B,R$ optimally tuned.}
   \label{result:theta_choice}
\end{figure}

Now we turn to optimal choices for the subsample size. Provided that $B,R$ is properly chosen as above, we examine the behavior of the variance estimation error as the subsample size varies. As we have discussed in Section \ref{sec:psvb}, subsampling is preferred when the input data size is relatively large, and thus we consider input data sizes $\geq 500$ for our M/M/1 queue and computer network, and for each considered data size we plot the variance estimation error versus the subsample size in Figure \ref{result:theta_choice}. We see that a too large size such as $120$ always leads to a larger estimation error than moderate sizes like $30$, whereas a too small size around $5$ can lift the error by even more in some cases, which is consistent with the theoretical insight from the bound \eqref{gross_error}. Therefore, in general we recommend the use of a subsample size $\theta \min_in_i$ between $20$ and $40$ to optimize the estimation accuracy. Figure \ref{result:theta_choice} shows that, under the suggested subsample size, the relative rmse is as low as $0.2$-$0.5$ across all the cases.

\subsection{Comparisons with the Variance Bootstrap and the Percentile Bootstrap}\label{sec: comparison}
We compare our subsampling method with the standard variance bootstrap and the percentile bootstrap, under the same total budget of $1500$ simulation runs. In addition to the relative rmse of the input variance estimate, we also report the actual coverage probability and width of the CI constructed by plugging in the input variance estimate. To estimate all these performance metrics, we construct $1000$ $95\%$-level CIs for the target performance measures, each from an independently generated input data set. The ``splitting'' approach that splits the total budget into $R_v=1000,R_e=500$ is adopted for the subsampling approach and the variance bootstrap, whereas for the percentile bootstrap all the $1500$ simulation runs are used for the resamples. As suggested in Section \ref{sec: configuration}, we use the parameter values $\theta=\frac{30}{\min_in_i},B=100,R=10$ in our method in all the cases, whereas for the other two methods we vary the parameter configurations over a reasonable range constrained by the simulation budget and then report the best results generated by these considered configurations. In particular, the parameters for the variance bootstrap are chosen to minimize the mean square error of the input variance estimate from four combinations, ``$B=25,R=40$'', ``$B=50,R=20$'', ``$B=100,R=10$'', ``$B=200,R=5$'', and those for the percentile bootstrap are chosen to achieve the best the coverage accuracy from four combinations, ``$B=50,R=30$'', ``$B=100,R=15$'', ``$B=300,R=5$'', ``$B=1500,R=1$''. Note that these give an upper hand to our competing alternatives in the comparisons.

Tables \ref{result:mm1_1} and \ref{result:mm1_2} summarize the experimental results for the M/M/1 queue when the true arrival rate is $0.5$ and $0.9$ respectively, and Table \ref{result:network} shows those for the computer network. The shorthand ``PSVB'' stands for proportionate subsampled variance bootstrap, i.e., our subsampling approach. For each method, the ``coverage estimate'' column displays estimates of the actual coverage probability based on $1000$ independent CIs, and the ``CI width'' column shows their average width. The second column of each table shows the ratio between the input standard error $\sigma_I$ and the simulation standard error $\sigma_S$ for different input data sizes in our ``splitting'' approach. A ratio close to or greater than $1$ means that the input noise is a major source of uncertainty relative to the simulation noise, thus indicating the need to be taken into account in output analysis.



\begin{table}[h]
\begin{center}
\begin{tabular}{|l|l|lll|lll|ll|}
\hline
\multirow{2}{*}{$\min_in_i$}&\multirow{2}{*}{$\frac{\sigma_I}{\sigma_S}$}&\multicolumn{3}{c|}{PSVB}&\multicolumn{3}{c|}{variance bootstrap}&\multicolumn{2}{c|}{percentile bootstrap}\\\cline{3-10}
&&\multicolumn{1}{l|}{\tabincell{l}{relative\\rmse}}&\multicolumn{1}{l|}{\tabincell{l}{coverage\\ estimate}}&\multicolumn{1}{l|}{CI width}&\multicolumn{1}{l|}{\tabincell{l}{relative\\rmse}}&\multicolumn{1}{l|}{\tabincell{l}{coverage\\ estimate}}&\multicolumn{1}{l|}{CI width}&\multicolumn{1}{l|}{\tabincell{l}{coverage\\ estimate}}&\multicolumn{1}{l|}{CI width}\\\hline
$30$&$7.74$&$0.73$&$84.3\%$&$0.422$&$0.73$&$84.3\%$&$0.422$&$91.9\%$&$0.467$\\\cline{1-1}
$100$&$3.77$&$0.55$&$92.5\%$&$0.251$&$0.80$&$88.6\%$&$0.248$&$98.8\%$&$0.356$\\\cline{1-1}
$300$&$2.13$&$0.44$&$94.8\%$&$0.156$&$1.04$&$85.6\%$&$0.148$&$99.9\%$&$0.307$\\\cline{1-1}
$1000$&$1.15$&$0.38$&$95.0\%$&$0.103$&$2.48$&$89.4\%$&$0.111$&$100\%$&$0.285$\\\cline{1-1}
$2000$&$0.79$&$0.38$&$95.9\%$&$0.087$&$5.43$&$92.8\%$&$0.107$&$100\%$&$0.280$\\\hline
\end{tabular}
\end{center}
\caption{Results for the M/M/1 queue with arrival rate $0.5$ and service rate $1$.}
\label{result:mm1_1}
\end{table}

\begin{table}[h]
\begin{center}
\begin{tabular}{|l|l|lll|lll|ll|}
\hline
\multirow{2}{*}{$\min_in_i$}&\multirow{2}{*}{$\frac{\sigma_I}{\sigma_S}$}&\multicolumn{3}{c|}{PSVB}&\multicolumn{3}{c|}{variance bootstrap}&\multicolumn{2}{c|}{percentile bootstrap}\\\cline{3-10}
&&\multicolumn{1}{l|}{\tabincell{l}{relative\\rmse}}&\multicolumn{1}{l|}{\tabincell{l}{coverage\\ estimate}}&\multicolumn{1}{l|}{CI width}&\multicolumn{1}{l|}{\tabincell{l}{relative\\rmse}}&\multicolumn{1}{l|}{\tabincell{l}{coverage\\ estimate}}&\multicolumn{1}{l|}{CI width}&\multicolumn{1}{l|}{\tabincell{l}{coverage\\ estimate}}&\multicolumn{1}{l|}{CI width}\\\hline
$30$&$11.12$&$0.59$&$81.4\%$&$0.609$&$0.59$&$81.4\%$&$0.609$&$94.6\%$&$0.639$\\\cline{1-1}
$100$&$6.22$&$0.42$&$89.9\%$&$0.372$&$0.63$&$88.6\%$&$0.386$&$97.2\%$&$0.446$\\\cline{1-1}
$300$&$3.46$&$0.32$&$92.6\%$&$0.225$&$0.71$&$87.0\%$&$0.225$&$99.3\%$&$0.348$\\\cline{1-1}
$1000$&$1.86$&$0.27$&$93.3\%$&$0.137$&$1.21$&$86.3\%$&$0.137$&$100\%$&$0.307$\\\cline{1-1}
$2000$&$1.30$&$0.24$&$95.0\%$&$0.108$&$2.19$&$90.7\%$&$0.119$&$100\%$&$0.294$\\\cline{1-1}
$4000$&$0.91$&$0.23$&$94.9\%$&$0.089$&$3.61$&$91.2\%$&$0.106$&$100\%$&$0.288$\\\hline
\end{tabular}
\end{center}
\caption{Results for the M/M/1 queue with arrival rate $0.9$ and service rate $1$.}
\label{result:mm1_2}
\end{table}

\begin{table}[h]
\begin{center}
\begin{tabular}{|l|l|lll|lll|ll|}
\hline
\multirow{2}{*}{$\min_in_i$}&\multirow{2}{*}{$\frac{\sigma_I}{\sigma_S}$}&\multicolumn{3}{c|}{PSVB}&\multicolumn{3}{c|}{variance bootstrap}&\multicolumn{2}{c|}{percentile bootstrap}\\\cline{3-10}
&&\multicolumn{1}{l|}{\tabincell{l}{relative\\rmse}}&\multicolumn{1}{l|}{\tabincell{l}{coverage\\ estimate}}&\multicolumn{1}{l|}{\tabincell{l}{CI width\\($\times 10^{-4}$)}}&\multicolumn{1}{l|}{\tabincell{l}{relative\\rmse}}&\multicolumn{1}{l|}{\tabincell{l}{coverage\\ estimate}}&\multicolumn{1}{l|}{\tabincell{l}{CI width\\($\times 10^{-4}$)}}&\multicolumn{1}{l|}{\tabincell{l}{coverage\\ estimate}}&\multicolumn{1}{l|}{\tabincell{l}{CI width\\($\times 10^{-4}$)}}\\\hline
$30$&$12.60$&$0.74$&$92.0\%$&$19.3$&$0.74$&$92.0\%$&$19.3$&$95.2\%$&$22.0$\\\cline{1-1}
$150$&$5.36$&$0.41$&$94.3\%$&$8.85$&$0.53$&$91.3\%$&$8.50$&$98.3\%$&$11.2$\\\cline{1-1}
$750$&$2.35$&$0.32$&$94.2\%$&$4.27$&$0.94$&$86.9\%$&$3.88$&$100\%$&$7.97$\\\cline{1-1}
$1800$&$1.53$&$0.28$&$95.3\%$&$3.03$&$1.63$&$87.1\%$&$3.01$&$100\%$&$7.34$\\\hline
\end{tabular}
\end{center}
\caption{Results for the computer network.}
\label{result:network}
\end{table}

We compare the approaches based on Tables \ref{result:mm1_1}-\ref{result:network}. Firstly, our subsampling approach significantly outperforms the variance bootstrap in terms of estimation accuracy of the input variance. The estimates generated by our approach have a smaller relative error than those by the variance bootstrap in all considered cases, and the gap becomes more significant as the data size grows larger. In particular, as the data size grows from $30$ to thousands, the estimation error keeps decreasing from $0.7$ to $0.25$ in our approach, whereas in variance bootstrap it keeps increasing from $0.7$ to larger than $1$, a level that makes the estimate too crude to be useful. These demonstrate the computational advantage and dictate the use of subsampling especially when the input data size is relatively large. Note that the same budget of $1000$ simulation runs are used in input variance estimation for all considered data sizes and that the estimation accuracy seems much better for large data sizes than for small sizes, and one may wonder whether more simulation runs should be used for small data sizes to further improve the estimation accuracy. It turns out that the estimation errors are mostly due to the inadequacy of the input data rather than the simulation budget, hence a budget of $1000$ is already large enough and further increasing the budget does not bring much benefit. For instance, in the case of data size $30$ in Table \ref{result:mm1_1}, the relative error of the input variance estimate remains as large as $0.69$ even if the simulation budget is increased by 10 times.


Secondly, thanks to the high accuracy in the input variance estimates, our subsampling approach generates accurate CIs whose coverage probabilities quickly approach the nominal level $95\%$ as the input data size grows. In contrast, the CIs using the variance bootstrap exhibit under-coverage, and the percentile bootstrap CIs significantly over-cover the truth. We see that the coverage of the variance bootstrap is below $90\%$ in most considered cases, and in the very few cases where the CIs happen to have relatively good coverages, the intervals are much wider than those by our subsampling approach. For example, in the case of data size $2000$ in Table \ref{result:mm1_1}, the variance bootstrap gives a fairly accurate coverage $92.8\%$, but on average the interval is $1.23$ times as wide as that by our method. This shows that the better estimates of the input variance using subsampling translate to better CIs significantly compared to using the variance bootstrap, in terms of both coverage accuracy and width. The percentile bootstrap CIs show an overly high coverage probability close to $100\%$ and are $2$-$3$ times wider than those by subsampling for all considered input data sizes except $30$. The over-coverage issue of the percentile CIs arises because the order statistics capture only the input noise but not the simulation noise in the resampled performance measures, a phenomenon that has been discussed in \cite{barton2007presenting,barton2018revisiting}. When one can afford a sufficiently large budget of simulation relative to the input data size, the simulation noise can be made negligible so that the CIs have the correct coverage. However, when simulation resources are relatively limited (e.g., when data size $\geq 100$ in Tables \ref{result:mm1_1}-\ref{result:network}), the CIs are unnecessarily widened by the extra simulation noise that leads to over-coverage. We also notice that the percentile bootstrap CIs do show more accurate coverage than the other two methods when the input data size is $30$, which may suggest that the percentile bootstrap is the preferred approach to constructing CIs in small data cases. However, this outperformance is a result of optimally choosing the parameters $B,R$ in hindsight. In our experiments, this best parameter set varies from one case to another, and the actual coverage under different configurations varies in a range of $8\%$.

Thirdly, results across different input data sizes show that, the advantages of subsampling in both input variance estimation and CI construction are most significant in situations with relatively large input data size. Note that one may argue in such situations input uncertainty is negligible. However, whether this is indeed the case relates to the error tolerance of the decision-maker and the magnitude of the target performance measure itself. For the large data sizes we consider, the input noise appears still relatively substantial. For instance, when the input data size is $2000$ in Table \ref{result:mm1_2}, the average width of the CIs as a measure of the input uncertainty and simulation uncertainty combined amounts to as much as $57\%$ of the target tail probability, and that the input uncertainty serves as a major component of the total uncertainty (a ratio of $1.3$ relative to the simulation uncertainty).

Lastly, in situations with small input data size like $30$ the CI coverage clearly falls below $95\%$ in Tables \ref{result:mm1_1} and \ref{result:mm1_2}. This under-coverage phenomenon may appear to stem from the nonlinear effect of the performance measure that is inadequately captured by the Gaussian-approximation-based CI given in \eqref{CLT1}. The real reason, as our experiments suggest, turns out to be the insufficient accuracy of the input variance estimates. In fact, if the true input variance $\sigma_I^2$ (which can be accurately estimated by repeatedly generating independent input data sets) is plugged into \eqref{CLT1} to construct CIs, the coverage probability under the data size $30$ rises to $94\%$-$95\%$ for both the M/M/1 queue and the computer network. This indicates a positive impact of an accurate input variance estimate on the CI quality, a point that we will discuss further momentarily.

\subsection{Constructing CI via Input Variance and Comparisons of the Splitting and Non-Splitting Approaches}\label{sec: split vs nonsplit}

We study in more depth the relation between the input variance estimation accuracy and CI quality, and compare the splitting approach for CI construction that has been used in previous subsections, with the alternate non-splitting approach described at the beginning of this section. Finally, we provide practical budget allocation strategies for the splitting approach.

\begin{figure}[h] 
   \begin{subfigure}{0.5\textwidth}
      \centering
      \includegraphics[width=2.5in]{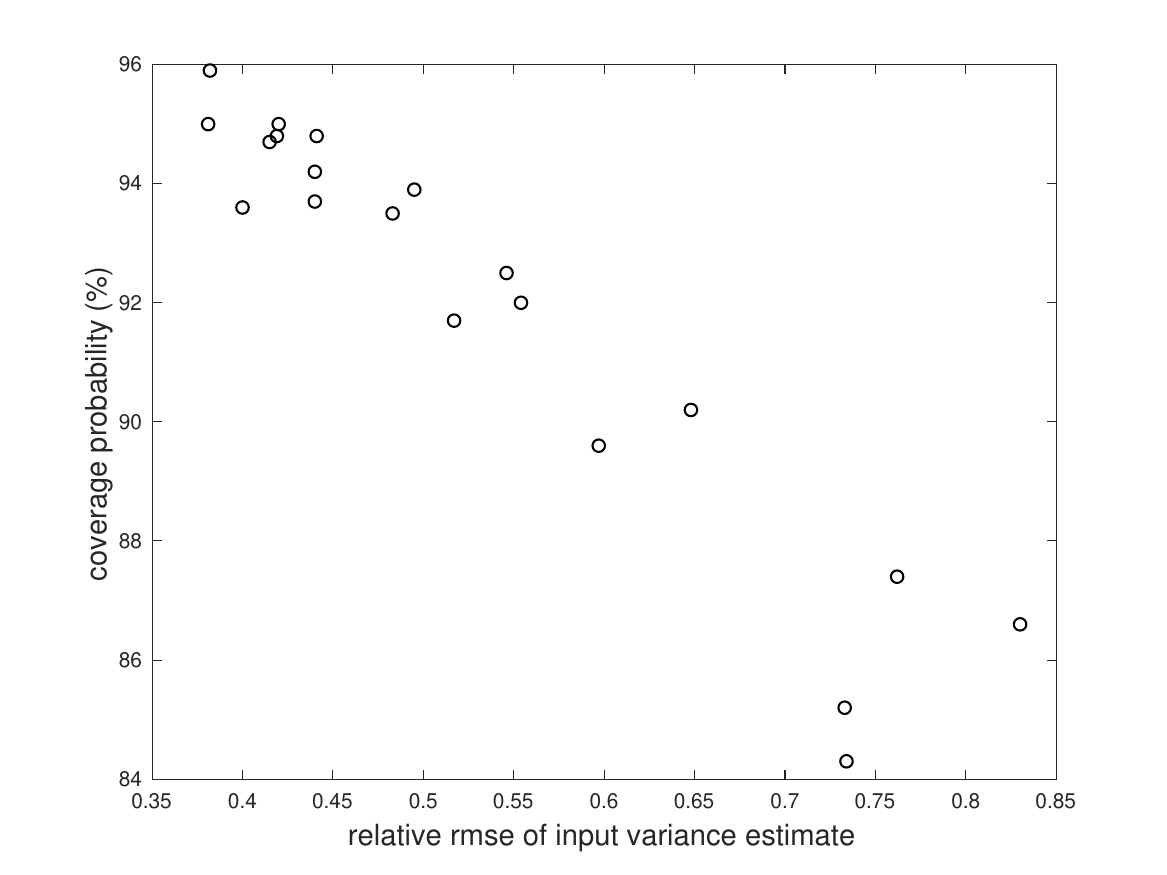}
      \caption{M/M/1 queue with arrival rate $0.5$, $\theta \min_in_i=30$.}
      \label{mse_to_cp:mm1}
   \end{subfigure}
   \begin{subfigure}{0.5\textwidth}
      \centering
      \includegraphics[width=2.5in]{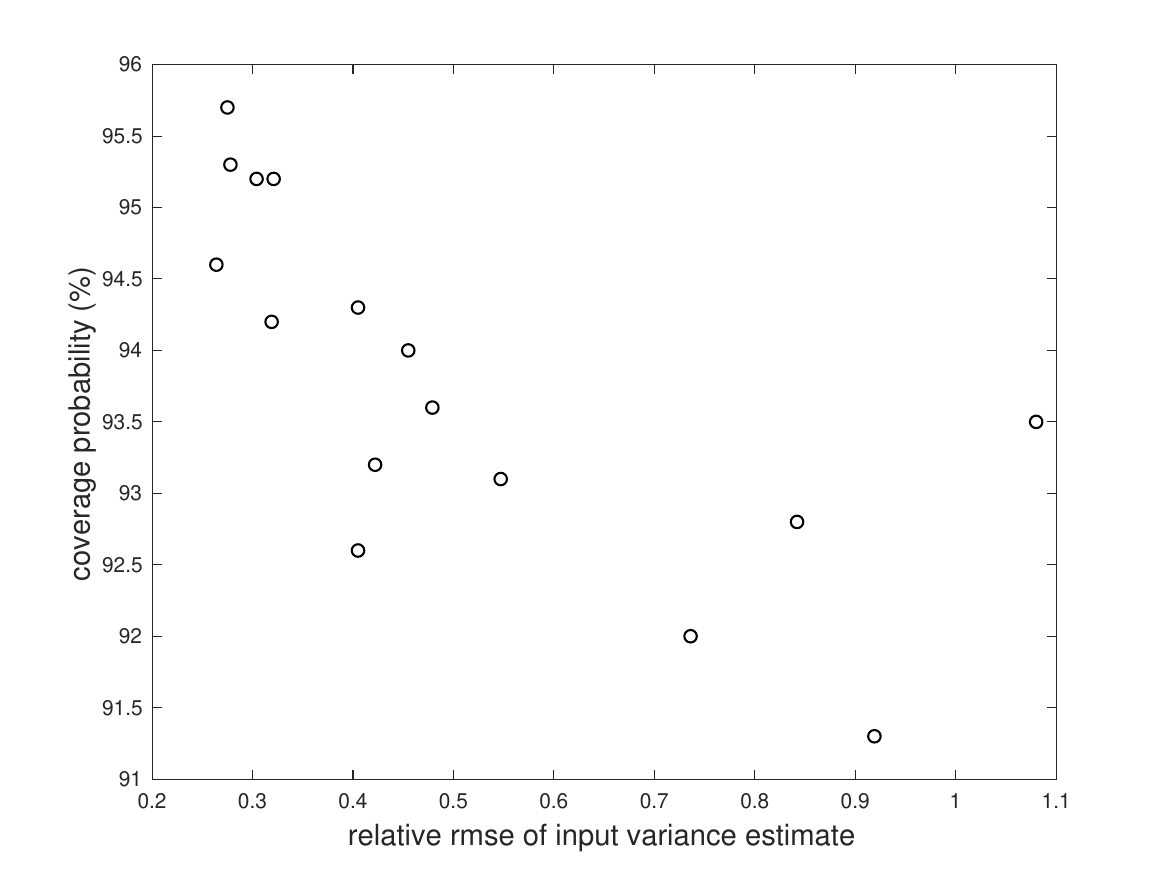}
      \caption{Computer network, $\theta \min_in_i=30$.}
      \label{mse_to_cp:network}
   \end{subfigure}
   \caption{Monotonicity between coverage accuracy and input variance estimation accuracy.}
   \label{result:mse_to_cp}
\end{figure}

First, to see how the estimation accuracy of the input variance affects the coverage accuracy of the CIs, we use the splitting approach to compute $95\%$-level CIs, with $1000$ simulation runs assigned to input variance estimation and another $500$ runs to point estimator evaluation. Figure \ref{mse_to_cp:mm1} plots the coverage probability versus the relative rmse when the subsample size $\theta \min_in_i$ is chosen $30$ in the M/M/1 queue example, where each point corresponds to a particular combination of the data size $\min_in_i$, the outer replication size $B$, and the inner replication size $R$. Figure \ref{mse_to_cp:network} plots the same for the computer network example with subsample size $30$. Both figures clearly show that, the more accurately the input variance is estimated, the closer to the nominal level $95\%$ the coverage probability will be. Accurate estimation of the input variance thus appears to play a crucial role in the construction of accurate CIs.

\begin{figure}[h] 
   \begin{subfigure}{0.5\textwidth}
      \centering
      \includegraphics[width=2.5in]{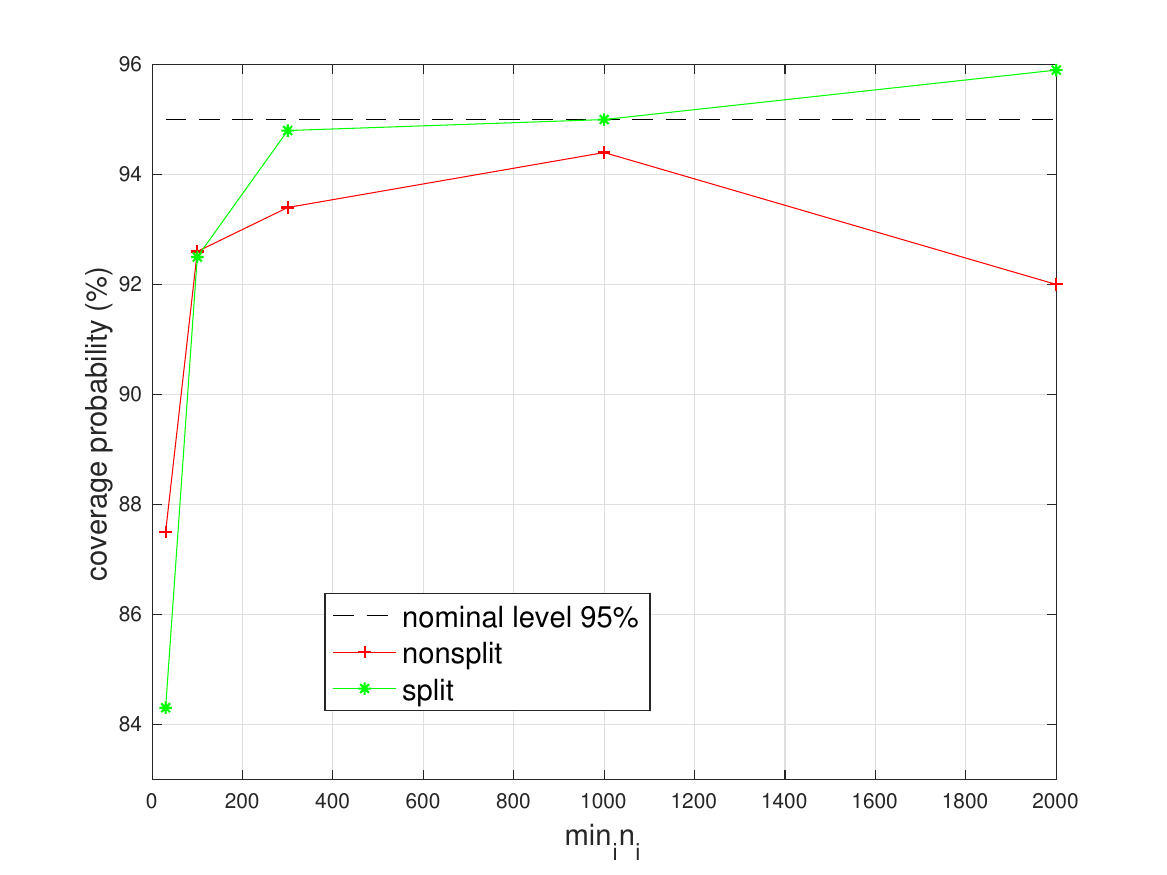}
      \caption{M/M/1 queue with arrival rate $0.5$.}
      \label{nonsplit:mm1_1}
   \end{subfigure}
   \begin{subfigure}{0.5\textwidth}
      \centering
      \includegraphics[width=2.5in]{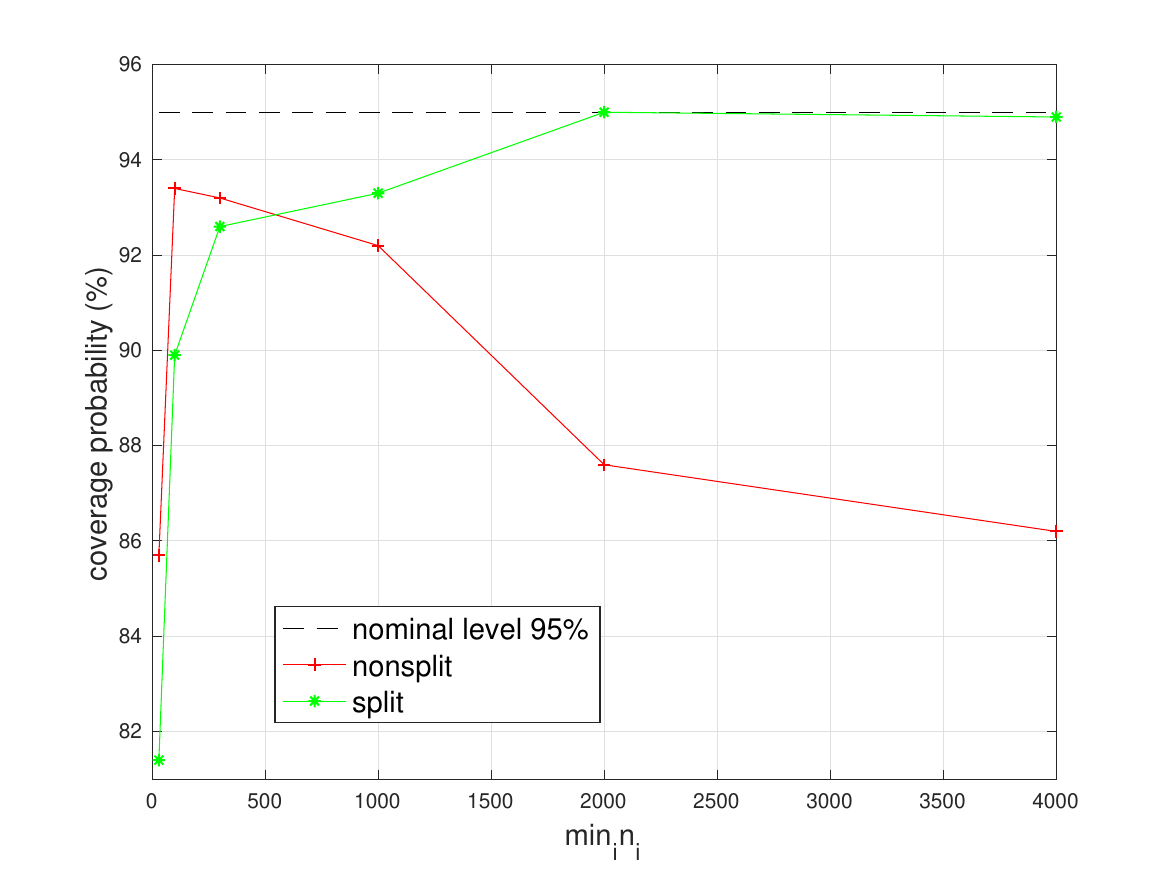}
      \caption{M/M/1 queue with arrival rate $0.9$.}
      \label{nonsplit:mm1_2}
   \end{subfigure}\\
   \begin{subfigure}{\textwidth}
      \centering
      \includegraphics[width=2.5in]{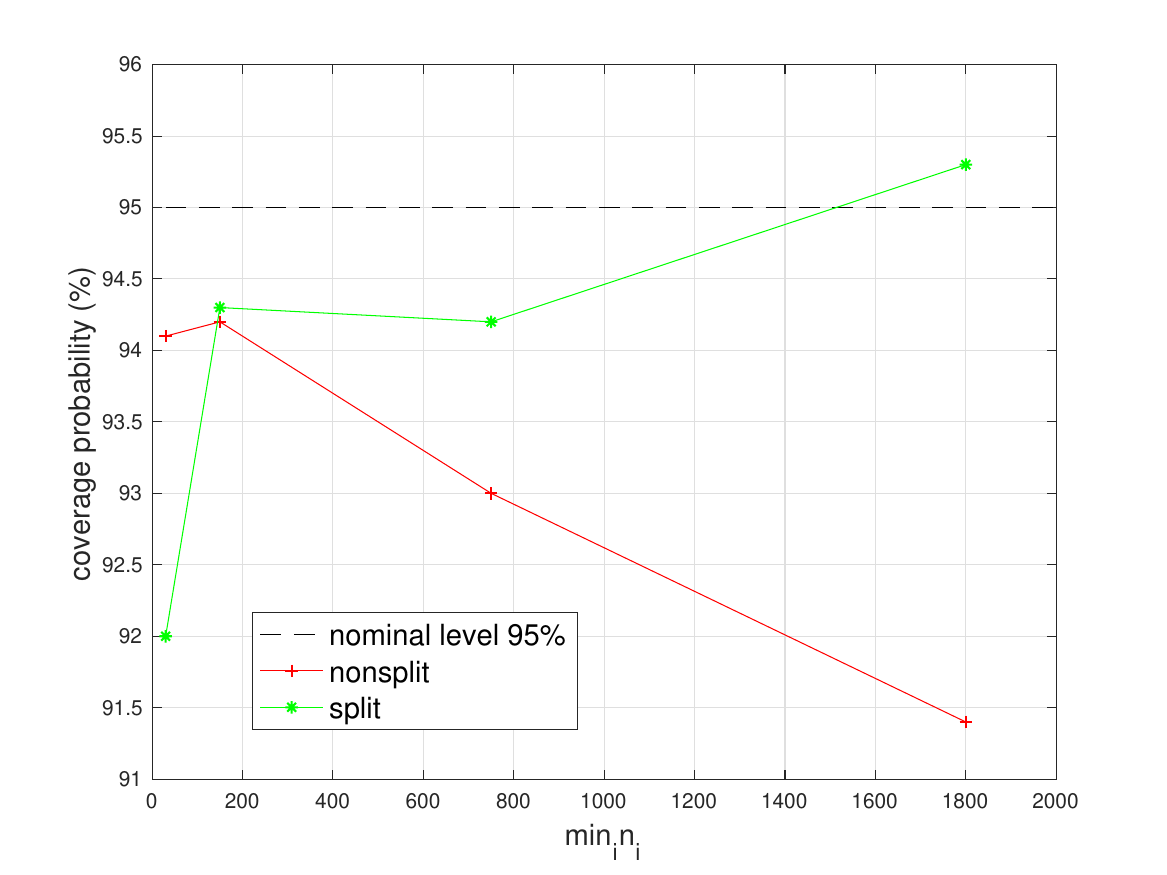}
      \caption{Computer network.}
      \label{nonsplit:network}
   \end{subfigure}
   \caption{Coverage comparison under the splitting and non-splitting approaches.}
   \label{result:nonsplit}
\end{figure}

Next we compare the splitting and non-splitting approaches under the same total budget of $1500$ simulation runs. Like in the splitting approach, we use a subsample size $\theta\min_in_i=30$ for our non-splitting approach, but use $B=75,B=20$ to consume all the $1500$ simulation runs. We find that the CIs generated from the two approaches have similar lengths, but the non-splitting approach underperforms in terms of coverage accuracy. Each plot in Figure \ref{result:nonsplit} shows the coverage probabilities of the non-splitting CIs versus the splitting ones for each of the considered example systems, as the input data size grows from $30$ to thousands. We see that when the data size is relatively small (e.g., below $500$), the two approaches generate CIs with similar coverage accuracy. When the data size grows larger, however, the coverage probability of the non-splitting CIs keeps dropping in all the three examples, especially in the M/M/1 queue with arrival rate $0.9$ where a drop towards $86\%$ is observed, whereas the splitting CIs exhibit almost exact $95\%$ coverage. A possible cause of the undercoverage is the overly small subsample size compared to the input data size, which leads to a high bias in the point estimator. With a subsample size $s$, the bias of the non-splitting point estimator $\mathbb E_*[\bar\psi]$ with respect to the truth $\psi(F_1,\ldots,F_m)$ can be as large as $O(1/s)$. Given that the input standard error is $\Theta(1/\sqrt n)$, $\mathbb E_*[\bar\psi]$ has a negligible bias only when the subsample size is large enough, namely when $s=\omega(\sqrt n)$, indicating that a small subsample size relative to the data size can corrupt the CI. In our experiment, we find that the (supposedly unobservable) bias can be as large as $25\%$ of the CI width when the input data size is $2000$ in the M/M/1 queue with arrival rate $0.9$, and that artificially removing the bias from the point estimator can improve the coverage to a similar level achieved by the splitting approach. Because of the bias and the consequent under-coverage issue, we caution the use of the non-splitting approach, that it should only be used when a relatively large subsample size is adopted.

\begin{figure}[h] 
   \centering
   \includegraphics[width=2.5in]{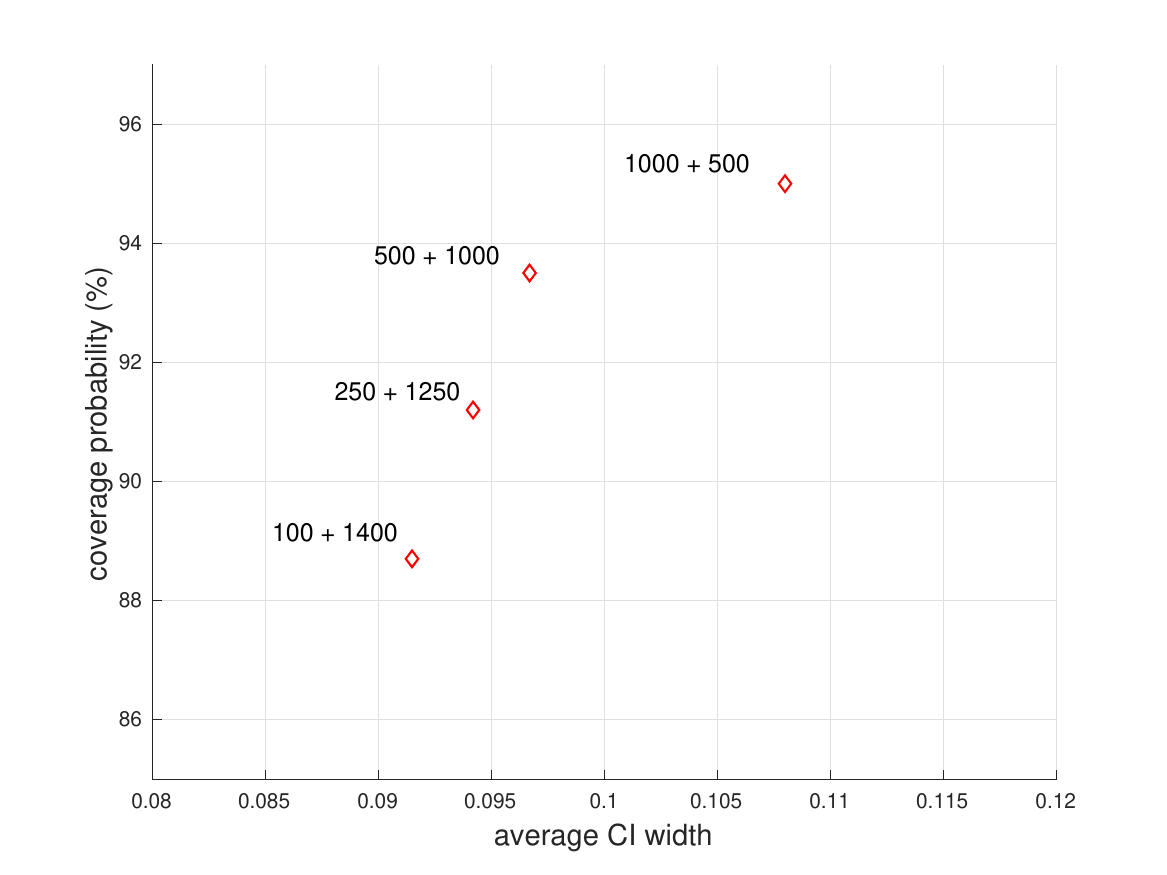}
   \caption{Coverage probability versus CI width, under different budget splits in the form of ``$R_v+R_e$''.}
   \label{result:split strategy}
\end{figure}

Since the splitting approach is recommended, next we explore strategies of splitting a given budget. Our goal is to generate shortest possible CIs that have a sufficiently accurate coverage probability. As in the beginning of the section, denote by $R_v$ the number of simulation runs used to estimate the input variance, and by $R_e$ to construct the point estimator. Under a fixed total budget $R_v+R_e=1500$, we try four different splits $R_v=100,250,500,1000$ (accordingly $R_e=1500-R_v$), and for each split the subsample size is fixed at $\theta \min_in_i=30$ and several choices of $B,R$ are tested among which the one with the best coverage probability is reported. Figure \ref{result:split strategy} plots the coverage probability versus the CI width for the four considered splits, where the M/M/1 queue with arrival rate $0.9$ is considered and input data size is $2000$. We notice that the split controls a tradeoff between the coverage accuracy and the CI width. The more simulation runs one allocates to input variance estimation, the more accurate but wider CIs one would obtain, because the input variance is more accurately estimated while the point estimator becomes more noisy. The plot suggests that allocating $500$-$1000$ replications to variance estimation achieves a good balance of accuracy and width, in the sense that the intervals from the split ``500+1000'' or ``$1000+500$'' are only slightly wider than those by other splits and that allocating less (say $250$) to variance estimation results in a considerable drop in coverage probability from the nominal level $95\%$. The results from Tables \ref{result:mm1_1}-\ref{result:network}, where the split ``$1000+500$'' is used, also validates the effectiveness of such a strategy. Therefore, for a given simulation budget, we recommend that the user allocate $500$-$1000$ replications to input variance estimation with our subsampling approach and all the remaining budget to the construction of the point estimator.

Lastly, to validate the various guidelines proposed in this section regarding the choices of the subsample size $\theta\min_in_i$, outer size $B$ and inner size $R$, as well as budget allocation strategies for the splitting approach to CI construction, we test their effectiveness and robustness under different configurations of the computer network. Specifically, under a fixed total simulation budget of $1500$ runs, we vary the channel capacity, the transmission speed of the channels, and the arrival rates of messages for the computer network (see Appendix \ref{sec:network configuration details} for these configuration details), otherwise keeping the same setting as stated at the beginning of this section, and apply PSVB with budget split ``$1000+500$'', subsample size $\theta\min_in_i=30$, and $B=100, R=10$ to compute input variance estimates and CIs. The standard variance bootstrap is also tested as a benchmark, with the $B$ and $R$ chosen in hindsight from four candidate combinations, ``$B=25,R=40$'', ``$B=50,R=20$'', ``$B=100,R=10$'', and ``$B=200,R=5$'', to minimize the mean squared error of the input variance estimate.
\begin{figure}[h]
    \centering
    \includegraphics[width=\textwidth]{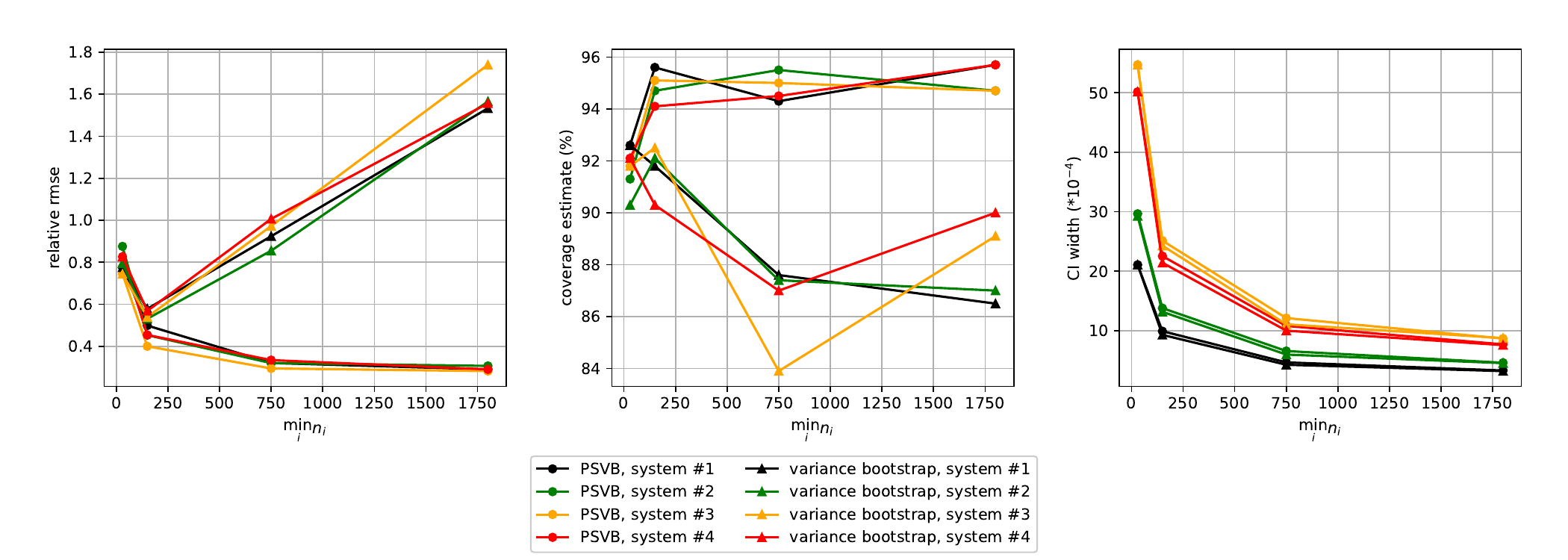}
    \caption{Comparison of PSVB and variance bootstrap under various configurations of the computer network.}
    \label{result:sensitivity}
\end{figure}
Plots of relative rmse, CI coverage estimate, and CI width against the input data size $\min_in_i$ are shown in Figure \ref{result:sensitivity}, where each line corresponds to either PSVB or the variance bootstrap applied to one of the four differently configured computer networks. The phenomena that we have observed in Tables \ref{result:mm1_1}-\ref{result:network} still persist for all the four computer networks. As the input data size grows, the estimation accuracy of the input variance improves to a level of $0.3$ in relative rmse for PSVB, but deteriorates significantly for the variance bootstrap. Accordingly the CI coverage of PSVB stays within a $1\%$ margin around the nominal level $95\%$, whereas the variance bootstrap shows a significantly lower coverage than the nominal level because of inaccurate input variance estimates. All these demonstrate that the proposed guidelines for using PSVB deliver superior and robust performance across different systems, and therefore can be used as a default algorithmic configuration in practice. However, the relatively low accuracy of the variance estimates and CIs appears again in our approach when the input data size is limited (around $30$). The same limitation arises for the variance bootstrap. This reconciles with our observation from Section \ref{sec: comparison} that subsampling is most beneficial for cases with moderately large input data sizes where other approaches like the variance bootstrap start to become computationally demanding.

\section{Conclusion}\label{sec:conclusion}
We have explained how estimating input variances in stochastic simulation can require large computation effort when using conventional bootstrapping. This arises as the bootstrap involves a two-layer sampling, which adds up to a total effort of larger order than the data size in order to achieve relative consistency. To alleviate this issue, we have proposed a subsampling method that leverages the relation between the structure of input variance and the estimation error from the two-layer sampling, so that the resulting total effort can be reduced to being independent of the data size. We have presented the theoretical results in this effort reduction, and the optimal choices of the subsample ratio and simulation budget allocation in terms of the data size and the budget. We have also demonstrated numerical results to support our theoretical findings, and provided guidelines in using our  proposed methods to estimate input variances and also construct output CIs. Future work comprises a more comprehensive investigation of our subsampling scheme, including its generalization to input processes with serial dependence and potentially non-smooth performance measures such as quantiles and other risk measures.
\ACKNOWLEDGMENT{We gratefully acknowledge support from the National Science Foundation under grants CMMI-1542020, CMMI-1523453 and CAREER CMMI-1653339/1834710. A preliminary conference version of this work, \cite{lam2018subsampling}, has appeared in the Proceedings of the Winter Simulation Conference 2018.}


\bibliographystyle{informs2014} 
\bibliography{bagging_ref} 

\ECSwitch
\ECHead{Proofs of Statements}
We first verify the proposed assumptions for the special case of finite-horizon performance measures in Section \ref{sec: verify finite horizon}. Section \ref{sec: input variance approximation} then proves results on the validity of the input variance decomposition \eqref{var decompose}. Section \ref{sec: consistency and monte carlo error} proves the consistency of the proposed input variance estimate and analyzes its Monte Carlo error in relation to the parameters $B,R$. Lastly, Section \ref{sec: overall error and subsample size} further analyzes the statistical error to obtain the overall error of the input variance estimate, and derives the optimal choices for $\theta,B,R$ that minimizes the overall error. In all the proofs, we write $a\approx b$ to mean $a/b\stackrel{p}{\to}1$.
\section{Finite-Horizon Performance Measures}\label{sec: verify finite horizon}
In this section, we show that Assumptions \ref{first-differentiability}-\ref{anova_mu4_sim} and \ref{third-differentiability}-\ref{3smoothness_empirical} hold for the finite-horizon performance measure \eqref{finite_horizon:pm}, thereby proving Theorems \ref{finitetime_justification} and \ref{finite_horizon_3smoothness}. We first prove Assumptions \ref{first-differentiability} and \ref{third-differentiability}, then present the useful Lemma \ref{error:key} which will later be used to prove all other assumptions.


\proof{Proof of Assumptions \ref{first-differentiability} and \ref{third-differentiability}.}
The finite horizon structure allows the following expansion of the performance measure $\psi(P_1^{\nu_1},\ldots,P_m^{\nu_m})$ around the input models $P_1,\ldots,P_m$
\begin{align*}
&\psi(P_1^{\nu_1},\ldots,P_m^{\nu_m})\\
=&\int h(\mathbf x_1,\ldots,\mathbf x_m)\prod_{i=1}^m\prod_{t=1}^{T_i}d(\nu_i(Q_i-P_{i})+P_{i})(x_{i,t})\\
=&\psi(P_{1},\ldots,P_{m})+\sum_{d=1}^T\sum_{\sum_{i=1}^m\lvert\mathcal T_i\rvert=d}\prod_{i=1}^m\nu_i^{\lvert\mathcal T_i\rvert}\int h(\mathbf x_1,\ldots,\mathbf x_m)\prod_{i=1}^m\prod_{t\notin \mathcal T_i}dP_{i}(x_{i,t})\prod_{i=1}^m\prod_{t\in \mathcal T_i}d(Q_i-P_i)(x_{i,t})\\
=&\psi(P_{1},\ldots,P_{m})+\sum_{d=1}^T\sum_{\sum_{i=1}^m\lvert\mathcal T_i\rvert=d}\prod_{i=1}^m\nu_i^{\lvert\mathcal T_i\rvert}\int h_{\mathcal T_1,\ldots,\mathcal T_m}(\mathbf x_{1,\mathcal T_1},\ldots,\mathbf x_{m,\mathcal T_m})\prod_{i=1}^m\prod_{t\in \mathcal T_i}d(Q_i-P_i)(x_{i,t})
\end{align*}
where $T=\sum_{i=1}^mT_i$ is the total run length, each $\mathcal T_i=\{\mathcal T_i(1),\ldots,\mathcal T_i(\lvert \mathcal T_i\rvert)\}$ is an ordered subset of $\{1,2,\ldots,T_i\}$, and
\begin{equation}\label{h1}
h_{\mathcal T_1,\ldots,\mathcal T_m}(\mathbf x_{1,\mathcal T_1},\ldots,\mathbf x_{m,\mathcal T_m})=\mathbb E_{P_1,\ldots,P_m}[h(\mathbf X_1,\ldots,\mathbf X_m)\vert X_i(t)=x_{i,t}\text{ for }i,t\in \mathcal T_i].
\end{equation}
Here each $\mathbf x_{i,\mathcal T_i}:=(x_{i,t})_{t\in\mathcal T_i}$. Expressing terms with $d=1,2,3$ in a more explicit form gives
\begin{align}
\notag&\psi(P_1^{\nu_1},\ldots,P_m^{\nu_m})\\
\notag=&\psi(P_{1},\ldots,P_{m})+\sum_{i=1}^m\nu_i\int \tilde{g}_i(x)d(Q_i-P_i)(x)+\sum_{i_1\leq i_2}\nu_{i_1}\nu_{i_2}\int \tilde{g}_{i_1i_2}(x,y)d(Q_{i_1}-P_{i_1})(x)d(Q_{i_2}-P_{i_2})(y)\\
\notag&+\sum_{i_1\leq i_2\leq i_3}\nu_{i_1}\nu_{i_2}\nu_{i_3}\int \tilde{g}_{i_1i_2i_3}(x,y,z)d(Q_{i_1}-P_{i_1})(x)d(Q_{i_2}-P_{i_2})(y)d(Q_{i_3}-P_{i_3})(z)\\
&+\sum_{d=4}^T\sum_{\sum_{i=1}^m\lvert\mathcal T_i\rvert=d}\prod_{i=1}^m\nu_i^{\lvert\mathcal T_i\rvert}\int h_{\mathcal T_1,\ldots,\mathcal T_m}(\mathbf x_{1,\mathcal T_1},\ldots,\mathbf x_{m,\mathcal T_m})\prod_{i=1}^m\prod_{t\in \mathcal T_i}d(Q_i-P_i)(x_{i,t}).\label{multiple_integral}
\end{align}
where
\begin{align*}
&\tilde{g}_i(x)=\sum_{1\leq t\leq T_i}\mathbb E_{P_1,\ldots,P_m}[h(\mathbf X_1,\ldots,\mathbf X_m)\vert X_i(t)=x]\\
&\tilde{g}_{i_1i_2}(x,y)=\begin{cases}
\sum_{1\leq t_1<t_2\leq T_i}\mathbb E_{P_1,\ldots,P_m}[h(\mathbf X_1,\ldots,\mathbf X_m)\vert X_i(t_1)=x,X_i(t_2)=y],&\text{ if }i_1=i_2=i\\
\sum_{t_1=1}^{T_{i_1}}\sum_{t_2=1}^{T_{i_2}}\mathbb E_{P_1,\ldots,P_m}[h(\mathbf X_1,\ldots,\mathbf X_m)\vert X_{i_1}(t_1)=x,X_{i_2}(t_2)=y],&\text{ if }i_1<i_2
\end{cases}\\
&\tilde{g}_{i_1i_2i_3}(x,y,z)=\begin{cases}
\sum_{1\leq t_1<t_2<t_3\leq T_i}\mathbb E_{P_1,\ldots,P_m}[h\vert X_i(t_1)=x,X_i(t_2)=y,X_i(t_3)=z],&\text{ if }i_1=i_2=i_3=i\\
\sum_{1\leq t_1<t_2\leq T_{i}}\sum_{t_3=1}^{T_{i_3}}\mathbb E_{P_1,\ldots,P_m}[h\vert X_{i}(t_1)=x,X_{i}(t_2)=y,X_{i_3}(t_3)=z],&\text{ if }i_1=i_2=i<i_3\\
\sum_{t_1=1}^{T_{i_1}}\sum_{1\leq t_2< t_3\leq T_{i}}\mathbb E_{P_1,\ldots,P_m}[h\vert X_{i_1}(t_1)=x,X_{i}(t_2)=y,X_{i}(t_3)=z],&\text{ if }i_1<i_2=i_3=i\\
\sum_{t_1=1}^{T_{i_1}}\sum_{t_2=1}^{T_{i_2}}\sum_{t_3=1}^{T_{i_3}}\mathbb E_{P_1,\ldots,P_m}[h\vert X_{i_1}(t_1)=x,X_{i_2}(t_2)=y,X_{i_3}(t_3)=z],&\text{ if }i_1<i_2<i_3
\end{cases}.
\end{align*}
Since each signed measure $Q_i-P_i$ in the product measure in \eqref{multiple_integral} has zero total measure, adding to the integrand a function that is independent of at least one of the integration variables does not change the integral value. Hence one can replace $\tilde{g}$'s by the following centered versions for $i_1\leq i_2\leq i_3$
\begin{align*}
\tilde{g}^c_i(x)&=\tilde{g}_i(x)-\mathbb E[\tilde{g}_i(X_i)]\\
\tilde{g}^c_{i_1i_2}(x,y)&=\tilde{g}_{i_1i_2}(x,y)-\mathbb E[\tilde{g}_{i_1i_2}(X_{i_1},y)]-\mathbb E[\tilde{g}_{i_1i_2}(x,X_{i_2})]+\mathbb E[\tilde{g}_{i_1i_2}(X_{i_1},X_{i_2}')]\\
\nonumber \tilde{g}^c_{i_1i_2i_3}(x,y,z)&=\tilde{g}_{i_1i_2i_3}(x,y,z)-\mathbb E[\tilde{g}_{i_1i_2i_3}(X_{i_1},y,z)]-\mathbb E[\tilde{g}_{i_1i_2i_3}(x,X_{i_2},z)]-\mathbb E[\tilde{g}_{i_1i_2i_3}(x,y,X_{i_3})]\\
\nonumber&\hspace{3ex}+\mathbb E[\tilde{g}_{i_1i_2i_3}(X_{i_1},X_{i_2}',z)]+\mathbb E[\tilde{g}_{i_1i_2i_3}(X_{i_1},y,X_{i_3}')]+\mathbb E[\tilde{g}_{i_1i_2i_3}(x,X_{i_2},X_{i_3}')]\\
&\hspace{3ex}-\mathbb E[\tilde{g}_{i_1i_2i_3}(X_{i_1},X_{i_2}',X_{i_3}'')]
\end{align*}
where $X_i,X_i',X_i''$ denote independent variables distributed under $F_i$, and replace the function $h_{\mathcal T_1,\ldots,\mathcal T_m}$ by
\begin{align}
\nonumber&h^c_{\mathcal T_1,\ldots,\mathcal T_m}(\mathbf x_{1,\mathcal T_1},\ldots,\mathbf x_{m,\mathcal T_m})\\
\nonumber=&h_{\mathcal T_1,\ldots,\mathcal T_m}-\sum_{i,t\in\mathcal T_i}\int h_{\mathcal T_1,\ldots,\mathcal T_m}dP_i(x_{i,t})+\sum_{(i_1,t_1)<(i_2,t_2),t_1\in\mathcal T_{i_1},t_2\in\mathcal T_{i_2}}\int h_{\mathcal T_1,\ldots,\mathcal T_m}dP_{i_1}(x_{i_1,t_1})dP_{i_2}(x_{i_2,t_2})+\cdots\\
&+(-1)^{\sum_{i=1}^m\lvert\mathcal T_i\rvert}\int h_{\mathcal T_1,\ldots,\mathcal T_m} \prod_{i=1}^m\prod_{t\in\mathcal T_i}dP_i(x_{i,t})\label{centered_h}
\end{align}
where the order $(i_1,t_1)<(i_2,t_2)$ is defined as either $i_1<i_2$, or $i_1=i_2$ but $t_1<t_2$. This leads to the new Taylor expansion
\begin{align}
\notag&\psi(P_1^{\nu_1},\ldots,P_m^{\nu_m})\\
\notag=&\psi(P_{1},\ldots,P_{m})+\sum_{i=1}^m\nu_i\int \tilde{g}^c_i(x)d(Q_i-P_i)(x)+\sum_{i_1\leq i_2}\nu_{i_1}\nu_{i_2}\int \tilde{g}^c_{i_1i_2}(x,y)d(Q_{i_1}-P_{i_1})(x)d(Q_{i_2}-P_{i_2})(y)\\
\notag&+\sum_{i_1\leq i_2\leq i_3}\nu_{i_1}\nu_{i_2}\nu_{i_3}\int \tilde{g}^c_{i_1i_2i_3}(x,y,z)d(Q_{i_1}-P_{i_1})(x)d(Q_{i_2}-P_{i_2})(y)d(Q_{i_3}-P_{i_3})(z)\\
&+\sum_{d=4}^T\sum_{\sum_{i=1}^m\lvert\mathcal T_i\rvert=d}\prod_{i=1}^m\nu_i^{\lvert\mathcal T_i\rvert}\int h^c_{\mathcal T_1,\ldots,\mathcal T_m}(\mathbf x_{1,\mathcal T_1},\ldots,\mathbf x_{m,\mathcal T_m})\prod_{i=1}^m\prod_{t\in \mathcal T_i}d(Q_i-P_i)(x_{i,t}).\label{centered_expansion}
\end{align}
Note that now all the integrands above have zero marginal means due to centering, e.g.
\begin{equation}\label{zeromarginal}
\int h^c_{\mathcal T_1,\ldots,\mathcal T_m}(\mathbf x_{1,\mathcal T_1},\ldots,\mathbf x_{m,\mathcal T_m})dP_{i}(x_{i,t})=0\text{ for all }i\text{ and }t\in\mathcal T_i.
\end{equation}
However, the functions $\tilde{g}^c_i,\tilde{g}^c_{i_1i_2},\tilde{g}^c_{i_1i_2i_3}$ are not necessarily symmetric under permutations as required in Assumption \ref{third-differentiability}, so we perform the following symmetrization to find the influence functions
\begin{align*}
g_i(x)&:=\tilde{g}_i^c(x)\\
g_{ii}(x_1,x_2)&:=\tilde{g}^c_{ii}(x_1,x_2)+\tilde{g}^c_{ii}(x_2,x_1)\\
g_{i_1i_2}(x_1,x_2)=g_{i_2i_1}(x_2,x_1)&:=\tilde{g}^c_{i_1i_2}(x_1,x_2)\text{ for }i_1<i_2\\
g_{iii}(x_1,x_2,x_3)&:=\sum_{\pi}\tilde{g}^c_{iii}(x_{\pi(1)},x_{\pi(2)},x_{\pi(3)})\\
g_{i_1i_1i_2}(x_1,x_2,x_3)=g_{i_1i_2i_1}(x_1,x_3,x_2)=g_{i_2i_1i_1}(x_3,x_1,x_2)&:=\tilde{g}^c_{i_1i_1i_2}(x_{1},x_{2},x_{3})+\tilde{g}^c_{i_1i_1i_2}(x_{2},x_{1},x_{3})\text{ for }i_1<i_2\\
g_{i_1i_2i_2}(x_1,x_2,x_3)=g_{i_2i_1i_2}(x_2,x_1,x_3)=g_{i_2i_2i_1}(x_2,x_3,x_1)&:=\tilde{g}^c_{i_1i_2i_2}(x_{1},x_{2},x_{3})+\tilde{g}^c_{i_1i_2i_2}(x_{1},x_{3},x_{2})\text{ for }i_1<i_2\\
\text{for all $\pi$ let }g_{i_{\pi(1)}i_{\pi(2)}i_{\pi(3)}}(x_{\pi(1)},x_{\pi(2)},x_{\pi(3)})&:=\tilde{g}^c_{i_1i_2i_3}(x_{1},x_{2},x_{3})\text{ for }i_1<i_2<i_3
\end{align*}
where the dependence on $P_1,\ldots,P_m$ is suppressed and $\pi$ denotes any permutation of $(1,2,3)$. Then one can check that $g_{i_1i_2}$ and $g_{i_1i_2i_3}$ not only retain the property of zero marginal means, but also satisfy the symmetry condition in Assumption \ref{third-differentiability}. Permutation symmetry implies that
\begin{align}
\notag&\psi(P_1^{\nu_1},\ldots,P_m^{\nu_m})\\
\notag=&\psi(P_{1},\ldots,P_{m})+\sum_{i=1}^m\nu_i\int g_i(x)d(Q_i-P_i)(x)+\frac{1}{2}\sum_{i_1,i_2}\nu_{i_1}\nu_{i_2}\int g_{i_1i_2}(x,y)d(Q_{i_1}-P_{i_1})(x)d(Q_{i_2}-P_{i_2})(y)\\
\notag&+\frac{1}{6}\sum_{i_1, i_2,i_3}\nu_{i_1}\nu_{i_2}\nu_{i_3}\int g_{i_1i_2i_3}(x,y,z)d(Q_{i_1}-P_{i_1})(x)d(Q_{i_2}-P_{i_2})(y)d(Q_{i_3}-P_{i_3})(z)\\
&+\sum_{d=4}^T\sum_{\sum_{i=1}^m\lvert\mathcal T_i\rvert=d}\prod_{i=1}^m\nu_i^{\lvert\mathcal T_i\rvert}\int h^c_{\mathcal T_1,\ldots,\mathcal T_m}(\mathbf x_{1,\mathcal T_1},\ldots,\mathbf x_{m,\mathcal T_m})\prod_{i=1}^m\prod_{t\in \mathcal T_i}d(Q_i-P_i)(x_{i,t}).\label{remainder:assumption11}
\end{align}
Since the integrals are all finite under Assumption \ref{moment8:finite}, the first-order and third-order remainders of the above expansion are respectively of order $O\big(\sum_{i=1}^m\nu_i^2\big)$ and $O\big(\big(\sum_{i=1}^m\nu_i^2\big)^2\big)$, leading to Assumptions \ref{first-differentiability} and \ref{third-differentiability}.\Halmos
\endproof

We continue to verify other assumptions, for which we use the following lemma.
\begin{lemma}\label{error:key}
Suppose Assumption \ref{moment8:finite} holds with positive and even $k$. For each $i$ let $\widetilde F_i\in\{F_i,\widehat F_i\}$ be either the $i$-th true or empirical input model. Then the following bounds hold uniformly for every $(\widetilde F_1,\ldots,\widetilde F_m)\in\prod_{i=1}^m\{F_i,\widehat F_i\}$ and arbitrary input data size $n_i$
\begin{align}
&\max_{I_1,\ldots,I_m}\mathbb E_{\widehat F_1,\ldots,\widehat F_m}[h^{k}(\mathbf X_{1,I_1},\ldots,\mathbf X_{m,I_m})]=O_p(1)\label{bound1}\\
&\mathbb E_{F_{1},\ldots,F_{m}}\big[\big(\psi(\widetilde F_1,\ldots,\widetilde F_m)-\psi(F_1,\ldots,F_m)\big)^k\big]\leq C_1\mathcal M \big(\sum_{i=1}^m\frac{1}{\sqrt{n_i}}\big)^k\label{bound2}\\
&\mathbb E_{F_{1},\ldots,F_{m}}\big[\big(\psi(\widetilde F_1,\ldots,\widetilde F_m)-\psi(F_1,\ldots,F_m)-\sum_{i=1}^m\int g_i(x)d(\widetilde F_i-F_i)(x)\big)^k\big]\leq C_2\mathcal M \big(\sum_{i=1}^m\frac{1}{\sqrt{n_i}}\big)^{2k}\label{bound3}
\end{align}
where the influence functions $g_i$'s are now under the true input models $F_1,\ldots,F_m$. Each empirical influence function $\hat g_i$ satisfies
\begin{align}
&\mathbb E_{F_1,\ldots,F_m}[(g_i(X_{i,1})-\hat{g}_i(X_{i,1}))^k]\leq C_3\mathcal M \big(\sum_{i=1}^m\frac{1}{\sqrt{n_i}}\big)^k\label{bound4}\\
&\mathbb E_{F_1,\ldots,F_m}[(\hat{g}_i(X_{i,1})-g_i(X_{i,1})-\sum_{i'=1}^m\int g_{ii'}(X_{i,1},x)d(\widehat F_{i'}-F_{i'})(x)+\int g_{i}(x)d(\widehat F_{i}-F_{i})(x))^k]\leq C_4\mathcal M \big(\sum_{i=1}^m\frac{1}{\sqrt{n_i}}\big)^{2k}\label{bound5}
\end{align}
Here $C_1,C_2,C_3,C_4$ are constants that only depend on $k,m$ and $T:=\sum_{i=1}^mT_i$, and
$$\mathcal M:=\max_{I_1,\ldots,I_m}\mathbb E_{F_1,\ldots,F_m}[h^k(\mathbf X_{1,I_1},\ldots,\mathbf X_{m,I_m})]<\infty.$$
\end{lemma}
\textit{Proof.} The first bound is the most straightforward. By rewriting the expectation $\mathbb E_{\widehat F_1,\ldots,\widehat F_m}[\cdot]$ as a sum, one can see that for a particular choice of $I_1,\ldots,I_m$
\begin{equation*}
\mathbb E_{F_1,\ldots,F_m}\Big[\mathbb E_{\widehat F_1,\ldots,\widehat F_m}[h^k(\mathbf X_{1,I_1},\ldots,\mathbf X_{m,I_m})]\Big]\leq \mathcal M.
\end{equation*}
Therefore $\mathbb E_{\widehat F_1,\ldots,\widehat F_m}[h^k(\mathbf X_{1,I_1},\ldots,\mathbf X_{m,I_m})]=O_p(1)$ for each $I_1,\ldots,I_m$. Since there are finitely many of them, the maximum is also bounded in probability. This proves the first bound.

To explain the other bounds, we put $\psi(\widetilde F_1,\ldots,\widetilde F_m)$ in the form of the expansion \eqref{centered_expansion} with $\nu_i=1,P_i=F_i,Q_i=\widetilde F_i$ to get
\begin{align*}
&\psi(\widetilde F_1,\ldots,\widetilde F_m)\\
=&\psi(F_{1},\ldots,F_{m})+\sum_{d=1}^T\sum_{\sum_{i=1}^m\lvert\mathcal T_i\rvert=d}\int h^c_{\mathcal T_1,\ldots,\mathcal T_m}(\mathbf x_{1,\mathcal T_1},\ldots,\mathbf x_{m,\mathcal T_m})\prod_{i=1}^m\prod_{t\in\mathcal T_i}d\widetilde F_i(x_{i,t})\\
=&\psi(F_{1},\ldots,F_{m})+\sum_{i=1}^m\int \big(\sum_{t=1}^{T_i}\mathbb E_{F_1,\ldots,F_m}[h(\mathbf X_1,\ldots,\mathbf X_m)\vert X_i(t)=x]-T_i\psi(F_1,\ldots,F_m)\big)d(\widetilde F_i-F_i)(x)\\
&+\sum_{d=2}^T\sum_{\sum_{i=1}^m\lvert\mathcal T_i\rvert=d}\int h^c_{\mathcal T_1,\ldots,\mathcal T_m}(\mathbf x_{1,\mathcal T_1},\ldots,\mathbf x_{m,\mathcal T_m})\prod_{i=1}^m\prod_{t\in\mathcal T_i}d\widetilde F_i(x_{i,t})
\end{align*}
where $\int h^c_{\mathcal T_1,\ldots,\mathcal T_m}(\mathbf x_{1,\mathcal T_1},\ldots,\mathbf x_{m,\mathcal T_m})dF_{i}(x_{i,t})=0\text{ for all }i\text{ and }t\in\mathcal T_i$, according to the property of zero marginal means \eqref{zeromarginal}. To obtain a moment bound for $h^c_{\mathcal T_1,\ldots,\mathcal T_m}$, observe that by Assumption \ref{moment8:finite} and Jensen's inequality any conditional expectation of the performance function $h$ has a $k$-th moment at most $\mathcal M$. Since $h^c_{\mathcal T_1,\ldots,\mathcal T_m}$ is the sum of several conditional expectations of $h$, one can apply Minkowski inequality to establish that for any $I_i=(I_i(1),\ldots,I_i(\lvert\mathcal T_i\rvert))\in \{1,2,\ldots,\lvert\mathcal T_i\rvert\}^{\lvert\mathcal T_i\rvert}$, $i=1,\ldots,m$
\begin{equation}\label{momentbd}
\mathbb E_{F_1,\ldots,F_m}[(h^{c}_{\mathcal T_1,\ldots,\mathcal T_m}(\mathbf X_{1,\mathcal T_1(I_1)},\ldots,\mathbf X_{m,\mathcal T_m(I_m)}))^k]\leq 2^{k\sum_{i=1}^m\lvert\mathcal T_i\rvert}\mathcal M.
\end{equation}
Again by Minkowski inequality
\begin{align*}
&\mathbb E_{F_1,\ldots,F_m}\big[\big(\psi(\widetilde F_1,\ldots,\widetilde F_m)-\psi(F_1,\ldots,F_m)\big)^k\big]\\
\leq &\Big(\sum_{d=1}^T\sum_{\sum_{i=1}^m\lvert\mathcal T_i\rvert=d}\Big(\mathbb E_{F_1,\ldots,F_m} \Big[\Big(\int h^c_{\mathcal T_1,\ldots,\mathcal T_m}(\mathbf x_{1,\mathcal T_1},\ldots,\mathbf x_{m,\mathcal T_m})\prod_{i=1}^m\prod_{t\in\mathcal T_i}d\widetilde F_i(x_{i,t})\Big)^k\big]\Big)^{\frac{1}{k}}\Big)^k\\
= &\Big(\sum_{d=1}^T\sum_{\sum_{i=1}^m\lvert\mathcal T_i\rvert=d}\prod_{i=1}^m\prod_{t\in\mathcal T_i}\mathbf{1}(\widetilde F_{i,t}=\widehat F_{i})\Big(\mathbb E_{F_1,\ldots,F_m} \Big[\Big(\frac{1}{\prod_{i=1}^mn_i^{\lvert\mathcal T_i\rvert}}\sum_{J_1,\ldots,J_m} h^c_{\mathcal T_1,\ldots,\mathcal T_m}(\mathbf X_{1,J_1},\ldots,\mathbf X_{m,J_m})\Big)^k\big]\Big)^{\frac{1}{k}}\Big)^k\\
\leq&\Big(\sum_{d=1}^T\sum_{\sum_{i=1}^m\lvert\mathcal T_i\rvert=d}\Big(\mathbb E_{F_1,\ldots,F_m} \Big[\Big(\frac{1}{\prod_{i=1}^mn_i^{\lvert\mathcal T_i\rvert}}\sum_{J_1,\ldots,J_m} h^c_{\mathcal T_1,\ldots,\mathcal T_m}(\mathbf X_{1,J_1},\ldots,\mathbf X_{m,J_m})\Big)^k\big]\Big)^{\frac{1}{k}}\Big)^k
\end{align*}
where each $J_i=(J_i(1),\ldots,J_i(\lvert\mathcal T_i\rvert))\in \{1,2,\ldots,n_i\}^{\lvert\mathcal T_i\rvert}$ and $\mathbf X_{i,J_i}=(X_{i,J_i(1)},\ldots,X_{i,J_i(\lvert\mathcal T_i\rvert)})$. Note that
\begin{align*}
&\mathbb E_{F_1,\ldots,F_m} \Big[\Big(\frac{1}{\prod_{i=1}^mn_i^{\lvert\mathcal T_i\rvert}}\sum_{J_1,\ldots,J_m} h^c_{\mathcal T_1,\ldots,\mathcal T_m}(\mathbf X_{1,J_1},\ldots,\mathbf X_{m,J_m})\Big)^k\big]\\
=&\frac{1}{\prod_{i=1}^mn_i^{k\lvert\mathcal T_i\rvert}}\sum_{J_1^1,\ldots,J_m^1}\cdots\sum_{J_1^k,\ldots,J_m^k}\mathbb E_{F_1,\ldots,F_m}[h^c_{\mathcal T_1,\ldots,\mathcal T_m}(\mathbf X_{1,J_1^1},\ldots,\mathbf X_{m,J_m^1})\cdots h^c_{\mathcal T_1,\ldots,\mathcal T_m}(\mathbf X_{1,J_1^k},\ldots,\mathbf X_{m,J_m^k})].
\end{align*}
By \eqref{zeromarginal} the expectation on the right hand side is zero if some data point $X_{i,j}$ appears only once. Therefore the number of non-zero expectations is bounded above by $C(k,m,\sum_{i=1}^m\lvert\mathcal T_i\rvert)\prod_{i=1}^mn_i^{k\lvert\mathcal T_i\rvert/2}$, where $C(k,m,\sum_{i=1}^m\lvert\mathcal T_i\rvert)$ is some constant that only depends on $k,m,\sum_{i=1}^m\lvert\mathcal T_i\rvert$. Moreover, from \eqref{momentbd} each expectation satisfies the following by generalized H\"{o}lder's inequality
\begin{align*}
\lvert\mathbb E_{F_1,\ldots,F_m}[h^c_{\mathcal T_1,\ldots,\mathcal T_m}(\mathbf X_{1,J_1^1},\ldots,\mathbf X_{m,J_m^1})\cdots h^c_{\mathcal T_1,\ldots,\mathcal T_m}(\mathbf X_{1,J_1^k},\ldots,\mathbf X_{m,J_m^k})]\rvert\leq 2^{k\sum_{i=1}^m\lvert\mathcal T_i\rvert}\mathcal M.
\end{align*}
Hence
\begin{align*}
&\mathbb E_{F_1,\ldots,F_m}\big[\big(\psi(\widehat F_1,\ldots,\widehat F_m)-\psi(F_1,\ldots,F_m)\big)^k\big]\\
\leq & \Big(\sum_{d=1}^T\sum_{\sum_{i=1}^m\lvert\mathcal T_i\rvert=d}\Big(C(k,m,\sum_{i=1}^m\lvert\mathcal T_i\rvert)\prod_{i=1}^mn_i^{-k\lvert\mathcal T_i\rvert/2}2^{k\sum_{i=1}^m\lvert\mathcal T_i\rvert}\mathcal M\Big)^{\frac{1}{k}}\Big)^k\\
=&\Big(\sum_{d=1}^T\sum_{\sum_{i=1}^m\lvert\mathcal T_i\rvert=d}C'(k,m,d)\prod_{i=1}^mn_i^{-\lvert\mathcal T_i\rvert/2}\mathcal M^{\frac{1}{k}}\Big)^k\\
\leq&\Big(\sum_{d=1}^TC'(k,m,d)\big(\sum_{i=1}^m\frac{T_i}{\sqrt{n_i}}\big)^d\Big)^k\mathcal M\leq C_1(k,m,T)\mathcal M\Big(\sum_{i=1}^m\frac{1}{\sqrt{n_i}}\Big)^k.
\end{align*}
This gives the second bound.

The third bound can be established by the same argument, but considering only the remainders for which $d\geq 2$.

We then prove the bounds on influence functions. According to the representation of $g_i(P_1,\ldots,P_m;\cdot)$ in the proof of Assumptions \ref{first-differentiability} and \ref{third-differentiability}, the empirical influence function $\hat{g}_i$ is
\begin{equation*}
\hat{g}_i(x)=\sum_{t=1}^{T_i}\mathbb E_{\widehat F_1,\ldots,\widehat F_m}[h(\mathbf X_1,\ldots,\mathbf X_m)\vert X_i(t)=x]-T_i\psi(\widehat F_1,\ldots,\widehat F_m).
\end{equation*}
First we derive the following Taylor expansion for each conditional expectation
\begin{align}
\notag&\mathbb E_{\widehat F_1,\ldots,\widehat F_m}[h(\mathbf X_1,\ldots,\mathbf X_m)\vert X_{i}(t)=X_{i,1}]\\
\notag=&\int h(\mathbf x_1,\ldots,\mathbf x_m)\prod_{t'\neq t}d\widehat F_{i}(x_{i,t'})\prod_{i'\neq i}\prod_{t'=1}^{T_{i'}}d\widehat F_{i'}(x_{i',t'})\Big\vert_{x_{i,t}=X_{i,1}}\\
\notag=&\mathbb E_{F_{1},\ldots,F_{m}}[h(\mathbf X_1,\ldots,\mathbf X_m)\vert X_{i}(t)=X_{i,1}]+\\
\notag&+\sum_{d=1}^{T-1}\sum_{\sum_{i'=1}^m\lvert\mathcal T_{i'}\rvert=d,t\notin\mathcal T_{i}}\int h(\mathbf x_1,\ldots,\mathbf x_m)\prod_{t'\notin \mathcal T_{i},t'\neq t}dF_{i}(x_{i,t'})\prod_{i'\neq i}\prod_{t'\notin \mathcal T_{i'}}dF_{i'}(x_{i',t'})\prod_{i'=1}^m\prod_{t'\in \mathcal T_{i'}}d(\widehat F_{i'}-F_{i'})(x_{i',t'})\Big\vert_{x_{i,t}=X_{i,1}}\\
\notag=&\mathbb E_{F_{1},\ldots,F_{m}}[h(\mathbf X_1,\ldots,\mathbf X_m)\vert X_{i}(t)=X_{i,1}]\\
\notag&+\sum_{d=1}^{T-1}\sum_{\sum_{i'=1}^m\lvert\mathcal T_{i'}\rvert=d,t\notin\mathcal T_{i}}\int h_{(i,t),\mathcal T_1,\ldots,\mathcal T_m}(\mathbf x_{1,\mathcal T_1},\ldots,\mathbf x_{i,\mathcal T_{i}\cup \{t\}},\ldots,\mathbf x_{m,\mathcal T_m})\prod_{i'=1}^m\prod_{t'\in \mathcal T_{i'}}d(\widehat F_{i'}-F_{i'})(x_{i',t'})\Big\vert_{x_{i,t}=X_{i,1}}\\
\notag=&\mathbb E_{F_{1},\ldots,F_{m}}[h(\mathbf X_1,\ldots,\mathbf X_m)\vert X_{i}(t)=X_{i,1}]\\
&+\sum_{t'=1,t'\neq t}^{T_i}\int \mathbb E_{F_{1},\ldots,F_{m}}[h(\mathbf X_1,\ldots,\mathbf X_m)\vert X_{i}(t)=X_{i,1},X_i(t')=x_{i,t'}]d(\widehat F_{i}-F_{i})(x_{i,t'})\label{g_remainder1}\\
&+\sum_{i'\neq i}\sum_{t'=1}^{T_{i'}}\int \mathbb E_{F_{1},\ldots,F_{m}}[h(\mathbf X_1,\ldots,\mathbf X_m)\vert X_{i}(t)=X_{i,1},X_{i'}(t')=x_{i',t'}]d(\widehat F_{i'}-F_{i'})(x_{i',t'})\label{g_remainder2}\\
\notag&+\sum_{d=2}^{T-1}\sum_{\sum_{i'=1}^m\lvert\mathcal T_{i'}\rvert=d,t\notin\mathcal T_{i}}\int h_{(i,t),\mathcal T_1,\ldots,\mathcal T_m}(\mathbf x_{1,\mathcal T_1},\ldots,\mathbf x_{i,\mathcal T_{i}\cup \{t\}},\ldots,\mathbf x_{m,\mathcal T_m})\prod_{i'=1}^m\prod_{t'\in \mathcal T_{i'}}d(\widehat F_{i'}-F_{i'})(x_{i',t'})\Big\vert_{x_{i,t}=X_{i,1}}
\end{align}
where each $\mathcal T_{i'}=\{\mathcal T_{i'}(1),\ldots,\mathcal T_{i'}(\lvert \mathcal T_{i'}\rvert)\}$ is still an ordered subset of $\{1,2,\ldots,T_{i'}\}$ but $t\notin \mathcal T_{i}$, and the function $h_{(i,t),\mathcal T_1,\ldots,\mathcal T_m}$ resembles \eqref{h1} except that the expectation is now further conditioned on $X_{i}(t)=x_{i,t}$. Introduce the counterpart of \eqref{centered_h}
\begin{align*}
&h^c_{(i,t),\mathcal T_1,\ldots,\mathcal T_m}(\mathbf x_{1,\mathcal T_1},\ldots,\mathbf x_{i,\mathcal T_{i}\cup \{t\}},\ldots,\mathbf x_{m,\mathcal T_m})\\
=&h_{(i,t),\mathcal T_1,\ldots,\mathcal T_m}-\sum_{i',t'\in\mathcal T_{i'}}\int h_{(i,t),\mathcal T_1,\ldots,\mathcal T_m}dF_{i'}(x_{i',t'})+\sum_{(i'_1,t'_1)<(i'_2,t'_2),t'_1\in\mathcal T_{i'_1},t'_2\in\mathcal T_{i'_2}}\int h_{(i,t),\mathcal T_1,\ldots,\mathcal T_m}dF_{i'_1}(x_{i'_1,t'_1})dF_{i'_2}(x_{i'_2,t'_2})\\
&+\cdots+(-1)^{\sum_{i'=1}^m\lvert\mathcal T_{i'}\rvert}\int h_{(i,t),\mathcal T_1,\ldots,\mathcal T_m} \prod_{i'=1}^m\prod_{t'\in\mathcal T_{i'}}dF_{i'}(x_{i',t'})
\end{align*}
then we have the following parallel property of \eqref{zeromarginal}
\begin{equation*}
\int h^c_{(i,t),\mathcal T_1,\ldots,\mathcal T_m}(\mathbf x_{1,\mathcal T_1},\ldots,\mathbf x_{i,\mathcal T_{i}\cup \{t\}},\ldots,\mathbf x_{m,\mathcal T_m})dF_{i'}(x_{i',t'})=0\text{ for all }i'\text{ and }t'\in\mathcal T_{i'}
\end{equation*}
and by comparing the first order remainders \eqref{g_remainder1} and \eqref{g_remainder2} of $\hat g_i$ with the second order influence functions $g_{i_1i_2}$ it is easy to establish that
\begin{align}
\notag&\hat g_i(X_{i,1})-g_i(X_{i,1})\\
\notag=&\sum_{t=1}^{T_i}\big(\mathbb E_{\widehat F_1,\ldots,\widehat F_m}[h(\mathbf X_1,\ldots,\mathbf X_m)\vert X_{i}(t)=X_{i,1}]-\mathbb E_{F_{1},\ldots,F_{m}}[h(\mathbf X_1,\ldots,\mathbf X_m)\vert X_{i}(t)=X_{i,1}]\big)\\
\notag&\hspace{2ex}-T_i(\psi(\widehat F_1,\ldots,\widehat F_m)-\psi( F_1,\ldots, F_m))\\
=&\sum_{i'=1}^m\int g_{ii'}(X_{i,1},x)d(\widehat F_{i'}-F_{i'})(x)-\int g_{i}(x)d(\widehat F_{i}-F_{i})(x)\label{g_first_order}\\
\notag&+\sum_{t=1}^{T_i}\sum_{d=2}^{T-1}\sum_{\sum_{i'=1}^m\lvert\mathcal T_{i'}\rvert=d}\int h^c_{(i,t),\mathcal T_1,\ldots,\mathcal T_m}(\mathbf x_{1,\mathcal T_1},\ldots,\mathbf x_{i,\mathcal T_{i}\cup \{t\}},\ldots,\mathbf x_{m,\mathcal T_m})\prod_{i'=1}^m\prod_{t'\in\mathcal T_{i'}}d\widehat F_{i'}(x_{i',t'})\Big\vert_{x_{i,t}=X_{i,1}}\\
&-T_i\sum_{d=2}^T\sum_{\sum_{i'=1}^m\lvert\mathcal T_{i'}\rvert=d}\int h^c_{\mathcal T_1,\ldots,\mathcal T_m}(\mathbf x_{1,\mathcal T_1},\ldots,\mathbf x_{m,\mathcal T_m})\prod_{i'=1}^m\prod_{t'\in\mathcal T_{i'}}d\widehat F_{i'}(x_{i',t'})\label{g_second_order}
\end{align}
By a similar technique used to bound the remainder of $\psi(\widetilde F_1,\ldots,\widetilde F_m)$, we can establish that the remainder \eqref{g_second_order} has a $k$-th moment of order $O\big(\mathcal M \big(\sum_{i=1}^m\frac{1}{\sqrt{n_i}}\big)^{2k}\big)$, and the first order term \eqref{g_first_order} has a $k$-th moment of order $O\big(\mathcal M \big(\sum_{i=1}^m\frac{1}{\sqrt{n_i}}\big)^{k}\big)$. This completes the proof.\Halmos
\endproof

With Lemma \ref{error:key} we now prove the other assumptions:

\proof{Proof of Assumption \ref{smoothness_truth}.}The moment bound on the remainder, i.e.~$\mathbb E[\epsilon^2]=o(n^{-1})$, comes from the bound \eqref{bound3} in Lemma \ref{error:key} with $\widetilde F_i=\widehat F_i$ for all $i$ and $k=2$. The non-degeneracy condition on the influence functions is exactly Assumption \ref{var_pos:finite}, whereas the finiteness of fourth order moments of $g_i$ easily follows because $g_i$ is simply a sum of $T_i$ conditional expectations of the performance function $h$ and each of the conditional expectations has finite fourth order moment by Assumption \ref{moment8:finite} and Jensen's inequality.\Halmos
\endproof

\proof{Proof of Assumption \ref{smoothness_empirical}.}The convergence of $\hat g_i$ to $g_i$ in fourth order moment is a direct consequence of the bound \eqref{bound4} in Lemma \ref{error:key} with $k=4$.
The moment condition on the remainder $\epsilon^*$ can be argued as follows. We treat the empirical distributions $\widehat F_1,\ldots,\widehat F_m$ as the truth, and the resampled distributions $\widehat F_{s_1,1}^*,\ldots,\widehat F_{s_m,m}^*$ as the input data, then apply the third bound \eqref{bound3} in Lemma \ref{error:key} with $k=4$ to get $\mathbb E_*[({\epsilon}^*)^4]\leq C_2\widehat{\mathcal M} \big(\sum_{i=1}^m\frac{1}{\sqrt{s_i}}\big)^{8}$, where $\widehat{\mathcal M}=\max_{I_1,\ldots,I_m}\mathbb E_{\widehat F_1,\ldots,\widehat F_m}[h^{4}(\mathbf X_{1,I_1},\ldots,\mathbf X_{m,I_m})]$ is $O_p(1)$ by the first bound \eqref{bound1} in Lemma \ref{error:key} with $k=4$. Therefore $\mathbb E_*[({\epsilon}^*)^4]=O_p((\sum_{i=1}^m\frac{1}{s_i})^4)=o_p(s^{-2})$.\Halmos
\endproof

\proof{Proof of Assumption \ref{var_emp}.}It suffices to show that $\mathbb E_{\widehat F_1,\ldots,\widehat F_m}[h^2]\stackrel{p}{\to}\mathbb E_{F_{1},\ldots,F_{m}}[h^2]$ and $\mathbb E_{\widehat F_1,\ldots,\widehat F_m}[h]\stackrel{p}{\to}\mathbb E_{F_{1},\ldots,F_{m}}[h]$. The latter convergence follows from the second bound \eqref{bound2} of Lemma \ref{error:key} with $k=2$ and $\widetilde F_i=\widehat F_i$ for all $i$. Since Assumption \ref{moment8:finite} holds with $k=4$ for the function $h$, it also holds with $k=2$ for the squared function $h^2$. One can apply the same bound from Lemma \ref{error:key} with $k=2$ to $h^2$ and then conclude the former convergence.\Halmos
\endproof

\proof{Proof of Assumption \ref{var_sim}.}We write $\bar\tau^2=\tau^2(\overline F_1,\ldots,\overline F_m)$ for short. First rewrite
\begin{align*}
(\bar\tau^2-\hat{\tau}^2)^2&=\big(\mathbb E_{\overline F_1,\ldots,\overline F_m}[h^2]-\mathbb E_{\widehat F_1,\ldots,\widehat F_m}[h^2]-\big((\mathbb E_{\overline F_1,\ldots,\overline F_m}[h])^2-(\mathbb E_{\widehat F_1,\ldots,\widehat F_m}[h])^2\big)\big)^2\\
&\leq 2\big(\mathbb E_{\overline F_1,\ldots,\overline F_m}[h^2]-\mathbb E_{\widehat F_1,\ldots,\widehat F_m}[h^2]\big)^2+2\big((\mathbb E_{\overline F_1,\ldots,\overline F_m}[h])^2-(\mathbb E_{\widehat F_1,\ldots,\widehat F_m}[h])^2\big)^2\\
&\leq 2\big(\mathbb E_{\overline F_1,\ldots,\overline F_m}[h^2]-\mathbb E_{\widehat F_1,\ldots,\widehat F_m}[h^2]\big)^2+4\big(\mathbb E_{\overline F_1,\ldots,\overline F_m}[h]-\mathbb E_{\widehat F_1,\ldots,\widehat F_m}[h]\big)^4\\
&\hspace{5ex}+16(\mathbb E_{\widehat F_1,\ldots,\widehat F_m}[h])^2\big(\mathbb E_{\overline F_1,\ldots,\overline F_m}[h]-\mathbb E_{\widehat F_1,\ldots,\widehat F_m}[h]\big)^2.
\end{align*}
Applying Lemma \ref{error:key} to $h^2$ ($k=2$) with the true distributions being $\widehat F_1,\ldots,\widehat F_m$ we get
\begin{align*}
\mathbb E_*[\big(\mathbb E_{\overline F_1,\ldots,\overline F_m}[h^2]-\mathbb E_{\widehat F_1,\ldots,\widehat F_m}[h^2]\big)^2]\leq C_1\widehat{\mathcal M} \big(\sum_{i=1}^m\frac{1}{\sqrt{s_i}}\big)^2=O_p\big(\sum_{i=1}^m\frac{1}{s_i}\big)
\end{align*}
where $\widehat{\mathcal M}=\max_{I_1,\ldots,I_m}\mathbb E_{\widehat F_1,\ldots,\widehat F_m}[h^{4}(\mathbf X_{1,I_1},\ldots,\mathbf X_{m,I_m})]=O_p(1)$. Another application of Lemma \ref{error:key} to $h$ with $k=4$ gives
\begin{align*}
\mathbb E_*[\big(\mathbb E_{\overline F_1,\ldots,\overline F_m}[h]-\mathbb E_{\widehat F_1,\ldots,\widehat F_m}[h]\big)^4]\leq C_1\widehat{\mathcal M} \big(\sum_{i=1}^m\frac{1}{\sqrt{s_i}}\big)^4=O_p\big(\sum_{i=1}^m\frac{1}{s_i^2}\big)
\end{align*}
which implies that $\mathbb E_*[\big(\mathbb E_{\overline F_1,\ldots,\overline F_m}[h]-\mathbb E_{\widehat F_1,\ldots,\widehat F_m}[h]\big)^2]=O_p\big(\sum_{i=1}^m\frac{1}{s_i}\big)$ as a consequence of Cauchy Schwartz inequality. Therefore in sum $\mathbb E_*[(\bar\tau^2-\hat{\tau}^2)^2]=O_p\big(\sum_{i=1}^m\frac{1}{s_i}\big)=o_p(1)$.\Halmos
\endproof

\proof{Proof of Assumption \ref{anova_mu4_sim}.}Note that $\mu_4(\overline F_1,\ldots,\overline F_m)\leq C\mathbb E_{\overline F_1,\ldots,\overline F_m}[h^4]$ for some absolute constant $C>0$, therefore
\begin{align*}
\mathbb E_*[\mu_4(\overline F_1,\ldots,\overline F_m)]\leq C\mathbb E_*[\mathbb E_{\overline F_1,\ldots,\overline F_m}[h^4]]\leq C\max_{I_1,\ldots,I_m}\mathbb E_{\widehat F_1,\ldots,\widehat F_m}[h^4(\mathbf X_{1,I_1},\ldots,\mathbf X_{m,I_m})]=O_p(1)
\end{align*}
where the last equality is due to the first bound \eqref{bound1} in Lemma \ref{error:key}.\Halmos
\endproof

\proof{Proof of Assumption \ref{3smoothness_truth}.}The third order remainder $\epsilon_3$, or equivalently the sum over $d\geq 4$ in \eqref{remainder:assumption11} with each $\nu_i=1$, consists of integrals under the product of at least four signed measures of the form $\widehat F_i-F_i$. Therefore, by employing the technique used in proving the second and third bounds \eqref{bound2}\eqref{bound3} in Lemma \ref{error:key}, one can show that $\mathbb E[\epsilon_3^2]=O(n^{-4})$. The details are omitted since they highly resemble those of Lemma \ref{error:key}. The fourth moments of $g_{i_1i_2}$ and $g_{i_1i_2i_3}$ are finite, because each of them is a finite sum of conditional expectations of $h$ which have finite fourth order moments due to Assumption \ref{moment8:finite} with $k=4$ and Jensen's inequality.\Halmos
\endproof

\proof{Proof of Assumption \ref{3smoothness_empirical}.}For the third order remainder of the resampled performance measure, one can derive the bound $\mathbb E_*[(\epsilon_3^*)^2]=O_p(s^{-4})$ in a similar way as in showing the bound \eqref{bound3} in Lemma \ref{error:key}. The details are omitted to avoid repetition. Moreover, some straightforward modifications of the proof for the bound \eqref{bound4} in Lemma \ref{error:key} lead to $O(n^{-1})$ upper bounds for the the mean squared errors of second and third order influence functions. The remainder in the Taylor expansion of the first order empirical influence function satisfies $\mathbb E[\epsilon_g^2]=O(n^{-2})$ due to the bound \eqref{bound5} in Lemma \ref{error:key} with $k=2$.\Halmos
\endproof

\section{Proofs of Propositions \ref{input_var_order} and \ref{input_var:tight}}\label{sec: input variance approximation}
This section proves results concerning the validity of the additive decomposition \ref{var decompose} of the input variance. We first prove Proposition \ref{input_var_order}, and then provide the key Lemma \ref{anova_decom} that will be used in the proof of Proposition \ref{input_var:tight} as well as many results in Section \ref{sec: overall error and subsample size}.
\proof{Proof of Proposition \ref{input_var_order}.}Following the expansion \eqref{taylor_expansion} we can write
\begin{align*}
\mathrm{Var}[\psi(\widehat F_1,\ldots,\widehat F_m)]&=\mathrm{Var}[\sum_{i=1}^m\frac{1}{n_i}\sum_{j=1}^{n_i} g_{i}(X_{i,j})]+\mathrm{Var}[\epsilon]+2\mathrm{Cov}(\sum_{i=1}^m\frac{1}{n_i}\sum_{j=1}^{n_i} g_{i}(X_{i,j}),\epsilon)\\
&=\sum_{i=1}^m\frac{\sigma_i^2}{n_i}+o(n^{-1})+O\Big(\sqrt{\mathrm{Var}[\sum_{i=1}^m\frac{1}{n_i}\sum_{j=1}^{n_i} g_{i}(X_{i,j})]\mathrm{Var}[\epsilon]}\Big)\\
&=\sum_{i=1}^m\frac{\sigma_i^2}{n_i}+o(n^{-1}).
\end{align*}
This completes the proof.\Halmos
\endproof

The following important lemma on variance decomposition plays a crucial role in our analysis.
\begin{lemma}[ANOVA decomposition, adapted from \cite{efron1981jackknife}]\label{anova_decom}
Let $Y_i,i=1,\ldots,n$ be independent but not necessarily identically distributed random variables, and $\phi(y_{1},\ldots,y_{n})$ be a function such that $\mathbb E[\phi^2(Y_{1},\ldots,Y_{n})]< \infty$, then there exist functions $\phi_{i_1,\ldots,i_k}$ for $1\leq i_1<\cdots<i_k\leq n$ and $k\leq n$ such that
\begin{align*}
&\phi(Y_{1},\ldots,Y_{n})\\
=&\mu+\sum_{i=1}^n\phi_i(Y_{i})+\sum_{i_1<i_2}\phi_{i_1,i_2}(Y_{i_1},Y_{i_2})+\cdots+\sum_{i_1<\cdots<i_k}\phi_{i_1,\ldots,i_k}(Y_{i_1},\ldots,Y_{i_k})+\cdots+\phi_{1,\ldots,n}(Y_{1},\ldots,Y_{n})
\end{align*}
where
\begin{align*}
\mu&=\mathbb E[\phi(Y_1,\ldots,Y_n)]\\
\phi_i(y)&=\mathbb E[\phi(Y_1,\ldots,Y_n)\vert Y_i=y]-\mu\\
\phi_{i_1,i_2}(y_1,y_2)&=\mathbb E[\phi(Y_1,\ldots,Y_n)\vert Y_{i_1}=y_1,Y_{i_2}=y_2]-\phi_{i_1}(y_1)-\phi_{i_2}(y_2)-\mu\\
&\vdots
\end{align*}
Moreover, the $2^n-1$ random variables in the decomposition have mean zero and are mutually uncorrelated.
\end{lemma}

With this lemma, we can prove Proposition \ref{input_var:tight}:
\proof{Proof of Proposition \ref{input_var:tight}.}The proof of Proposition \ref{input_var} derives the following expression for input variance
\begin{align*}
\mathrm{Var}[\psi(\widehat F_1,\ldots,\widehat F_m)]&=\mathrm{Var}[\sum_{i=1}^m\frac{1}{n_i}\sum_{j=1}^{n_i} g_{i}(X_{i,j})]+\mathrm{Var}[\epsilon]+2\mathrm{Cov}(\sum_{i=1}^m\frac{1}{n_i}\sum_{j=1}^{n_i} g_{i}(X_{i,j}),\epsilon)
\end{align*}
where the covariances can be simplified to
\begin{align*}
\mathrm{Cov}(\sum_{i=1}^m\frac{1}{n_i}\sum_{j=1}^{n_i} g_{i}(X_{i,j}),{\epsilon})&=\sum_{i=1}^m\frac{1}{n_i}\sum_{k=1}^{n_i}\mathbb E[g_i(X_{i,j})({\epsilon}-\mathbb E[\epsilon])]\\
&=\sum_{i=1}^m\frac{1}{n_i}\sum_{j=1}^{n_i}\mathbb E[g_i(X_{i,j})(\mathbb E[{\epsilon}\vert X_{i,j}]-\mathbb E[{\epsilon}])]\\
&=\sum_{i=1}^m\mathbb E[g_i(X_{i,1})(\mathbb E[{\epsilon}\vert X_{i,1}]-\mathbb E[{\epsilon}])].
\end{align*}
Using the cubic expansion in Assumption \ref{3smoothness_truth} and the vanishing marginal expectations of influence functions we have
\begin{align}
\nonumber\mathbb E[{\epsilon}\vert X_{i,1}]-\mathbb E[{\epsilon}]=&\frac{1}{2n_i^2}(g_{ii}(X_{i,1},X_{i,1})-\mathbb E[g_{ii}(X_{i},X_{i})])+\frac{1}{6n_i^3}(g_{iii}(X_{i,1},X_{i,1},X_{i,1})-\mathbb E[g_{iii}(X_{i},X_{i},X_{i})])\\
\nonumber&+\frac{n_i-1}{2n_i^3}\mathbb E_{X_i}[g_{iii}(X_{i,1},X_{i},X_{i})]+\sum_{i'\neq i}\frac{1}{2n_in_{i'}}\mathbb E_{X_{i'}}[g_{ii'i'}(X_{i,1},X_{i'},X_{i'})]\\
&+\mathbb E[\epsilon_3\vert X_{i,1}]-\mathbb E[\epsilon_3].\label{remainder-expand}
\end{align}
Each term except the last in \eqref{remainder-expand} has a second moment of order $O(n^{-4})$. To argue the last term $\mathbb E[\epsilon_3\vert X_{i,1}]-\mathbb E[\epsilon_3]$ also has a second moment of order at most $O(n^{-4})$, note that $\epsilon_3$ is a symmetric statistic hence by Lemma \ref{anova_decom} $\mathrm{Var}[\mathbb E[\epsilon_3\vert X_{i,1}]]\leq \mathrm{Var}[\epsilon_3]/n_i$ and $\mathrm{Var}[\epsilon_3]=o(n^{-3})$ by assumption, hence $\mathrm{Var}[\mathbb E[{\epsilon_3}\vert X_{i,1}]]=o(n^{-4})$. This leads to
\begin{equation*}
\mathrm{Var}[\mathbb E[{\epsilon}\vert X_{i,1}]]=O(n^{-4}).
\end{equation*}
Using Cauchy Schwartz inequality we conclude $\mathrm{Cov}(\sum_{i=1}^m\frac{1}{n_i}\sum_{j=1}^{n_i} g_{i}(X_{i,j}),{\epsilon})=O(n^{-2})$. On the other hand, one can easily show $\mathrm{Var}[\epsilon]=O(n^{-2})$ by using the same technique in the proof of Lemma \ref{error:key} to bound each term in the cubic expansion. This leads to the desired conclusion.\Halmos
\endproof

\section{Proofs for Results in Section \ref{sec:psvb theory} and Section \ref{sec:main guarantees}}\label{sec: consistency and monte carlo error}
We now prove the consistency of our proportionate subsampled bootstrap variance $\sigma_{SVB}^2$ (Theorem \ref{consis:AV}), and derive the mean square error of the Monte Carlo estimate $\hat{\sigma}_{SVB}^2$ relative to $\sigma_{SVB}^2$ (Lemma \ref{MC_MSE:anova}). These results will then be used to prove Theorems \ref{consis:anova} and \ref{opt_allocation_detail:anova}. Theorem \ref{sub ratio}, Corollaries \ref{consis:anova optimal}-\ref{rough sub var} are consequences of Theorem \ref{consis:anova}. Theorem \ref{opt_allocation:anova} is a consequence of Theorem \ref{opt_allocation_detail:anova}.

Recall that $\sigma_i^2=\mathrm{Var}_{F_i}[g_i(X_i)]$ is the variance of the $i$-th influence function. For its empirical counterpart $\hat g_i$ we denote by $\hat\sigma_i^2:=\mathrm{Var}_{\widehat F_i}[\hat g_i(X_i)]$ its variance under the empirical input models. Under the convergence condition $\mathbb E [(\hat g_i-g_i)^4(X_{i,1})]\to 0$ in Assumption \ref{smoothness_empirical}, the convergence of $\hat\sigma_i^2$ to $\sigma_i^2$ follows from
\begin{eqnarray*}
\Big\lvert\hat\sigma_i^2-\frac{1}{n_i}\sum_{j=1}^{n_i}g_i^2(X_{i,j})\Big\rvert&=&\Big\lvert\frac{1}{n_i}\sum_{j=1}^{n_i}\hat g_i^2(X_{i,j})-\frac{1}{n_i}\sum_{j=1}^{n_i}g_i^2(X_{i,j})\Big\rvert\\
&\leq&\frac{2}{n_i}\sqrt{\sum_{j=1}^{n_i}g_i^2(X_{i,j})\sum_{j=1}^{n_i}(\hat g_i-g_i)^2(X_{i,j})}+\frac{1}{n_i}\sum_{j=1}^{n_i}(\hat g_i-g_i)^2(X_{i,j})=o_p(1)
\end{eqnarray*}
and that $\sum_{j=1}^{n_i}g_i^2(X_{i,j})/n_i\stackrel{p}{\to}\sigma_i^2$. For convenience we denote by
\begin{equation*}
\psi^*=\psi(\widehat F_{s_1,1}^*,\ldots,\widehat F_{s_m,m}^*),\;\hat\psi^*=\hat\psi(\widehat F_{s_1,1}^*,\ldots,\widehat F_{s_m,m}^*)
\end{equation*}
the expected value and a single simulation replication, respectively, of the performance measure under the resampled input models, and by
\begin{equation*}
\hat\tau_{*}^2=\tau^2(\widehat F_{s_1,1}^*,\ldots,\widehat F_{s_m,m}^*),\;\hat\mu^*_4=\mu_4(\widehat F_{s_1,1}^*,\ldots,\widehat F_{s_m,m}^*)
\end{equation*}
the variance and central fourth moment of a single Monte Carlo replication $\hat\psi^*$ conditioned on the resampled input models.


\proof{Proof of Theorem \ref{consis:AV}.}Let $s_i=\lfloor\theta n_i\rfloor$. Following the expansion \eqref{taylor_expansion_empirical} with each $\overline F_i=\widehat F_{s_i,i}^*$ we have
\begin{align*}
\mathrm{Var}_*[\psi^*]&=\mathrm{Var}_*[\sum_{i=1}^m\frac{1}{s_i}\sum_{k=1}^{s_i} \hat{g}_{i}(X_{i,k}^*)+\epsilon^*]\\
&=\mathrm{Var}_*[\sum_{i=1}^m\frac{1}{s_i}\sum_{k=1}^{s_i} \hat{g}_{i}(X_{i,k}^*)]+\mathrm{Var}_*[{\epsilon}^*]+2\mathrm{Cov}_*(\sum_{i=1}^m\frac{1}{s_i}\sum_{k=1}^{s_i} \hat{g}_{i}(X_{i,k}^*),{\epsilon}^*)\\
&=\sum_{i=1}^m\frac{\hat{\sigma}_i^2}{s_i}+\mathrm{Var}_*[{\epsilon}^*]+O\Big(\sqrt{\sum_{i=1}^m\frac{\hat{\sigma}_i^2}{s_i}\mathrm{Var}_*[{\epsilon}^*]}\Big)\\
&=\sum_{i=1}^m\frac{\hat{\sigma}_i^2}{\ceil{\theta n_i}}+\mathrm{Var}_*[{\epsilon}^*]+O\Big(\sqrt{\sum_{i=1}^m\frac{\hat{\sigma}_i^2}{\ceil{\theta n_i}}\mathrm{Var}_*[{\epsilon}^*]}\Big).
\end{align*}
Hence
\begin{align}
\nonumber\sigma_{SVB}^2&=\theta\mathrm{Var}_*[\psi^*]=\sum_{i=1}^m\frac{\hat{\sigma}_i^2}{\ceil{\theta n_i}/\theta}+\theta\mathrm{Var}_*[{\epsilon}^*]+O\Big(\sqrt{\sum_{i=1}^m\frac{\hat{\sigma}_i^2}{\ceil{\theta n_i}/\theta}\theta\mathrm{Var}_*[{\epsilon}^*]}\Big)\\
&=\sum_{i=1}^m(\frac{\hat{\sigma}_i^2}{n_i}+O(\frac{\hat{\sigma}_i^2}{n_i^2\theta}))+\theta\mathrm{Var}_*[{\epsilon}^*]+O\Big(\sqrt{\sum_{i=1}^m(\frac{\hat{\sigma}_i^2}{n_i}+O(\frac{\hat{\sigma}_i^2}{n_i^2\theta}))\theta\mathrm{Var}_*[{\epsilon}^*]}\Big).\label{anova_error}
\end{align}
The convergence $\hat\sigma_i^2\stackrel{p}{\to}\sigma_i^2$ and that $\theta=\omega(1/n)$ allow us to conclude
\begin{align*}
\frac{1}{\theta}\sum_{i=1}^m\frac{\hat{\sigma}_i^2}{n_i^2}=o_p(\sum_{i=1}^m\frac{\sigma_i^2}{n_i}),\;\theta \mathrm{Var}_*[{\epsilon}^*]=\theta o_p(\sum_{i=1}^m\frac{1}{\ceil{\theta n_i}})=o_p(\sum_{i=1}^m\frac{1}{ n_i})
\end{align*}
therefore $\sigma_{SVB}^2=\sum_{i=1}^m\frac{\sigma_i^2}{n_i}+o_p(\sum_{i=1}^m\frac{\sigma_i^2}{n_i})$.\Halmos
\endproof

\textit{Proof of Lemma \ref{MC_MSE:anova}:} Define $w:=\hat{\psi}^*-\psi^*$ and $\delta:=\psi^*-\mathbb E_*[\psi^*]$. Unbiasedness is well known, see e.g.~\cite{searle2009variance}. The variance of $\hat{\sigma}_{SVB}^2/\theta$ has been derived in \cite{sun2011efficient} as
\begin{align*}
\frac{1}{\theta^2}\mathrm{Var}_*[\hat{\sigma}_{SVB}^2]=&\frac{1}{B}(\mathbb E_*[\delta^4]-(\mathbb E_*[\delta^2])^2)+\frac{2}{B(B-1)}(\mathbb E_*[\delta^2])^2+\frac{2}{B^2R^2(B-1)}(\mathbb E_*[w^2])^2\\
&+\frac{2(B+1)}{B^2R(B-1)}\mathbb E_*[\delta^2]\mathbb E_*[w^2]+\frac{2}{B^2R^3}\mathbb E_*[w^4]+\frac{4B+2}{B^2R}\mathbb E_*[\delta^2w^2]\\
&+\frac{2(BR^2+R^2-4R+3)}{B^2R^3(R-1)}\mathbb E_*[(\mathbb E[w^2\vert \widehat F_{s_1,1}^*,\ldots,\widehat F_{s_m,m}^*])^2]+\frac{4}{B^2R^2}\mathbb E_*[\delta w^3].
\end{align*}
Applying Jensen's inequality (or generalized Holder's inequality) gives
\begin{align*}
&\mathbb E_*[(\mathbb E[w^2\vert \widehat{F}_1^*,\ldots,\widehat{F}_m^*])^2]\leq \mathbb E_*[w^4]\\
&\mathbb E_*[\delta^2w^2]\leq (\mathbb E_*[\delta^4]\mathbb E_*[w^4])^{1/2},\;\lvert \mathbb E_*[\delta w^3]\rvert\leq (\mathbb E_*[\delta^4](\mathbb E_*[w^4])^3)^{1/4}
\end{align*}
The convergence condition $\mathbb E [(\hat g_i-g_i)^4(X_{i,1})]\to 0$ implies that $\frac{1}{n_i}\sum_{j=1}^{n_i}\hat g_i^4(X_{i,j})=\frac{1}{n_i}\sum_{j=1}^{n_i}g_i^4(X_{i,j})+o_p(1)=O_p(1)$. Together with the moment condition $\mathbb E_*[(\epsilon^*-\mathbb E_*[\epsilon^*])^4]=o_p(s^{-2})$, we get
\begin{align*}
\mathbb E_*[\delta^4]=3\big(\sum_{i=1}^m\frac{\hat{\sigma}_i^2}{s_i}\big)^2+o_p\big(\big(\sum_{i=1}^m\frac{1}{s_i}\big)^2\big),\mathbb E_*[\delta^2]=\sum_{i=1}^m\frac{\hat{\sigma}_i^2}{s_i}+o_p\big(\sum_{i=1}^m\frac{1}{s_i}\big),\mathbb E_*[w^4]=\mathbb E_*[\mu_4^*]=O_p(1).
\end{align*}
Hence the leading terms of the mean squared error can be identified as
\begin{align*}
\frac{1}{\theta^2}\mathrm{Var}_*[\hat{\sigma}_{SVB}^2]&\approx \frac{1}{B}(\mathbb E_*[\delta^4]-(\mathbb E_*[\delta^2])^2)+\frac{4}{BR}\mathbb E_*[\delta^2w^2]+\frac{2}{BR^2}\mathbb E_*[(\mathbb E[w^2\vert \widehat F_{s_1,1}^*,\ldots,\widehat F_{s_m,m}^*])^2]\\
&\approx \frac{2}{B}\big(\sum_{i=1}^m\frac{\hat{\sigma}_i^2}{s_i}\big)^2+\frac{4\hat{\tau}^2}{BR}\sum_{i=1}^m\frac{\hat{\sigma}_i^2}{s_i}+\frac{2\hat{\tau}^4}{BR^2}=\frac{2}{B}\big(\sum_{i=1}^m\frac{\hat{\sigma}_i^2}{s_i}+\frac{\hat{\tau}^2}{R}\big)^2.
\end{align*}
Here $a\approx b$ means $a/b\stackrel{p}{\to}1$ as aforementioned. Therefore the variance can be expressed as
\begin{equation*}
\mathrm{Var}_*[\hat{\sigma}_{SVB}^2]=\frac{2}{B}\big(\sum_{i=1}^m\frac{\hat{\sigma}_i^2}{n_i}+\frac{\hat{\tau}^2\theta}{R}\big)^2(1+o_p(1))=\frac{2}{B}\big(\sum_{i=1}^m\frac{\sigma_i^2}{n_i}+\frac{\tau^2\theta}{R}\big)^2(1+o_p(1))
\end{equation*}
where the second equality holds because of the convergence of $\hat \sigma_i^2,\hat\tau^2$ to $\sigma_i^2,\tau^2$.\Halmos
\endproof

\proof{Proof of Theorem \ref{consis:anova}.}Under the choice of $B,R,\theta$ stated in the theorem, we have $\mathrm{Var}_*[\hat{\sigma}_{SVB}^2]=o_p(1/n^2)$ hence $\hat\sigma_{SVB}^2-\sigma_{SVB}^2=o_p(1/n)$ on one hand. On the other hand we know the subsampling bootstrap variance estimate $\sigma_{SVB}^2$ is consistent for $\sigma_I^2$ and $\sigma_I^2=\Theta(1/n)$ hence $\sigma_{SVB}^2-\sigma_I^2=o_p(1/n)$. Now $\hat\sigma_{SVB}^2-\sigma_I^2=\hat\sigma_{SVB}^2-\sigma_{SVB}^2+\sigma_{SVB}^2-\sigma_I^2=o_p(1/n)$ from which consistency immediately follows.\Halmos
\endproof

\proof{Proof of Theorem \ref{opt_allocation_detail:anova}.}One can easily verify that such $B^*$ and $R^*$ minimize the mean squared error \eqref{mse:anova sub} under the constraint that $BR=N$ and $B=\omega(1)$. The mean square error \eqref{opt_mse:anova} then follows from evaluating \eqref{mse:anova sub} at $B^*,R^*$.\Halmos
\endproof

\proof{Proof of Corollary \ref{consis:anova optimal}.}It is obvious that when $B=\omega(1)$ and $R=\Omega(\theta n)$ the configuration \eqref{parameter:psvb} is satisfied hence the estimate $\hat \sigma_{SVB}^2$ is relatively consistent under such allocation. To show that a simulation budget $N=\omega(\theta n)$ is necessary for \eqref{parameter:psvb} to hold, note that multiplying the first two requirements in \eqref{parameter:psvb} gives that $B^2R^2=\omega((\theta n)^2)$, hence $BR=\omega(\theta n)$ must hold true.\Halmos
\endproof

\proof{Proof of Corollary \ref{rough sub var}.}This follows from letting $\theta =\omega(1/n)$ in Corollary \ref{consis:anova optimal} so that the required simulation budget $N=\omega(\theta n)=\omega(\omega(1))=\omega(1)$.\Halmos
\endproof

\proof{Proof of Theorem \ref{sub ratio}.}The requirement $\omega(1/n)\leq \theta$ is stipulated by \eqref{parameter:psvb}. If $\theta \leq o(N/n)\wedge 1$, then we have $\theta n=o(N)$, or equivalently $N/(\theta n)=\omega(1)$, so that we can afford a $B=\omega(1)$ when $R=\Omega(\theta n)$ to satisfy the first two requirements of \eqref{parameter:psvb}. Theorem \ref{consis:anova} then guarantees consistent variance estimation.\Halmos
\endproof

\proof{Proof of Theorem \ref{opt_allocation:anova}.}It follows from Theorem \ref{opt_allocation_detail:anova} by observing that $\tau^2=\Theta(1)$ and $\sum_{i=1}^m\sigma_i^2/n_i=\Theta(1/n)$.\Halmos
\endproof

\section{Proofs for Results in Section \ref{sec:opt_subsample} and Theorem \ref{optimal_allocation}}\label{sec: overall error and subsample size}
In this section we analyze the statistical error of $\sigma_{SVB}^2$ relative to the true input variance $\sigma_I^2$, therefore, combined with the Monte Carlo error $\hat{\sigma}_{SVB}^2-\sigma_{SVB}^2$ given in Lemma \ref{MC_MSE:anova}, provide the overall error of the estimate $\hat{\sigma}_{SVB}^2$, and then minimize the overall error to obtain the optimal choices for the parameters $\theta,B,R$. We first prove Lemma \ref{PSBV:error} using Lemma \ref{error:key} and Proposition \ref{input_var:tight} which have been presented in Section \ref{sec: input variance approximation}, then use Lemma \ref{PSBV:error} to conclude Theorem \ref{overall_error}. Lastly, Theorem \ref{optimal_allocation} is derived from Theorem \ref{overall_error}.

\proof{Proof of Lemma \ref{PSBV:error}.}The proof of Theorem \ref{consis:AV} derives the following expression for the proportionate subsampled bootstrap variance
\begin{align*}
\frac{\sigma_{SVB}^2}{\theta}=\sum_{i=1}^m\frac{\hat{\sigma}_i^2}{s_i}+\mathrm{Var}_*[{\epsilon}^*]+2\mathrm{Cov}_*(\sum_{i=1}^m\frac{1}{s_i}\sum_{k=1}^{s_i} \hat{g}_{i}(X_{i,k}^*),{\epsilon}^*).
\end{align*}
As is the case in the proof of Proposition \ref{input_var:tight}, the covariances can be simplified to
\begin{align*}
\mathrm{Cov}_*(\sum_{i=1}^m\frac{1}{s_i}\sum_{k=1}^{s_i} \hat{g}_{i}(X_{i,k}^*),{\epsilon}^*)=\sum_{i=1}^m\mathbb E_*[\hat{g}_i(X_{i,1}^*)(\mathbb E_*[{\epsilon}^*\vert X_{i,1}^*]-\mathbb E_*[{\epsilon}^*])].
\end{align*}
This leads to
\begin{equation*}
\sigma_{SVB}^2=\sum_{i=1}^m\frac{\theta\hat{\sigma}_i^2}{\lfloor\theta n_i\rfloor}+\theta\mathbb E_*[({\epsilon}^*-\mathbb E_*[{\epsilon}^*])^2]+2\theta\sum_{i=1}^m\mathbb E_*[\hat{g}_i(X_{i,1}^*)(\mathbb E_*[{\epsilon}^*\vert X_{i,1}^*]-\mathbb E_*[{\epsilon}^*])].
\end{equation*}
From the above expression of the variance estimator one can verify that it suffices to show the following three results
\begin{align}
&\sum_{i=1}^m\frac{\hat{\sigma}_i^2}{n_i}=\sigma_I^2+\mathcal Z+o_p(\frac{1}{n^{3/2}})\label{normal-error}\\
&\mathbb E_*[({\epsilon}^*-\mathbb E_*[{\epsilon}^*])^2]=\sum_{i,i'=1}^m\frac{1}{4s_{i}s_{i'}}\mathrm{Var}[g_{ii'}(X_{i},X_{i'}')]+o_p(\frac{1}{s^2})\label{remainder-var}\\
&\mathbb E_*[\hat{g}_i(X_{i,1}^*)(\mathbb E_*[{\epsilon}^*\vert X_{i,1}^*]-\mathbb E_*[{\epsilon}^*])]\label{IF-remainder-cov}\\
\nonumber=&\frac{1}{2s_i^2}\mathrm{Cov}(g_i(X_{i}),g_{ii}(X_{i},X_{i}))+\sum_{i'=1}^m\frac{1}{2s_is_{i'}}\mathrm{Cov}(g_i(X_{i}),\mathbb E_{X'_{i'}}[g_{ii'i'}(X_{i},X'_{i'},X'_{i'})])+o_p(\frac{1}{s^2}).
\end{align}
To see this, if the three equations hold then
\begin{align*}
\sigma_{SVB}^2&=\sum_{i=1}^m\frac{\theta\hat{\sigma}_i^2}{\theta n_i-\mathrm{frac}(\theta n_i)}+\sum_{i,i'=1}^m\frac{1}{4n_{i}s_{i'}}\mathrm{Var}[g_{ii'}(X_{i},X_{i'}')]+o_p(\frac{\theta}{s^2})\\
&\hspace{3ex}+\sum_{i=1}^m\frac{1}{n_is_i}\mathrm{Cov}(g_i(X_{i}),g_{ii}(X_{i},X_{i}))+\sum_{i,i'=1}^m\frac{1}{n_is_{i'}}\mathrm{Cov}(g_i(X_{i}),\mathbb E_{X'_{i'}}[g_{ii'i'}(X_{i},X'_{i'},X'_{i'})])+o_p(\frac{\theta}{s^2})\\
&=\sum_{i=1}^m\frac{\hat{\sigma}_i^2}{n_i}+\sum_{i=1}^m\frac{\mathrm{frac}(\theta n_i)\sigma_i^2}{n_is_i}+o_p(\frac{1}{ns})+\sum_{i,i'=1}^m\frac{1}{4n_{i}s_{i'}}\mathrm{Var}[g_{ii'}(X_{i},X_{i'}')]\\
&\hspace{3ex}+\sum_{i=1}^m\frac{1}{n_is_i}\mathrm{Cov}(g_i(X_{i}),g_{ii}(X_{i},X_{i}))+\sum_{i,i'=1}^m\frac{1}{n_is_{i'}}\mathrm{Cov}(g_i(X_{i}),\mathbb E_{X'_{i'}}[g_{ii'i'}(X_{i},X'_{i'},X'_{i'})])+o_p(\frac{1}{ns})\\
&=\sum_{i=1}^m\frac{\hat{\sigma}_i^2}{n_i}+\mathcal R+o_p(\frac{1}{ns})\\
&=\sigma_I^2+\mathcal Z+\mathcal R+o_p(\frac{1}{ns})+o_p(\frac{1}{n^{3/2}})
\end{align*}
where \eqref{remainder-var} and \eqref{IF-remainder-cov} are used in the first equality and \eqref{normal-error} used in the last equality.

Now we prove the above three equations \eqref{normal-error}-\eqref{IF-remainder-cov}. By the expansion of $\hat g_i$ from Assumption \ref{3smoothness_empirical} and the vanishing moment condition on the remainder $\epsilon_g$, we write
\begin{align}
\nonumber\hat\sigma_i^2&=\frac{1}{n_i}\sum_{j=1}^{n_i}g_i^2(X_{i,j})+\frac{2}{n_i}\sum_{j=1}^{n_i}g_i(X_{i,j})\Big(\sum_{i'=1}^m\frac{1}{n_{i'}}\sum_{j'=1}^{n_{i'}}g_{ii'}(X_{i,j},X_{i',j'})+\frac{1}{n_i}\sum_{j'=1}^{n_i}g_i(X_{i,j'})\Big)\\
&\hspace{17.75ex}+\frac{1}{n_i}\sum_{j=1}^{n_i}\Big(\sum_{i'=1}^m\frac{1}{n_{i'}}\sum_{j'=1}^{n_{i'}}g_{ii'}(X_{i,j},X_{i',j'})+\frac{1}{n_i}\sum_{j'=1}^{n_i}g_i(X_{i,j'})\Big)^2+o_p(\frac{1}{\sqrt n})\label{negligible_term}\\
\nonumber&=\frac{1}{n_i}\sum_{j=1}^{n_i}g_i^2(X_{i,j})+\frac{2}{n_i}\sum_{j=1}^{n_i}g_i(X_{i,j})\sum_{i'=1}^m\frac{1}{n_{i'}}\sum_{j'=1}^{n_{i'}}g_{ii'}(X_{i,j},X_{i',j'})+2\Big(\frac{1}{n_i}\sum_{j=1}^{n_i}g_i(X_{i,j})\Big)^2+o_p(\frac{1}{\sqrt n})\\
&=\frac{1}{n_i}\sum_{j=1}^{n_i}g_i^2(X_{i,j})+2\sum_{i'=1}^m\frac{1}{n_in_{i'}}\sum_{j=1}^{n_i}\sum_{j'=1}^{n_{i'}}g_i(X_{i,j})g_{ii'}(X_{i,j},X_{i',j'})+o_p(\frac{1}{\sqrt n}).\label{sigma_hat}
\end{align}
Note that the first term in line \eqref{negligible_term} has an expectation of order $O(1/n)$ hence can be absorbed into the $o_p(1/\sqrt n)$ term. Similarly the fourth line \eqref{sigma_hat} holds because $(\sum_{j=1}^{n_i}g_i(X_{i,j})/n_i)^2$ has an expectation of order $O(1/n)$. The second term in \eqref{sigma_hat} is a sum of $m$ V-statistics, each of which by standard results is well approximated by the Hajek projection
\begin{align*}
\frac{1}{n_in_{i'}}\sum_{j=1}^{n_i}\sum_{j'=1}^{n_{i'}}g_i(X_{i,j})g_{ii'}(X_{i,j},X_{i',j'})=\frac{1}{n_{i'}}\sum_{j'=1}^{n_{i'}}\mathbb E_{X_i}[g_i(X_{i})g_{ii'}(X_{i},X_{i',j'})]+O_p(\frac{1}{n}).
\end{align*}
The finite fourth moment condition of $g_i$ and $g_{i_1i_2}$ are used to ensure that the product $g_i(X_{i,j})g_{ii'}(X_{i,j},X_{i',j'})$ has a finite second moment so that the above approximation holds. Denoting
\begin{align*}
\mu_{1}^i=\frac{1}{n_i}\sum_{j=1}^{n_i}g_i^2(X_{i,j}),\;\mu^{ii'}_2=\frac{1}{n_{i'}}\sum_{j'=1}^{n_{i'}}\mathbb E_{X_i}[g_i(X_{i})g_{ii'}(X_{i},X_{i',j'})]
\end{align*}
we have
\begin{align*}
\sum_{i=1}^m\frac{\hat\sigma_i^2}{n_i}&=\sum_{i=1}^m\frac{\mu_{1}^i}{n_i}+2\sum_{i=1}^m\sum_{i'=1}^m\frac{\mu^{ii'}_2}{n_i}+o_p(\frac{1}{n^{3/2}}).
\end{align*}
Because of independence among input models the variance of the leading term takes the additive form $\sum_{i=1}^m\lambda_i^T\Sigma_i\lambda_i/n_i$ as described in the theorem. By Proposition \ref{input_var:tight} $\sigma_I^2=\sum_{i=1}^m\sigma_i^2/n_i+O(n^{-2})$ hence equation \eqref{normal-error} follows. To show \eqref{remainder-var}, we note that in the cubic expansion of Assumption \ref{3smoothness_empirical} the cubic term and the remainder $\epsilon_3^*$ both have a second moment of order $O_p(s^{-3})$. Therefore it suffices to consider the quadratic term. Since the second order influence function $\hat g_{i_1i_2}$ has vanishing marginal expected value, one can verify that
\begin{align*}
\mathrm{Var}_*\big[\sum_{i,i'=1}^m\frac{1}{s_{i}s_{i'}}\sum_{j=1}^{s_i}\sum_{j'=1}^{s_{i'}}\hat g_{ii'}(X^*_{i,j},X^*_{i',j'})\big]&=\sum_{i,i'=1}^m\frac{1}{s_is_{i'}n_in_{i'}}\sum_{j=1}^{n_i}\sum_{j'=1}^{n_{i'}}\hat g^2_{ii'}(X_{i,j},X_{i',j'})+O_p(\frac{1}{s^3})\\
&=\sum_{i,i'=1}^m\frac{1}{s_is_{i'}n_in_{i'}}\sum_{j=1}^{n_i}\sum_{j'=1}^{n_{i'}}g^2_{ii'}(X_{i,j},X_{i',j'})+o_p(\frac{1}{s^2})
\end{align*}
where the second equality follows from the convergence of $\hat{g}_{ii'}$ to $g_{ii'}$ as imposed in Assumption \ref{3smoothness_empirical}. Equation \eqref{remainder-var} then follows from consistency of the V-statistic $\sum_{j=1}^{n_i}\sum_{j'=1}^{n_{i'}}g^2_{ii'}(X_{i,j},X_{i',j'})/(n_in_{i'})$.

Let's continue to prove equation \eqref{IF-remainder-cov}. Denote by $X_i^*$ a generic resampled data point from the $i$-th input data set. Then one can check that
\begin{align*}
&\mathbb E_*[{\epsilon}^*\vert X_{i,1}^*=X_{i,j}]-\mathbb E_*[{\epsilon}^*]\\
=&\frac{1}{2s_i^2}(\hat{g}_{ii}(X_{i,j},X_{i,j})-\mathbb E_*[\hat{g}_{ii}(X_{i}^*,X_{i}^*)])+\frac{1}{6s_i^3}(\hat{g}_{iii}(X_{i,j},X_{i,j},X_{i,j})-\mathbb E_*[\hat{g}_{iii}(X_{i}^*,X_{i}^*,X_{i}^*)])\\
&+\frac{s_i-1}{2s_i^3}\mathbb E_*[\hat{g}_{iii}(X_{i,j},X_{i}^*,X_{i}^*)]+\sum_{i'\neq i}\frac{1}{2s_is_{i'}}\mathbb E_*[\hat{g}_{ii'i'}(X_{i,j},X_{i'}^*,X_{i'}^*)]+\mathbb E_*[\epsilon_3^*\vert X_{i,1}^*=X_{i,j}]-\mathbb E_*[\epsilon_3^*].
\end{align*}
Note that $\mathrm{Var}_*[\epsilon_3^*\vert X_{i,1}^*]=o_p(s^{-4})$ because of Assumption \ref{3smoothness_empirical} and Lemma \ref{anova_decom}. Hence
\begin{align*}
&\mathbb E_*[\hat{g}_i(X_{i,1}^*)(\mathbb E_*[{\epsilon}^*\vert X_{i,1}^*]-\mathbb E_*[{\epsilon}^*])]\\
=&\frac{1}{n_i}\sum_{j=1}^{n_i}\hat g_i(X_{i,j})(\frac{1}{2s_i^2}(\hat{g}_{ii}(X_{i,j},X_{i,j})-\mathbb E_*[\hat{g}_{ii}(X_{i}^*,X_{i}^*)])+\sum_{i'=1}^m\frac{1}{2s_is_{i'}}\mathbb E_*[\hat{g}_{ii'i'}(X_{i,j},X_{i'}^*,X_{i'}^*)])+o_p(\frac{1}{s^2})\\
=&\frac{1}{2s_i^2}\mathrm{Cov}_*(\hat g_i(X^*_{i}),\hat{g}_{ii}(X^*_{i},X^*_{i}))+\sum_{i'=1}^m\frac{1}{2s_is_{i'}}\mathrm{Cov}_*(\hat g_i(X^*_{i}),\mathbb E_{X_{i'}^{*\prime}}[\hat{g}_{ii'i'}(X^*_{i},X_{i'}^{*\prime},X_{i'}^{*\prime})])+o_p(\frac{1}{s^2})\\
=&\frac{1}{2s_i^2}\mathrm{Cov}(g_i(X_{i}),g_{ii}(X_{i},X_{i}))+\sum_{i'=1}^m\frac{1}{2s_is_{i'}}\mathrm{Cov}(g_i(X_{i}),\mathbb E_{X'_{i'}}[g_{ii'i'}(X_{i},X'_{i'},X'_{i'})])+o_p(\frac{1}{s^2})
\end{align*}
where the $o_p(1/s^2)$ term in the first equality comes from applying Cauchy Schwartz inequality, and the last equality holds since convergence of $\hat g_i,\hat g_{i_1i_2},\hat g_{i_1i_2i_3}$ to $g_i,g_{i_1i_2},g_{i_1i_2i_3}$ in mean squared error implies
\begin{align*}
&\mathrm{Cov}_*(\hat g_i(X^*_{i}),\hat{g}_{ii}(X^*_{i},X^*_{i}))\stackrel{p}{\to}\mathrm{Cov}(g_i(X_{i}),g_{ii}(X_{i},X_{i}))\\
&\mathrm{Cov}_*(\hat g_i(X^*_{i}),\mathbb E_{X_{i'}^{*\prime}}[\hat{g}_{ii'i'}(X^*_{i},X_{i'}^{*\prime},X_{i'}^{*\prime})])\stackrel{p}{\to}\mathrm{Cov}(g_i(X_{i}),\mathbb E_{X'_{i'}}[g_{ii'i'}(X_{i},X'_{i'},X'_{i'})]).
\end{align*}
This gives rise to the equation \eqref{IF-remainder-cov}.\Halmos
\endproof

\proof{Proof of Theorems \ref{optimal_allocation} and \ref{overall_error}.}We first show Theorem \ref{overall_error}. Under a given subsampling ratio $\theta$, we know from Lemma \ref{PSBV:error} and Theorem \ref{opt_allocation:anova} that under the optimal allocation $B^*=N/R^*$ and $R^*=\Theta(\theta n)$
\begin{align*}
\hat\sigma_{SVB}^2-\sigma_{SVB}^2&=\mathcal E_1+o_p\big(\sqrt{\frac{\theta}{Nn}}\big)\\
\sigma_{SVB}^2-\sigma_{I}^2&=\mathcal E_2+o_p\big(\frac{1}{n^{3/2}}+\frac{1}{\theta n^2}\big)
\end{align*}
where the errors $\mathcal E_1,\mathcal E_2$ satisfy $\mathbb E_*[\mathcal E_1]=0,\mathbb E[\mathcal E_1^2]=\Theta(\theta/(Nn))$ and $\mathbb E[\mathcal E_2^2]=\mathcal R^2+\sum_{i=1}^m\lambda_i^T\Sigma_i\lambda_i/n_i$. Letting $\mathcal E=\mathcal E_1+\mathcal E_2$, we have $\mathbb E[\mathcal E^2]=\mathbb E[\mathcal E_1^2]+\mathbb E[\mathcal E_2^2]$ because
\begin{equation*}
\mathbb E[\mathcal E_1\mathcal E_2]=\mathbb E_{\text{data}}[\mathbb E_*[\mathcal E_1\mathcal E_2]]=\mathbb E_{\text{data}}[\mathcal E_2\mathbb E_*[\mathcal E_1]]=0.
\end{equation*}
This gives Theorem \ref{overall_error}.

To prove Theorem \ref{optimal_allocation}, note that if $\mathcal R=\Theta ((ns)^{-1})$, and at least one of the $\Sigma_i$'s are positive definite, then $\sum_{i=1}^m\lambda_i^T\Sigma_i\lambda_i/n_i=\Theta(1/n^3)$ hence $\mathbb E[\mathcal E_2^2]= \Theta(1/n^3+1/(\theta^2 n^4))$. We have
\begin{equation*}
\hat\sigma_{SVB}^2-\sigma_I^2=\mathcal E+o_p\big(\sqrt{\frac{\theta}{Nn}}+\frac{1}{n^{3/2}}+\frac{1}{\theta n^2}\big)
\end{equation*}
where $\mathbb E[\mathcal E^2]= \Theta(\theta/(Nn)+1/n^3+1/(\theta^2 n^4))$. To minimize the leading term $\mathcal E$, just note that $\theta/(Nn)+1/(\theta^2 n^4)$ is minimized at $\theta^*=(2N)^{1/3}/n$ resulting in $\mathbb E[\mathcal E^2]=\Theta(1/(N^{2/3}n^2)+1/n^3)$. When $N> n^{3/2}$, we have $1/(N^{2/3}n^2)< 1/n^3$, hence as long as $\theta^*$ is chosen such that $\theta^*/(Nn)\leq 1/n^3$ and $1/(\theta^{*2} n^4)\leq 1/n^3$, or equivalently $1/\sqrt{n}\leq \theta^*\leq N/n^2\wedge 1$, then the error $\mathbb E[\mathcal E^2]=\Theta(1/n^3)$. This leads to the optimal subsample size \eqref{opt_sub:PSBV}. If the depicted conditions do not hold, we have $\mathbb E[\mathcal E_2^2]\leq\Theta(1/n^3+1/(\theta^2 n^4))$ in general, hence all upper bounds we just obtained for $\mathbb E[\mathcal E^2]$ could be loose in order, leading to \eqref{min var}.\Halmos
\endproof

\section{Computer Network Configuration Details}\label{sec:network configuration details}
\begin{table}[h]
    \centering
    \begin{tabular}{|c|c|c|}
    \hline
       &channel capacity (bits)& transmission speed (miles per second)  \\\hline
         system \#1&$275000$&$200000$\\\hline
         system \#2&$200000$&$125000$\\\hline
         system \#3&$150000$&$100000$\\\hline
         system \#4&$200000$&$125000$\\\hline
    \end{tabular}
    \caption{Channel capacity and transmission speed configurations.}
    \label{network configurations}
\end{table}

\begin{table}[h]
\begin{center}
\renewcommand{\arraystretch}{0.75}
\begin{tabular}{|c|cccc|}
\hline
\diagbox[width=20mm, height=13mm]{node $i$}{node $j$}&1&2&3&4\\\hline
1&n.a.&50&40&45\\
2&60&n.a.&55&25\\
3&70&25&n.a.&30\\
4&35&40&50&n.a.\\\hline
\end{tabular}
\end{center}
\caption{True arrival rates $\lambda_{i,j}$ for system \#1.}
\label{network1: true rates}
\end{table}

\begin{table}[h]
\begin{center}
\renewcommand{\arraystretch}{0.75}
\begin{tabular}{|c|cccc|}
\hline
\diagbox[width=20mm, height=13mm]{node $i$}{node $j$}&1&2&3&4\\\hline
1&n.a.&35&25&30\\
2&45&n.a.&40&10\\
3&55&10&n.a.&15\\
4&20&25&35&n.a.\\\hline
\end{tabular}
\end{center}
\caption{True arrival rates $\lambda_{i,j}$ for system \#2.}
\label{network2: true rates}
\end{table}

\begin{table}[h]
\begin{center}
\renewcommand{\arraystretch}{0.75}
\begin{tabular}{|c|cccc|}
\hline
\diagbox[width=20mm, height=13mm]{node $i$}{node $j$}&1&2&3&4\\\hline
1&n.a.&20&50&15\\
2&30&n.a.&25&35\\
3&40&35&n.a.&40\\
4&45&50&20&n.a.\\\hline
\end{tabular}
\end{center}
\caption{True arrival rates $\lambda_{i,j}$ for system \#3.}
\label{network3: true rates}
\end{table}

\begin{table}[h]
\begin{center}
\renewcommand{\arraystretch}{0.75}
\begin{tabular}{|c|cccc|}
\hline
\diagbox[width=20mm, height=13mm]{node $i$}{node $j$}&1&2&3&4\\\hline
1&n.a.&70&60&20\\
2&25&n.a.&25&30\\
3&80&10&n.a.&10\\
4&50&60&20&n.a.\\\hline
\end{tabular}
\end{center}
\caption{True arrival rates $\lambda_{i,j}$ for system \#4.}
\label{network4: true rates}
\end{table}

\end{document}